\documentclass[12pt]{article}
\usepackage[body={17cm, 23cm, centered}]{geometry}
\usepackage{epsfig,verbatim,cite}
\usepackage{lmodern}
\usepackage{braket}
\usepackage[T1]{fontenc}
\usepackage{mathtext,marvosym,textcomp}
\usepackage{mathtools}
\usepackage{slashed}
\usepackage[usenames,dvipsnames,svgnames,table]{xcolor}

\usepackage{tikz}
\usetikzlibrary{
  arrows.meta,
  calc,
  decorations.pathreplacing,
  decorations.markings,
  positioning,
  patterns,
  fadings
}
\usepackage[compat=1.1.0]{tikz-feynman}

\usepackage[ normalem]{ulem}
\usepackage{tocloft}
\usepackage{epsfig,amsmath,amsfonts,amssymb,arydshln}
\usepackage[makeroom]{cancel}
\usepackage{multirow}

 \usepackage{relsize}

 \usepackage[debug,pageanchor=false]{hyperref}
\hypersetup{colorlinks=true,linktocpage,breaklinks,
            urlcolor=blue,
            linkcolor=red,
            citecolor=blue
            }

\numberwithin{equation}{section}

\def\a{\alpha} \def\b{\beta} \def\g{\gamma} \def\d{\delta} \def\e{\epsilon}
\def\ve{\varepsilon} \def\z{\zeta} \def\h{\eta} \def\q{\theta}
  \def\k{\kappa} \def\l{\lambda} \def\m{\mu}
\def\n{\nu} \def\x{\xi} \def\p{\pi}  \def\r{\rho}
 \def\s{\sigma} \def\t{\tau}  \def\f{\phi}
\def\vf{\varphi} \def\c{\chi} \def\y{\psi} \def\w{\omega}

\def\G{\Gamma} 
\def\D{\Delta} 
   
   \def\L{\Lambda} 
   
 \def\S{\Sigma}  
\def\F{\Phi}   \def\W{\Omega}

\def\fr{\frac}  \def\dt{\partial}

\def\ph{\phantom}
\def\mc{\mathcal}

\def\mF{\mathcal{F}}
\def\mG{\mathcal{G}}

\def\ta{\tilde{\a}}
\def\tb{\tilde{b}}

\def\Tr{\mbox{Tr}}

\def\Tr{\mbox{Tr}}

\def\RR{\mathbb{R}}

\def\SS{\mathbb{S}}

\newcommand\bqa {\begin{eqnarray}}
\newcommand\eqa {\end{eqnarray}}

\newcommand{\bear}{\begin{array}}
\newcommand{\enar}{\end{array}}

\newcommand{\ol}[1]{\overline{#1}}

\newcommand{\be}{\begin{equation}}
\newcommand{\ee}{\end{equation}}
\newcommand{\bea}{\begin{eqnarray}}
\newcommand{\eea}{\end{eqnarray}}

\def\rmE{\mathrm{E}}

\def\rmSO{\mathrm{SO}}

\def\rmSL{\mathrm{SL}}

\def\rmSU{\mathrm{SU}}

\def\frso{\mathfrak{so}}

\usepackage{tikz}
\usetikzlibrary{arrows}
\usetikzlibrary{shapes.misc}
\usepackage{tikz-cd}

\usetikzlibrary{arrows,shapes,positioning, fit}
\tikzstyle{every picture}+=[remember picture]
\tikzstyle{na} = [baseline=-.5ex]
\tikzstyle{format} = [rounded rectangle,
                      thick,
                      minimum size=1cm,
                      draw=red!00!black!80,
                      top color=white,
                      bottom color=red!00!black!00,
                      on grid]
\tikzstyle{format1} = [rectangle,
                      thick,
                      minimum size=1cm,
                      draw=red!00!black!80,
                      top color=white,
                      bottom color=red!00!black!00,
                      on grid]
\tikzstyle{format0} = [rounded rectangle,
					  thick,
				      minimum size=1.2cm,
					  draw=blue!00!black!80,
					  top color=white,
                      bottom color=blue!00!black!00,
                      text centered]
\tikzstyle{formatd} = [rounded rectangle,
					  thick,
					  dashed,
				      minimum size=1cm,
					  draw=blue!50!black!80,
					  top color=white,
                      bottom color=blue!00!black!00,
                      text centered]
\tikzstyle{format1d} = [rounded rectangle,
                      thick,
                      dashed,
                      minimum size=2cm,
					  draw=blue!50!black!80,
					  top color=white,
                      bottom color=blue!00!black!00,                  
                      text centered]
                      
\tikzset{cross/.style={cross out, draw=black, minimum size=2*(#1-\pgflinewidth), inner sep=0pt, outer sep=0pt},
	cross/.default={5pt}}
\usepackage[ normalem]{ulem}
\usepackage{tocloft}

\begin{document}
\renewcommand{\contentsname}{}
\renewcommand{\refname}{\begin{center}References\end{center}}
\renewcommand{\abstractname}{\begin{center}\footnotesize{\bf Abstract}\end{center}} 

\begin{titlepage}
\ph{preprint}

\vfill

\begin{center}
   \baselineskip=16pt
   {\large \bf BRANES
   }
   \vskip 2cm
    Edvard T. Musaev\footnote{\tt emusaev@theor.jinr.ru}
       \vskip .6cm
             \begin{small}
                          {\it
                          $^\dagger$Bogoliubov Laboratory of Theoretical Physics, Joint Institute for Nuclear Research,\\ 6, Joliot Curie, 141980 Dubna, Russia 
                          } \\ 
\end{small}
\end{center}

\vfill 
\begin{center} 
\textbf{Abstract}
\end{center} 
\begin{quote}
In this review branes of string theory are described from three different perspectives: as endpoints of open string, as supergravity backgrounds with BPS properties, as dynamical objects with gauge invariant actions. Based on these descriptions various effects of brane interactions are reviewed: brane bound states, Hanany--Witten and Myers effects, supertubes. The review is based on the lecture course given at MIPT.
\end{quote} 

\vfill
\setcounter{footnote}{0}
\end{titlepage}

\tableofcontents

\setcounter{page}{2}

\section{Introduction}

Despite the name the string theory is not restricted to only a theory of strings, i.e. of fundamentally one-dimensional objects. Quite the opposite, strings are only a subset of objects representing degrees of freedom particularly convenient for perturbative description of the theory. Speaking about supersymmetric theories where GSO projection can be consistently performed to remove tachyonic excitation from the string spectrum and the conformal quantum anomaly is absent, we find five distinct string theories: Type I, Type IIA/B and heterotic string with $\rmSO(32)$ or $\rmE_8\times \rmE_8$ gauge group. In addition to one-dimensional fundamental strings, which we will denote F1, one finds Dirichlet $p$-dimensional branes, usually denoted Dp-branes, Neveu-Schwarz 5-branes (NS5), Kaluza-Klein monopole (KK5) and infinitely many so-called exotic branes \cite{deBoer:2012ma}. While the fundamental string usually provides perturbative degrees of freedom  encoding dynamics of the theory, branes generate and support a background in a non-perturbative fashion. One of the well known examples of such applications of Dp-branes is the AdS/CFT holography first suggested in \cite{Maldacena:1997re} and then developed in more details in \cite{Witten:1998qj,Gubser:1998bc}. Given the amazing power of holography for investigation of the strong coupling dynamics of quantum field theories, huge amount of various reviews and books is available (we refer the reader to \cite{Ammon:2015wua,Aharony:1999ti,Beisert:2010jr}). In the most well understood example of the AdS${}_5\times \SS^5$ AdS/CFT correspondence one considers dynamics of the system of a stack of $N$ D3-branes with open string attached to it and closed string probing the background. The holographic duality is based on equivalence between two different descriptions of the same system. At small coupling $g_s N \ll 1$ dynamics of the system can be effectively described in terms of closed string living on a flat space $\RR^{1,9}$ and open string attached to the D3-branes. At low energy the open string spectrum is that of $\mc{N}=4$ $d=4$ $\rmSU(N)$ super Yang-Mills theory, while for the closed string one has $d=10$ Type IIB supergravity on a flat background. We will discuss this perspective in Section \ref{sec:endpoints} in greater details. On the other hand at large coupling $g_s N \ll 1$ and low energy the stack of D3-branes provides a non-trivial background for closed strings, that is a solution to Type IIB field equations. Near the horizon, i.e. close to the D3-brane, we have closed string theory on AdS${}_5\times \SS^5$ background, while far from the brane the background is flat. Moreover, at low energy closed string near the horizon do not interact with that far from the brane. We will discuss this perspective in Section \ref{sec:solutions}.

Therefor, one concludes that at different coupling the same system is most effectively described by two different sets of degrees of freedom:
\begin{itemize}
    \item $\mc{N} = 4$ $d=4$ $\rmSU(N)$ super Yang–Mills theory and type IIB supergravity on $\RR^{1,9}$;
    \item Type IIB supergravity on AdS${}_5\times \SS^5$ and type IIB supergravity on $\RR^{1,9}$.
\end{itemize}
Since these descriptions are equivalent and both contain type IIB supergravity on $\RR^{1,9}$ the other bits must also be equivalent, that brings us to the actual AdS/CFT correspondence: $\mc{N} = 4$ $d=4$ $\rmSU(N)$ super Yang–Mills theory is equivalent to Type IIB string theory on AdS$_5\times \SS^5$.

Another huge area where branes play important role is string compactifications and in particular string cosmological model building. As we discuss in more details in Section \ref{sec:actions} branes are dynamical objects that carry charge under $p$-form fields of supergravity. Under a dimensional reduction from ten to four dimensions Dp-branes are in principle allowed to wrap non-trivial cycles of the internal 6d manifold (say a CY$_3$) supporting its finite size and providing masses to (a subset of) scalar field of the theory, known as moduli fields. In a general compactification of string theory on a CY manifold one typically ends up with a huge number (hundreds and thousands) of massless scalar fields, that is sharply disagrees with experimental data. Wrapping branes along non-trivial (homological) cycles of the internal manifold one introduces additional parameters proportional to the amount of flux generated by the brane along the cycle. In the lower dimensional theory these parameters give masses to the moduli fields, and the whole process is called moduli stabilization. For a concise review the reader is referred to \cite{Grana:2005jc,Blumenhagen:2005mu,Buchmuller:2018jsq,Marchesano:2024gul}. In this approach crucial is that Dp-branes are not simply flat surfaces where open string ends travel or solutions to supergravity equations, but actually are dynamical objects carrying certain charge. In Section \ref{sec:endpoints} we will calculate charge of a Dp-brane based on its interaction with closed strings.

Branes may interact with each other in a non-trivial way changing the charge of the resulting configuration, creating new branes and even changing dimensionality of the initial branes. In Section \ref{sec:interact} we consider various interaction effects, that include the Hanany--Witten and Myers effects, brane bound states, supertubes and brane intersections. Apparently, such complicated brane configurations are equally demanded in various model building constructions as single brane states. They also find other applications, e.g. supertubes are important in building microstate black hole geometries and horizonless supergravity solutions, the Hanany--Witten setup provides microscopic description for various gauge theories on brane intersections. We discuss more application of the brane interaction effects considered in the review in Conclusions section \ref{sec:concl}.

In terms of the string coupling constant $g_s$ the string tension is of order one $T_{F1}\sim 1$, tension of D-branes $T_{Dp}\sim g_s{}^{-1}$, NS5-branes and the KK5-monopole have tension $T_5 \sim g_s^{-2}$ and exotic branes have tension proportional to even larger negative powers of the coupling constant. Hence, if string degrees of freedom have been chosen for the perturbation theory, branes become non-perturbative objects whose dynamics becomes really complicated and hard to analyze. Given that of great value are effective descriptions accessing brane dynamics from various points of views. Thus, Dp-branes from the point of view of the theory of open strings are the set of points, where strings end. In terms of closed strings these are surfaces charged under fields corresponding to certain states of the excited closed string, and hence can emit or absorb closed strings. Integrating out degrees of freedom of the open string leaving only dynamics of its end-points, one arrives at an effective description of a Dp-brane in terms of a Nambu--Goto-like action. Alternatively, in the low-energy description of string theory given by 10-dimensional supergravity fundamental branes manifest themselves as supersymmetric (extremal) black-brane solutions, generalizing the (extremal) Reissner--Nordstr\"om black hole solution.

The idea of this manuscript is to review properties of Dp-branes, NS5-banes and the KK5-monopole from three different perspectives:
\begin{enumerate}
    \item as endpoints of opens strings (D-branes) or magnetic duals of the fundamental string (NS5 and KK5);
    \item as dynamical objects defined by an effective action;
    \item as solutions to supergravity equations.
\end{enumerate}
The review aims at collecting together these approaches to branes of string theory in a single text and presenting the relations between them in uniform notations. Before turning to the fundamental objects of string theory let us illustrate these ideas by the simple case of a charged massive (and massless) particle. 

Since in a theory of particles only we cannot consider them as endpoints of anything, let us proceed with the second description, that is by an action. For a massive relativistic particle in a flat background the kinetic action reads
\begin{equation}
    S_0 = - m  \int ds = -m \int d\t \sqrt{\h_{\m\n}\dt_\t X^\m \dt_\t X^\n }.
\end{equation}
This is a Nambu--Goto-like action and it can be modified in two ways. First, it is possible to rewrite it in the Polyakov form by introducing an auxiliary world-volume  field $e=e(\t)$:
\begin{equation}
    S_0^P = \int d\t \left(e^{-1} \h_{\m\n}\dt_\t X^\m \dt_\t X^\n + m^2 e \right).
\end{equation}
The field $e(\t)$ has the meaning of the 1d vielbein and its equations of motion are algebraic and read
\begin{equation}
    -e^{-2} \h_{\m\n}\dt_\t X^\m \dt_\t X^\n + m^2 =0.
\end{equation}
Substituting this back into the action $S_0^P$ one recovers the Nambu--Goto action $S_0$. One advantage of the Polyakov action over the Nambu--Goto formulation of the particle dynamics is that the action is quadratic in fields, thta makes quantization of the theory  more straightforward. Second: in the Polyakov formulation one is able to turn to the massless case, that is impossible in the Nambu--Goto picture. The price for that is that equation of motion for $e(\t)$ does not give the vielbein in terms of $X^\m(\t)$ instead imposing the light-like condition on the velocity vector
\begin{equation}
    \h_{\m\n}\dt_\t X^\m \dt_\t X^\n=0.
\end{equation}

Another modification of the action for the particle action
is adding a term responsible for interaction with an external field, that in the minimal case is a vector $A_\mu$:
\begin{equation}
    S_{int}= \int d\t A_\mu \dt_\t X^\m = \int A.
\end{equation}
Here $A=A_\mu dX^\m$ is a 1-form defined in the whole space-time. Certainly the vector field can be dynamical and itself enter in an action of, say, Maxwell theory
\begin{equation}
    S_A = -\fr1{4 e}\int d^Dx F_{\m\n}F^{\m\n}, 
\end{equation}
where $F_{\m\n} = \dt_\m A_\n - \dt_\n A_\m$. Coordinate of a point in the $D-$dimensional space-time is denoted by $x^\m$ not to confuse with $X^\m$ that is an embedding of the particle's world--volume. 

For full generality one may also consider dynamics of the particle in a general curved space-time with a metric $g_{\m\n}$ (instead of $\h_{\m\n}$). As before this metric might be dynamical and ruled by the standard Einstein--Hilbert action
\begin{equation}
    S_g = \fr{1}{4\k}\int d^D x \sqrt{-g}R[g], 
\end{equation}
where $R[g]$ is the Ricci scalar for the metric $g_{\m\n}$.
Altogether, the full action defining dynamics of the particle and the vector and gravitational fields interacting with it is given by
\begin{equation}
    S_{full} = S_g + S_A + S_0 + S_{int}.
\end{equation}
Solving equations for $g_{\m\n}$ and $A_\m$ for a particular particle movement one obtains the gravitational and electro-magnetic fields created by such a particle. For example for the particle being at rest at the origin the result will be the familiar Reissner--Nordstrom solution. It is important to mention, that the whole system of equations for the metric, the vector field and the embedding function $X^\m$ can be solved consistently only in the extremal case when $e=m$. 

We observe, that a massive charged particle can be described by a background field configuration or by the dynamical action $S_0 + S_{int}$ depending on the purpose. For Dp-branes one adds one more description --- as endpoints of open strings.  Below we give more details for each description of fundamental objects of string theory,  highlight relations between them and use these descriptions to discuss details of brane interactions effects.

\section{D-branes as endpoints of open strings}
\label{sec:endpoints}

Spectrum of the closed string contains a massless spin-2 excitation that is normally associated with the graviton of the corresponding supergravity theory. In addition the massless sector contains excitations of an anti-symmetric tensor (Kalb--Ramond field) $B_{\m\n}$, the dilaton $\f$ and a set of $p$-form fields whose particular content depends on the type of supersymmetry. For the Type II string the metric, the dilaton and the Kalb--Ramond field are referred to as the NS-NS (Neveu--Schwarz) fields, while the other $p$-form fields belong to the so-called R-R (Ramond) sector. The fundamental string is electrically charged under the Kalb--Ramond field and hence carry the NS-NS charge. The bosonic action can be written as
\begin{equation}
    S_1 = T_1 \int d^2 \s \dt_\a X^\m \dt_\b X^\n (\sqrt{-h} h^{\a\b} g_{\m\n} + 2\pi \a' B_{\m\n}),
\end{equation}
where $h_{\a\b}$ is the world-sheet metric and $\s^\a$ label world-sheet coordinates. The constant $\a'$ determines the characteristic length $l_s$ of the string
\begin{equation}
    \a' = l_s^2.
\end{equation}
In the RNS formalism for the superstring one adds  superpartners $\y^\m$ for the world-volume scalars $X^\m$. In \cite{Polchinski:1995mt} it has been shown that the objects of string theory carrying R-R charges are Dirichlet (D-)branes of the open string theory: surfaces where open strings end. In this Section we focus at this viewpoint and consider Dp-branes as endpoints of open strings.

\subsection{Closed and open string spectrum}

In light-cone quantization states of the string spectrum belong to representations of the little group that for the supersymmetric string is $\rmSO(8)$. Let us briefly remind the results of this procedure and list reference equations without digging into much details. For a detailed description of various quantizations of the superstring the reader is referred to \cite{Blumenhagen:2013fgp}.

Start with the mass spectrum for the open string that is given by
\begin{equation}
    \a' M^2 = \sum_{n>0}\a_{-n}{}^i \a_n{}^i + \sum_{r>0} r b_{-r}{}^i b_r{}^i + \fr{1}{4\p^2 \a'} \D X^2 + a, 
\end{equation}
where $\D X^2 $ denotes square of the distance between the open string  ends in the Dirichlet directions. For left (L) and right (R) moving states of the closed string one has
\begin{equation}
    \begin{aligned}
         \a' M_L^2 = \sum_{n>0}\ta_{-n}{}^i \ta_n{}^i + \sum_{r>0} r \tb_{-r}{}^i \tb_r{}^i + a,\\
         \a' M_R^2 = \sum_{n>0}\a_{-n}{}^i \a_n{}^i + \sum_{r>0} r b_{-r}{}^i b_r{}^i + a,
    \end{aligned}
\end{equation}
The bosonic $\a_n^i$ and fermionic ladder operators satisfy the standard commutation relations
\begin{equation}
    \begin{aligned}[]
        [\a_{n}^i,\a^j{}_m] = n \d_{n+m}\d^{ij}, && \{b_r{}^i,b_s{}^j\} = \d_{r+s}\d^{ij}
    \end{aligned}
\end{equation}
and the same for $\ta_m{}^i$ and $\tb_r{}^i$. Note that since this is the light-cone quantization the target space indices $i,j,\dots = 1,\dots,8$. Depending on boundary (for the open string) and periodicity (for the closed string) conditions for the world-volume fermionic fields one distinguishes two sectors: Neveu--Schwarz (NS) and Ramond (R). Since we are dealing with a first quantized theory these sectors must be considered separately, however in the full quantum string (field) theory strings from different sectors may transform into each other via interactions (like photons and electrons do in the quantum field theory of electrodynamics). 

In the \textbf{NS} sector the ground state $|0\rangle_{NS}$ is a scalar in the target space and is defined as
\begin{equation}
    \a_m{}^i|0\rangle_{NS}= b_r{}^i|0\rangle_{NS}=0, \quad m,n,\dots = 1,2,\dots, \quad r,s,\dots = \fr12,\fr32,\dots.
\end{equation}
Note that fermionic ladder operators are labeled by half integers. In the \textbf{R} sector the ground state is a right $|a\rangle_R$ or a left $|\dot{a}\rangle_R$ spinor of $\rmSO(8)$ and is defined as
\begin{equation}
    \a_m{}^i|a\rangle_{R}=\a_m{}^i|\dot{a}\rangle_{R}= b_r{}^i|a\rangle_{R}=b_r{}^i|\dot{a}\rangle_{R}=0, \quad m,n,r,s\dots = 1,2,\dots.
\end{equation}
Value of the constant $a$, that comes from the normal ordering of operators in Virasoro generators, also differs in the NS and R sectors:
\begin{equation}
    \begin{aligned}
        a_{NS} & = -\fr{1}{2},\\
        a_{R} & = 0.
    \end{aligned}
\end{equation}

For the open string starting and ending on the same surface (D-brane or a stack of D-branes) $\D X^2 = 0$ one finds the tachyonic state $|0\rangle_{NS}$ whose mass squared is negative
\begin{equation}
    \a' M^2 |0\rangle_{NS} = a_{NS} |0\rangle_{NS} = -\fr12 |0\rangle_{NS}.
\end{equation}
This states get projected by the GSO projection together with an infinite number of other states leaving only states with $M^2 >0$ that fall into supersymmetry multiplets. In \cite{Gliozzi:1976jf} it has been shown that the projected states do not reappear in interactions. Hence, we are left with the following superstring spectrum at the massless level
\begin{equation}
    \begin{aligned}
        b_{-1/2}{}^i &|0\rangle_{NS}, && \mbox{transforms in }\mathbf{8}_v \mbox{ of SO(8)},\\
        &|a\rangle_{R}, &&\mbox{transforms in }\mathbf{8}_s \mbox{ of SO(8)}.
    \end{aligned}
\end{equation}
The left spinor state also gets projected. This is nothing but the on-shell spectrum of $D=10$ $\mc{N}=1$ supersymmetric Yang--Mills theory (SYM) and is interpreted as the low energy effective description of dynamics of the open string whose ends satisfy Neumann boundary conditions in the whole 10D space-time.  

If in $9-p$ directions Dirichlet boundary conditions are imposed on both string ends, the little group SO(8) get broken to $\rmSO(p-1)\times \rmSO(9-p)$. Since the open ends are allowed to have momentum only along Neumann directions, the little group is $\rmSO(p-1)$ while $\rmSO(9-p)$ is the internal symmetry group. For concreteness set $p=3$, then the little group is SO(2) the internal symmetry group is SO(6). The massless multiplet decomposes into a 4D vector, six scalars and four spinors (gaugino states), that is the spectrum of the $D=4$ $\mc{N}=4$ SYM. This is interpreted as the effective low energy description of open strings ending on D3-branes, or equivalently as the low energy description of the D3-brane dynamics. The last statement will be discussed in more details in Section \ref{sec:actions}.

Certainly, two ends of an open string are not necessarily restricted to end on the same Dp-brane. They may have different boundary conditions and correspond to intersecting branes. We will discuss such systems in more details in Section \ref{sec:interact}. In all these examples one always finds a vector excitation in the spectrum that corresponds to the vector field interacting with the open string ends. 

The closed Type II string spectrum is square that of the open string and its massless sector after GSO projection resembles that of excitations of 10D Type IIA/B supergravity. The resulting type A or B supersymmetry depends on whether one keeps the opposite or the same target space spinor chirality states. As before the tachyon states gets projected and one ends up with the following states
\begin{equation}
\label{eq:massless_spectrum}
    \begin{aligned}
        & IIA: &&  \tb_{-1/2}{}^i |0\rangle_{NS} \otimes b_{-1/2}{}^j |0\rangle_{NS}  && \in \mathbf{1\oplus 28} \oplus \mathbf{35}_v,\\
        & &&   |\dot{a}\rangle_{R} \otimes  |b\rangle_{R}  &&\in \mathbf{8}_v \oplus \mathbf{56}_v,\\
       & &&  |a\rangle_{R} \otimes b_{-1/2}{}^j |0\rangle_{NS}  &&\in \mathbf{8}_c\oplus \mathbf{56}_c,\\
       & &&  \tb_{-1/2}{}^i |0\rangle_{NS} \otimes |\dot{a}\rangle_{R}  &&\in \mathbf{8}_s \oplus \mathbf{56}_s, \\
       & IIB: &&  \tb_{-1/2}{}^i |0\rangle_{NS} \otimes b_{-1/2}{}^j |0\rangle_{NS}  && \in \mathbf{1\oplus 28} \oplus \mathbf{35}_v,\\
        & &&   |a\rangle_{R} \otimes  |b\rangle_{R}  &&\in  \mathbf{1 \oplus 28} \oplus \mathbf{35}_s,\\
       & &&  |a\rangle_{R} \otimes b_{-1/2}{}^j |0\rangle_{NS}  &&\in \mathbf{8}_c\oplus \mathbf{56}_c,\\
       & &&  \tb_{-1/2}{}^i |0\rangle_{NS} \otimes |a\rangle_{R}  &&\in \mathbf{8}_c \oplus \mathbf{56}_c.
    \end{aligned}
\end{equation}
These states are indeed the massless sector of the Type IIA and Type IIB supergravities, whose on-shell field content is given by
\begin{equation}
\label{eq:massless_fields}
    \begin{aligned}
        & IIA: && \f,\, b_{\m\n},\, g_{\m\n}, C_\m,\, C_{\m\n\r}, \l, \, \y_\m \\
        & IIA: && \f,\, b_{\m\n},\, g_{\m\n}, C, \, C_{\m\n},\, C_{\m\n\r\s}, \l^+_{1,2}, \, \y^+_{\m,1,2}.
    \end{aligned}
\end{equation}
Here $\l$ and $\y_\m$ are dilatino and gravitino respectively and the $+$ superscript denotes positive chirality, the 4-form gauge field of the Type IIB theory is restricted to have a self-dual field strength.

We now observe, that the 10D supergravity in addition to the metric has the 2-form NS-NS gauge field $b_{\m\n}$ that is precisely the Kalb--Ramond potential coupled to the fundamental string. Naturally to expect objects that couple to the R-R gauge fields that are 1- and 3-forms in IIA and 0-, 2- and 4-forms in IIB. These are nothing but Dp-branes and this fact can be seen in multiple ways: computing charge of Dirichlet surfaces, coupling effective action to 10D supergravity, looking at S-duality transformations and investigating the corresponding supergravity solutions. In the rest of this section tension and charge of Dirichlet surfaces will be calculated.

Before moving further it is worth to mention that according to the massless spectrum we so far encountered only D0-, D2-branes in Type IIA and D1-, D3-branes in Type IIB (also D(-1)-brane coupled to the R-R 0-form). Certainly, there are Dp-branes with $p$ running all the way up to $p=9$. To arrive at the remaining Dp-branes one has to perform electro-magnetic duality, that will be discussed in more details later.

\subsection{Coupling of Dirichlet surfaces to the supergravity fields}

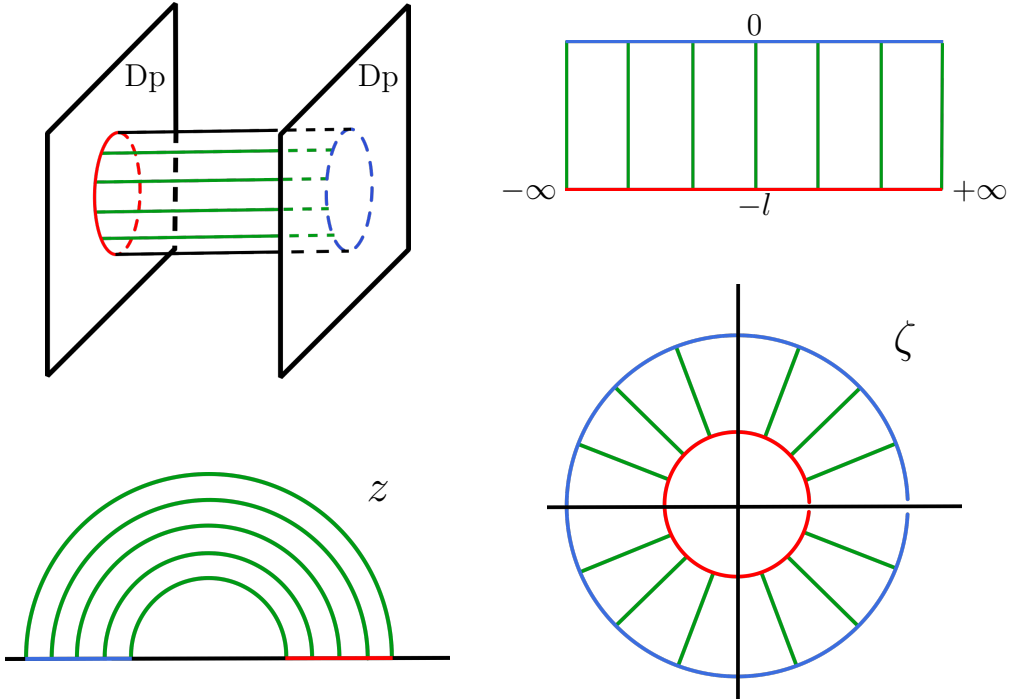
\begin{figure}
    \centering
    \begin{tikzpicture}
    \node at (0,0) (1) {\includegraphics[height=5cm]{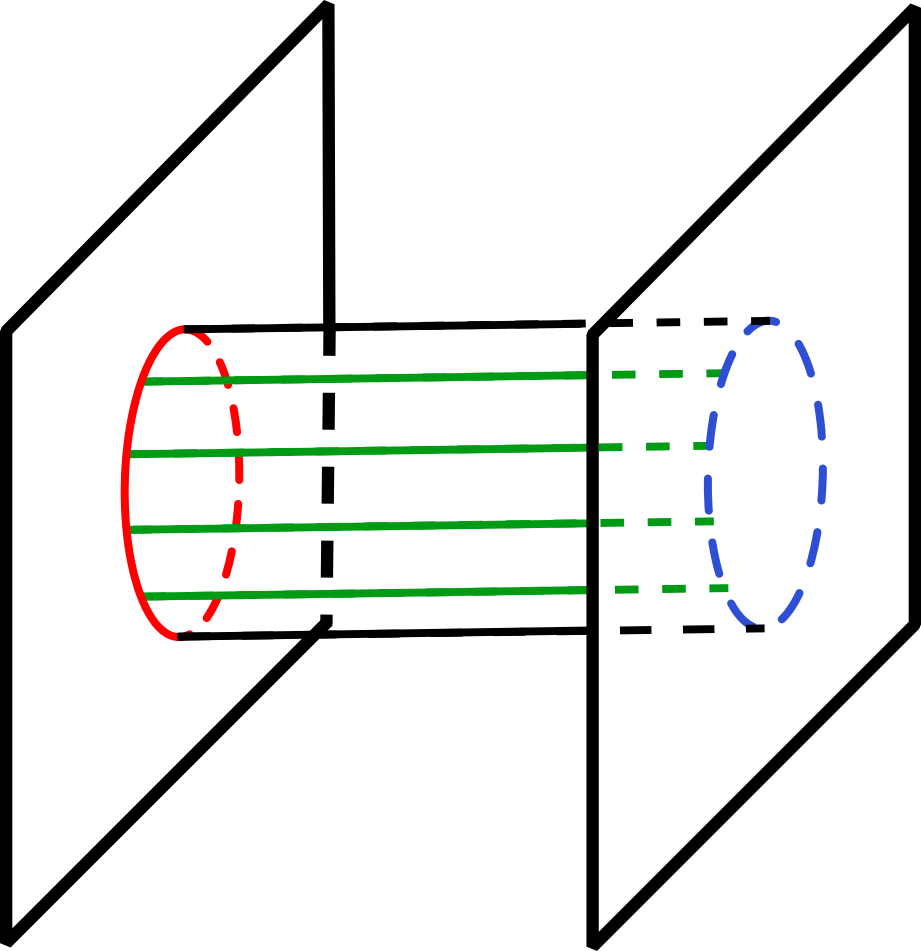} };
    \node at (-1.1,1.5) (dp) {Dp};
    \node at (2,1.5) (dp) {Dp};
    \node at (7,1) (2) {\includegraphics[height=2cm]{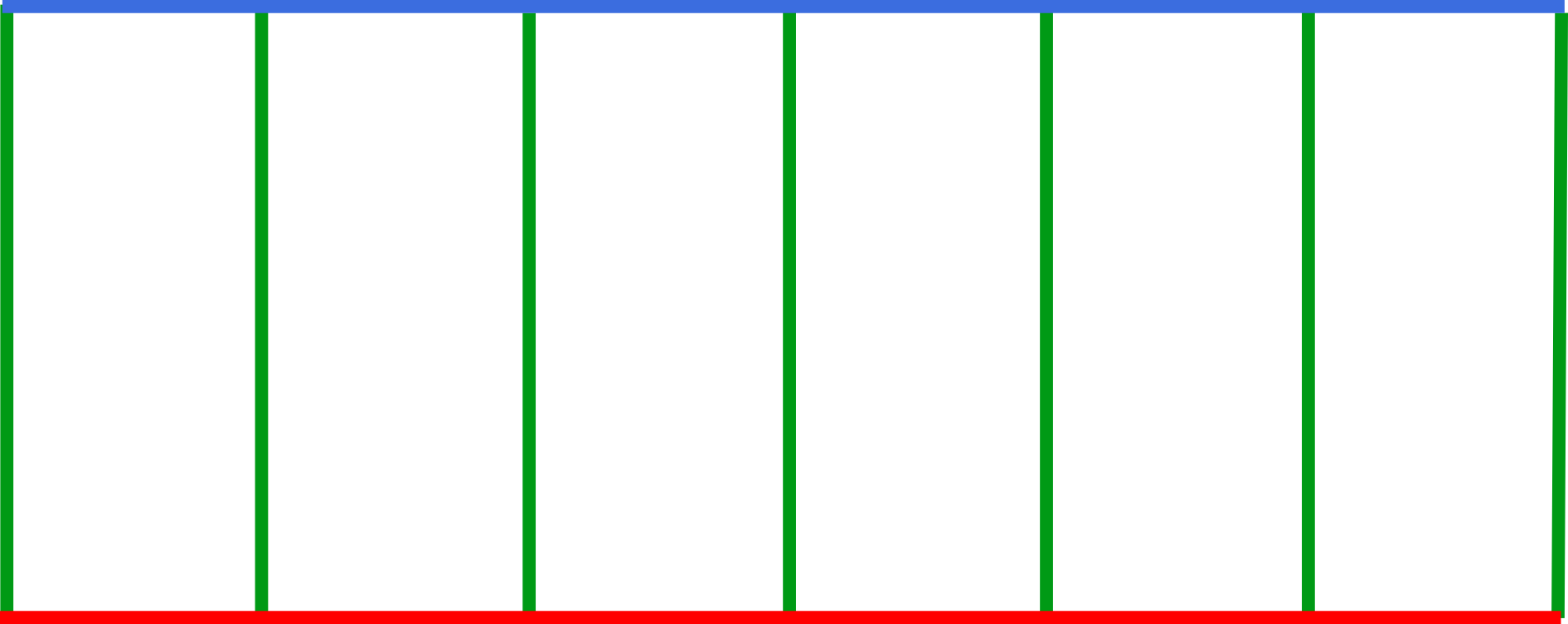}};
    \node at (4,0) (inf) {$-\infty$};
    \node at (10,0) (inf) {$+\infty$};
    \node at (7,2.2) (0) {$0$};
    \node at (7,-0.2) (l) {$-l$};
    \node at (7,-4) (4) {\includegraphics[height=5.5cm]{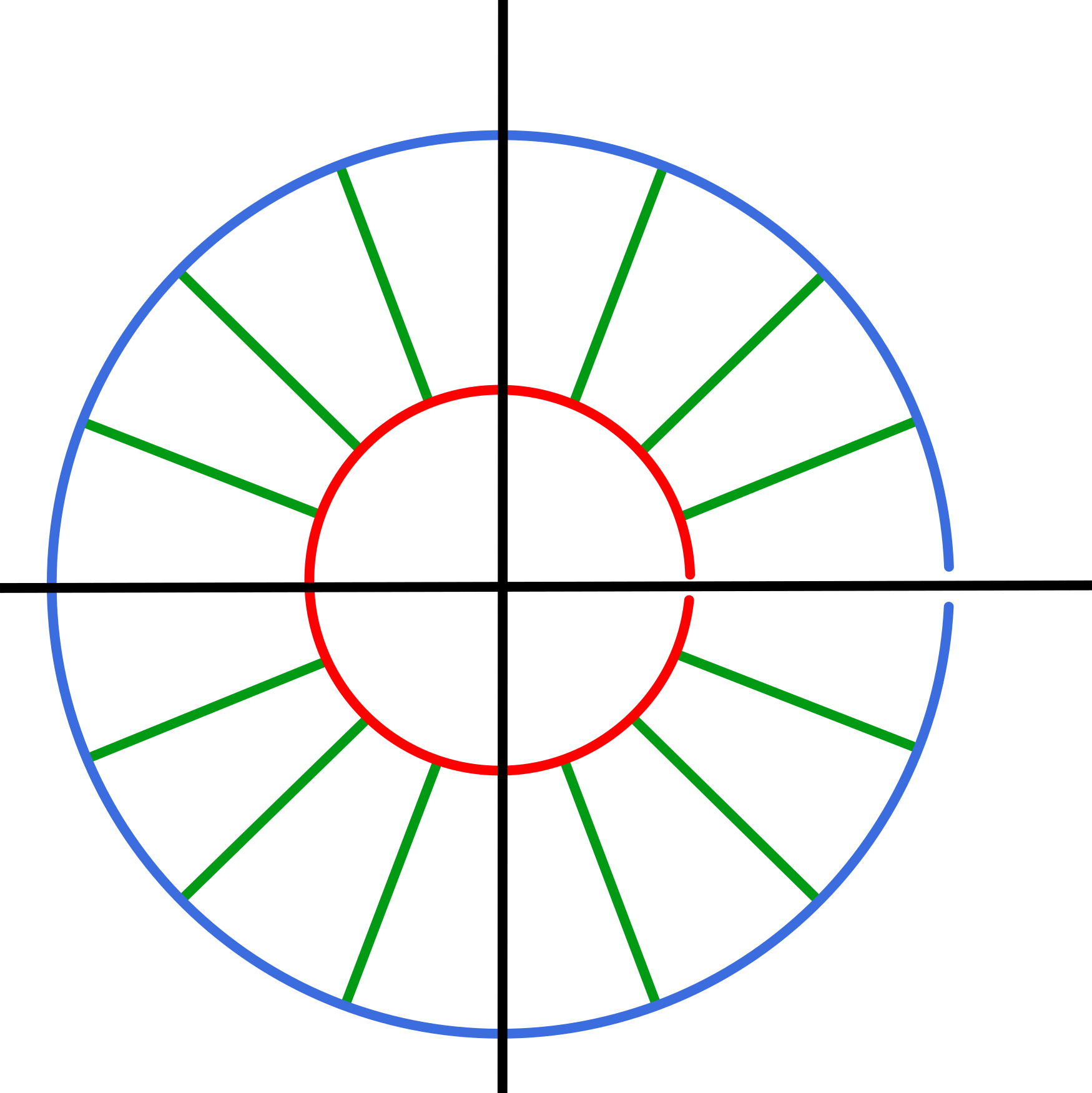}};
    \node at (9,-2) (zeta) {\Large$\zeta$};
    \node at (0,-5) (3) {\includegraphics[height=2.5cm]
    {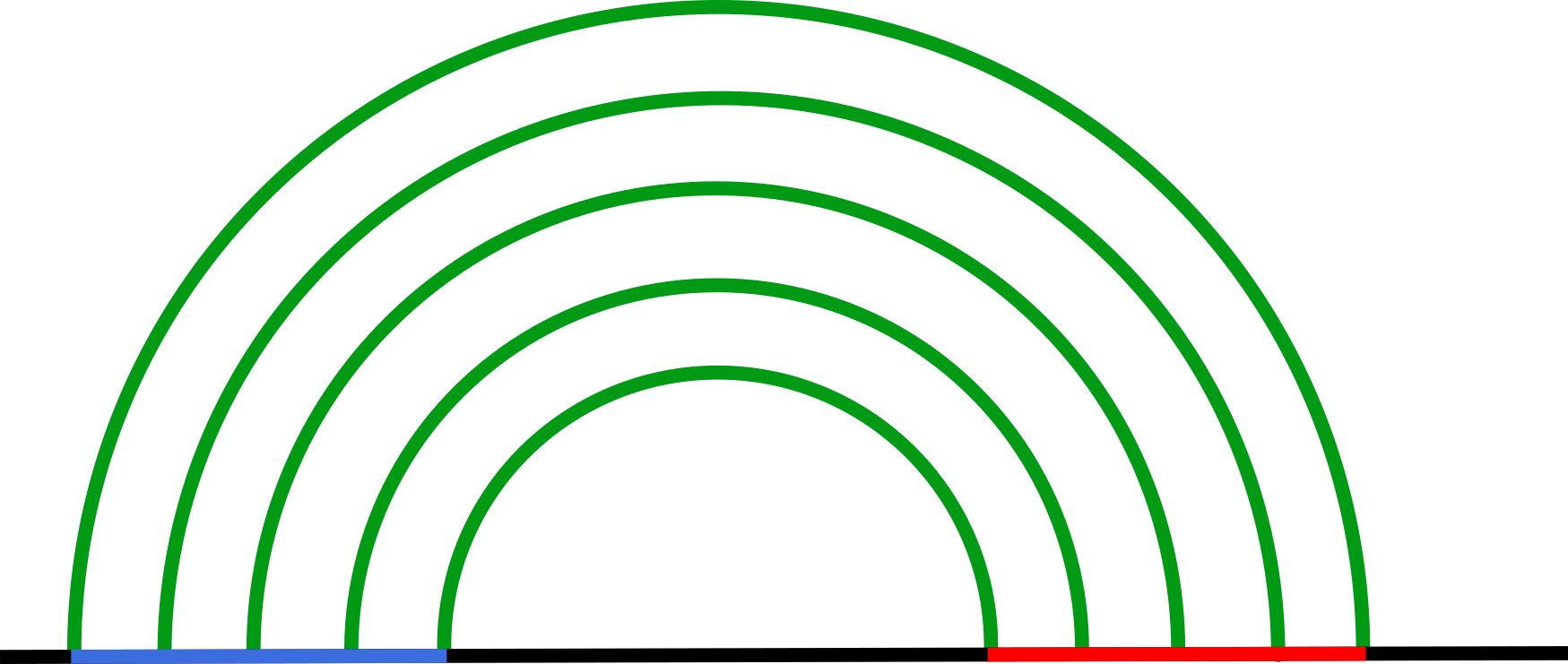} };
    \node at (2,-4) (z) {\Large$z$};
    \end{tikzpicture}
    \caption{Illustration of the open-closed string duality. For details see the text.}
    \label{fig:openclosed}
\end{figure}

From the point of view of the perturbative open string theory Dp-branes are surfaces of Dirichlet boundary conditions, meaning that they carry ends of open strings. The open-closed string duality states that the same world-sheet surface (upon the Wick rotation) can be understood from the two equivalent points of view: as an open or closed string diagram. In the former picture boundaries of the world-sheet lie on Dp-branes, in the latter these are identified with initial and final states of an open string. An example relevant to the present discussion is given by Fig. \ref{fig:openclosed} where three different mappings of the cylinder diagram are drawn. The cylinder can be understood as the 1-loop diagram in the theory of open strings stretched between two Dp-branes. In this case lines of constant $\t$ are drawn in green and boundaries of the string are drawn in blue and red. The cylinder can be mapped to an infinite strip of height $l$ if $\s \in [-l,0]$ parametrized by the complex coordinate $w = \t - i\s$. To turn to the radial quantization one maps the strip into a half annulus by
\begin{equation}
    z = e^{\p w/l}.
\end{equation}
In this case lines of constant $\t$ are given by semi-circles that end on boundaries of the open string. Gluing the boundaries gives back the cylinder and corresponds to taking trace in the partition function.

The map $\z = e^{2 \p i w/l}$ sends the strip into a full annulus cut open along the real axis. In this case the radial time runs along $\s$ and the red and blue circles correspond to initial and final states of the closed string. Physically this means that the 1-loop diagram of the open string between two Dp-branes can be understood as a tree-level propagation of a closed string state emitted by one brane and absorbed by another. Since at the massless level the closed string spectrum contains excitations of the metric and R-R fields we conclude that Dirichlet branes must be charged with respect to both. This implies that a Dp-brane has tension (coupling to the metric) and (on principle) a R-R charge.

To compute tension and R-R charge of a Dp-brane we need to evaluate the corresponding amplitude for which we need to know coupling of the Dp-brane to closed string states. There is an obvious subtlety here following from the fact that a Dp-brane is a non-perturbative object in the string perturbation theory. It follows directly from two observations. On the one hand in the quantum theory of open string Dp-branes appear as static objects, one the other hand they apparently must be dynamical, i.e. interact with and emit excitations of various closed string states receiving a back reaction. Therefore in the quantum mechanical theory of open/closed string it is not possible to directly derive coupling of Dp-branes and one has to use alternative approaches. In the closed string picture the amplitude must be given by and expression of the form
\begin{equation}
    \mc{A} = \langle \b | e^{-l H_{cl}} | \a \rangle,
\end{equation}
where $H_{cl} = L_0 + \bar{L}_0 + a$ is the Hamiltonian of the closed string and $|\a\rangle$ and $\langle \b|$ are closed string states emitted and absorbed by the branes. Since the evolution is defined by the closed string Hamiltonian these boundary states must be made of states in the closed string spectrum as
\begin{equation}
    | B \rangle = \sum_{I,\bar{J}}\a_{I,\bar{J}}|I\rangle \otimes |\bar{J}\rangle,
\end{equation}
where $I$ and $\bar{J}$ label the left and right moving closed string states. The coefficients $\a_{I,\bar{J}}$ measure the strength of coupling of a state $|I\rangle \otimes |\bar{J}\rangle$ to the Dp-brane and are nothing but the probability amplitudes for the brane to emit the state $|I\rangle \otimes |\bar{J}\rangle$.

Let us now construct the boundary state using the knowledge gained from considering the open/closed string duality. Clearly to define the boundary state one has to impose a condition on $|B\rangle$ and the only condition on the open string is the Dirichlet and Neumann boundary conditions. For the string end $\s=0$ these can be written as
\begin{equation}
    \begin{aligned}
        \mbox{Neumann:}&& \dt_\s X^{||} \big|_{\s=0} &= 0, && \y_+^{||}\big|_{\s=0} =\h \y_-^{||}\big|_{\s=0} && ,\\
       \mbox{Dirichlet:}&&  X^{\perp}\big|_{\s=0} &= x_0^{\perp}, && \y_+^{||}\big|_{\s=0} =-\h \y_-^{||}\big|_{\s=0} && ,
    \end{aligned}
\end{equation}
where $||$ and $\perp$ denote directions parallel and orthogonal to the Dp-brane respectively, $\y_{\pm}^{||}$ and $\y_{\pm}^{\perp}$ denote components of the fermionic fields and $\h=+1$ for the Ramond and $\n=-1$ for the NS sectors. 

Since the map $z \to \z$ from the open to the closed string effectively replaces $(\t,\s)\to (\s,\t)$ it is natural to define the boundary state as follows
\begin{equation}
    \begin{aligned}
        \mbox{Neumann:}&& \dt_\t X^{||} \big|_{\t=0} &= 0, && \y_+^{||}\big|_{\t=0} =\h \y_-^{||}\big|_{\t=0} && ,\\
       \mbox{Dirichlet:}&&  X^{\perp}\big|_{\t=0} &= x_0^{\perp}, && \y_+^{||}\big|_{\t=0} =-\h \y_-^{||}\big|_{\t=0} &&
    \end{aligned}
\end{equation}
where the subscript $closed$ emphasizes the fact that closed string operators are used. With the appropriate replacement $\s_{\pm} \to \pm \s_{\pm}$  in the oscillator expansion of the coordinate operators and of its superpartner we arrive at the following conditions (in the light-cone gauge)
\begin{equation}
    \begin{aligned}
        \left(\a_n{}^i + D^i{}_j \ta_{-n}{}^j \right)|Dp,\h\rangle & = 0,\\
        \left(b_n{}^i + i \h D^i{}_j \tb_{-n}{}^j \right)|Dp,\h\rangle & = 0,
    \end{aligned}
\end{equation}
where the block-diagonal matrix $D^i{}_j = (\mathbf{1}^{||},-\mathbf{1}^{\perp})$ has been introduced. The boundary state is now denoted as $|Dp,\h\rangle$ to emphasize that it describes the Dp-brane. It is pretty straightforward to arrive at the explicit form of $|Dp,\h\rangle$ in terms of the closed string oscillators. 

Consider first the simplest case of the set of only two bosonic oscillators $[\a, \a^\dagger]=1$ and $[\ta,\ta^\dagger]=1$. We are looking for a state that solves
\begin{equation}
    (\a \pm \ta^\dagger)|B\rangle =0.
\end{equation}
Recall that the state $|z\rangle = \exp\left(z \a^\dagger\right)|0\rangle$ is an eigenstate of the lowering operator: $\a |z\rangle = z|a\rangle$, where $z$ commutes with $\a$ and $\a^\dagger$. Then since the two sets of operators are independent it is easy to see that
\begin{equation}
    |B\rangle = N e^{\mp \ta^\dagger \a^\dagger}|0\rangle.
\end{equation}
Now it is straightforward to write the answer for the Dp-brane bound state that is
\begin{equation}
\label{eq:Dp_boundary}
    |Dp,\h\rangle = 
    \int dk_\perp \exp\left[- \sum_{n>0}\fr{1}{n}\a_{-n}{}^iD_{ij}\ta_{-n}{}^j - i \h  \sum_{r>0}b_{-r}{}^iD_{ij}\tb_{-r}{}^j\right]e^{i k_\perp x_0^{\perp}} |k_\perp\rangle,
\end{equation}
where $|k_\perp\rangle$ is simply the closed string ground state with momentum $k_\perp$ in the direction orthogonal to the Dp-brane. Certainly, in the R-R sector the ground state must satisfy a similar condition
\begin{equation}
    \left(b_0{}^\m + i \h D^\m{}_\n \tb_{0}{}^\n \right)|k_\perp,\h\rangle,
\end{equation}
that effectively halves the amount of allowed supersymmetries and restrict the R-R fields in the state. This is in consistency with the fact that a Dp-brane is a 1/2BPS object and hence carries only the Yang--Mills supermultiplet. This however is not the end of the story as also the GSO projection must be taken into account, i.e. the Dp-brane boundary state must be invariant under the GSO projection. The final result reads (for details of the calculation see \cite{Blumenhagen:2013fgp})
\begin{equation}
    |Dp\rangle = \fr{1}{\mathcal{N}}\Big(|Dp,1\rangle_{NS}-|Dp,-1\rangle_{NS} + i|Dp,1\rangle_{R}+ i |Dp,-1\rangle_{R}\Big),
\end{equation}
where the subscripts $NS$ and $R$ denote the NS-NS and R-R sectors of the strings. Note that the values $\h=+1$ and $\h=-1$ correspond to the NS and R sectors of the \textbf{open} string. The distinguishing between NS-NS and R-R sectors of the \textbf{closed} string appears here due to the GSO projection, that acts differently in different sectors. Technically this means that for $|Dp,\h\rangle_{NS}$ one takes $r$  half-integer and for $|Dp,\h\rangle_{R}$ the index $r \in \mathbb{Z}$.

The state $|k_\perp\langle$ in the expression \eqref{eq:Dp_boundary} is defined as a tensor product of ground states of the closed string in the NS-NS and R-R sectors. It is suggestive to analyze these separately in the massless sector to see which supergravity states can be emitted by a Dp-brane. Let us start with the NS-NS sector where $|k_\perp\rangle$ is defined as the corresponding vertex operator acting on the NS-NS ground state $|0\rangle_L\otimes|0\rangle_R$. Expanding the exponent and denoting $e^{i k_\perp x_0^\perp}|k_\perp\rangle = |x_0\rangle$ we have
\begin{equation}
    \begin{aligned}
    |Dp,\h \rangle & = |x_0\rangle + \left(- \sum_{n=1}^\infty \fr{1}{n}\a_{-n}{}^i D_{ij}\ta_{-n}{}^j - i \h \sum_{r=1/2}^{\infty}b_{-r}{^iD_{ij}\tb_{-r}{}^j}\right)|x_0\rangle + \dots \\
    & = |x_0\rangle  - i\h b_{-1/2}{}^i D_{ij}\tb_{-1/2}{}^j + \dots,
    \end{aligned}
\end{equation}
where in the last line we left only the linear contribution from the massless sector. The first term $|x_0\rangle$ is the tachyonic excitation, and the second term contains excitations of the dilaton $\f$, the metric $g_{ij}$ and of the Kalb--Ramond field. Contraction with $D_{ij}$ removes the latter and as a result the Dp-brane does not emit the B-field quanta, at least at the linear level. This is interpreted as the absence of the minimal coupling of a D-brane to the Kalb--Ramond field. The tachyonic excitation $|x_0\rangle$ is canceled in the combination $|Dp,1\rangle_{NS}-|Dp,-1\rangle_{NS}$.

The R-R sector is more tricky as its ground state is degenerate and transforms in a spinorial representation under the action of $b_0{}^\m$ and $\tb_0{}^\m$. Recall the gluing condition that is
\begin{equation}
    \left(b_0{}^\m + i \h D^\m{}_\n \tb_0{}^\n\right)|x_0\rangle=0.
\end{equation}
Such a state can be constructed by an action of $b_0^\m$ and $\tb_0{}^\m$  on a R-R ground state $|0\rangle$. The highest weight state is defined with respect to the action of raising and lowering fermionic oscillators. Since a Dirichlet brane breaks the full space-time Lorentz symmetry, to analyze representation structure of  emitted R-R states one must specify dimension of the brane and analyze Dirichlet and Neumann directions separately. For the sake of simplicity we stick to the case of the D0-brane, in which case symmetry of the transverse space is SO(8). 

Define raising and lowering operators of the corresponding Clifford algebra as usual
\begin{equation}
    \begin{aligned}
        b_A{}^{\pm} & = \fr{i}{\sqrt{2}}\left(b_0^{2A}\pm i b_0{}^{2A+1}\right),\\
        \tb_A{}^{\pm} & = \fr{i}{\sqrt{2}}\left(\tb_0^{2A}\pm i \tb_0{}^{2A+1}\right),
    \end{aligned}
\end{equation}
where $A=1,\dots,4$. Commutation relations are
\begin{equation}
    \{b_{A_1}{}^+,b_{A_2}{}^-\} = \d_{A_1A_2},
\end{equation}
and the same for $\tb_A{}^\pm$. The spinorial representations are constructed by acting by the lowering operators $b_A{}^-$ on the highest weight state defined as
\begin{equation}
    b_A{}^+|0\rangle_R = 0, \quad |0\rangle_R = \left|\fr12,\fr12,\fr12,\fr12\right\rangle, 
\end{equation}
where $1/2$ is the value of the spin projection for each $A=1,\dots 4$. Basis for the space of the spinorial representation $\mathbf{8}_s \oplus \mathbf{8}_c$ is then given by
\begin{equation}
    \begin{aligned}
        |0\rangle_R, && b_A{}^-|0\rangle_R \\
        b_{A_1}^-b_{A_2}^-|0\rangle_R, && b_{A_1}^-b_{A_2}^-b_{A_3}^-|0\rangle_R \\
        b_{A_1}^-b_{A_2}^-b_{A_2}^-b_{A_4}^-|0\rangle_R, &&  
    \end{aligned}
\end{equation}
Chirality of a state is simply given by the number of lowering operators acting on $|0\rangle_R$: states with the even number belong to $\mathbf{8}_s$, states with the odd number --- to $\mathbf{8}_c$. Certainly, the same construction has to be repeated for $\tb_A{}^\pm$.

The gluing condition can then be written in terms of raising and lowering operators as follows
\begin{equation}
    \left(b_A{}^\pm - i \h \tb_A{}^\pm\right)|D0,\h\rangle = 0.
\end{equation}
A straightforward check shows that in this case the state $|x_0\rangle_R$ is given by action of the corresponding vertex operator on
\begin{equation}
    \left(b_1{}^- - i \h \tb_1{}^-\right)\left(b_2{}^- - i \h \tb_2{}^-\right)\left(b_3{}^- - i \h \tb_3{}^-\right)\left(b_4{}^- - i \h \tb_4{}^-\right)|0\rangle_R.
\end{equation}
Since the D0-brane boundary state in the R-R sector is the symmetric combination of such states with $\h=\pm1$, only states with even powers of $\h$ survive, that gives
\begin{equation}
    \begin{aligned}
        &\e^{A_1A_2A_3A_4}\left(\tb_{A_1}{}^-\tb_{A_2}{}^-\tb_{A_3}{}^-\tb_{A_4}{}^- 
        - 6 b_{A_1}{}^-b_{A_2}{}^-\tb_{A_3}{}^-\tb_{A_4}{}^-
        + b_{A_1}{}^-b_{A_2}{}^-b_{A_3}{}^-b_{A_4}{}^-
        \right)|0\rangle_R.
    \end{aligned}
\end{equation}
This is precisely the singlet in the branching of $\mathbf{8}_c \otimes \mathbf{8}_c$. Therefore at the linear level we find that the D0-brane emits a R-R field that is scalar in the transverse directions. This will be true for any Dp-brane as we will explicitly see when analyzing supergravity solutions. 

In the remaining directions $0$ and $9$ the boundary conditions are mixed and hence we end up with the following gluing conditions
\begin{equation}
    \begin{aligned}
        \left(b_0{}^+ + i \tb_0{}^-\right)|x_0\rangle=0, \\
        \left(b_0{}^- + i \tb_0{}^+\right)|x_0\rangle=0,
    \end{aligned}
\end{equation}
where the raising and lowering operators are defined as
\begin{equation}
    \begin{aligned}
        b_0{}^\pm & = \fr{i}{\sqrt{2}}\left(b_0{}^0 \pm b_0{}^9\right),\\
        \tb_0{}^\pm & = \fr{i}{\sqrt{2}}\left(\tb_0{}^0 \pm \tb_0{}^9\right).
    \end{aligned}
\end{equation}
It is straightforward to see that in the $0-9$ directions the boundary state is given by
\begin{equation}
	\left(1 - i b_0{}^- \tb_0{}^-\right)|0\rangle_R,
\end{equation}
that belongs to the vector representation of $SO(1,1)$. Therefore we conclude that at linear level the D0-brane emits a R-R state that is vector along the Neumann directions and is scalar along the Dirichlet directions. Repeating the same calculations for other choices of $p$ one finds that the Dp-brane interacts with the R-R $C_{(p+1)}$--form and with the metric and the dilaton in the NS-NS sector. In the following Section we calculate strength of these interactions.


\subsection{Tension and R-R-charge of Dirichlet surfaces}

In previous section it has been shown that Dirichlet surfaces behave as sources emitting closed string states. However, no backreaction has been taken into account and the surfaces where assumed to be external currents realized as boundary states in the two-dimensional CFT. As we will see below this is natural in the string perturbation theory, where the string coupling constant $g_s$ is small, since the branes acquire infinite tension. In the full (yet unknown) quantum string field theory Dp-branes must be a dynamical object, that interacts with closed string states. Therefore one naturally expects the dynamics of a Dp-brane to be described by an action and restricting to only the massless level of the closed string states one may write 
\begin{equation}
    S_{Full} = S_{SUGRA} + S_{Dp}.
\end{equation}
This can be thought of as an approximation to dynamics of the Dp-brane interacting with closed strings where energies are low compared to $\a'$ (hence only the massless sector survives), while $g_s$ is not small. Now, calculating one particle exchange amplitude between two Dp-branes understood as objects charged with respect to supergravity fields, and comparing to the annulus diagram in the theory of open strings we will derive the value of the coupling.

Let us start with the first calculation in the NS-NS sector and restrict ourselves to the tree-level. Since the Dp-brane does not minimally couple to the Kalb--Ramond at the tree level the amplitude of interaction between two Dp-brane is given by the free propagator of the (transverse) graviton and the dilaton times the coupling constant. One must take into account that in the string frame, where the full action takes the form
\begin{equation}
\label{eq:sugra+brane}
    S_{Full} = \fr{1}{2 \tilde{\k}^2} \int d^{D}x \sqrt{-G} e^{-2\Phi} \big(R[G] + 4 \dt_\m \Phi \dt^\m \Phi \big) - T_p \int_{\S} d^{p+1} \x e^{-\Phi} \sqrt{-\G},
\end{equation}
the propagator will have terms non-diagonal in the basis of the dilaton and the graviton states. Here $\tilde{\k}$ denotes the gravitational coupling constant in D dimensions in the string frame, $T_p$ is tension of the Dp-brane, $\x^\a$ are coordinates on the world-volume $\S$ of the brane and $\G$ is its induced  volume. We will justify the factor $e^{-\Phi}$ later in Section \ref{sec:actions} when deriving the effective DBI action for a Dp-brane. To diagonalize the propagator we first expand the dilaton around its constant background value $\Phi = \Phi_0 + \phi$ and then rescale the metric as
\begin{equation}
    G_{\m\n} = e^{\fr{4\phi}{D-2}}g_{\m\n}.
\end{equation}
The gravitational part of the action takes the standard Einstein--Hilbert form and the full action becomes
\begin{equation}
    S_{Full} = \fr{e^{-2\Phi_0}}{2 \tilde{\k}^2} \int d^{D}x \sqrt{-g} \big(R[g] + 4 \dt_\m \phi \dt^\m \phi \big) - T_p e^{-\Phi_0} \int_{\S} d^{p+1} \x e^{\fr{4-D + 2p}{D-2}\f} \sqrt{-\g},
\end{equation}
where $\g$ is the world-volume metric induced by $g_{\m\n}$. In the leading order in field excitation $h_{\m\n}$ and $\vf$ defined as
\begin{equation}
    g_{\m\n} = \h_{\m\n} + 2 \tilde{\k} e^{\Phi_0}h_{\m\n} , \quad \f = \fr12\tilde{\k}\sqrt{D-2}e^{\Phi_0}\vf
\end{equation}
the action takes the following form
\begin{equation}
\label{eq:brane_linear}
    S_{Full} \approx S^{(0)} + \int d^Dx \left(\fr{1}{2}h^{\m\n}\Box h_{\m\n} - \fr14 h \Box h + \fr12 \vf \Box \vf\right) + T_p \tilde{\k}\int d^{p+1}\x \left(\h^{\a\b}h_{\a\b} + \fr{4-D+2p}{2\sqrt{D-2}}\vf\right).
\end{equation}
Here $S^{(0)}$ is the constant shift of the action coming from the determinant expansion of $S_{br}$ and after integrating by parts in the Einstein--Hilbert action, $h = \h^{\m\n}h_{\m\n}$ denotes trace of the graviton field. In addition to the standard gauge-fixing in the gravitational term we impose the following gauge fixing in the Dp-brane action
\begin{equation}
    X^\m = \big(X^\a(\x), X^m\big), \quad X^m=\mathrm{const}.
\end{equation}
The second requirement means that we do not take into account back reaction of the radiation to the brane. This is needed to compare the result to the open string calculation where Dp-branes are static.

The second integral in the linearized action \eqref{eq:brane_linear} encodes interaction of the graviton and the dilaton with the external classical current
\begin{equation}
    \begin{aligned}
        J^{\a\b}(x,z) = T_p \tilde{\k} \h^{\a\b}\int d^{p+1}\x \d^{\perp}(x - z(\x)),
    \end{aligned}
\end{equation}
where $z^\m=z^\m(\x)$ is the classical trajectory of the Dp-brane and $x^\m$ denotes coordinates of the point where the current is measured. The vacuum amplitude in the presence of the classical current is precisely the one-particle exchange amplitude (at the tree level). It is given by
\begin{equation}
    \tilde{\mc{A}}^{eff}_{Dp,Dp} = \int d^{D}x d^{D}y J^{\a\b}(x,z) J^{\g\d}(y,z') \left(G_{\a\b,\g\d}(x-y) + \fr{(4-D+2p)^2}{4(D-2)(p+1)^2}\h_{\a\b}\h_{\g\d}G(x-y)\right),
\end{equation}
where $G_{\a\b,\g\d}(x-y)$ and $G(x-y)$ are the free graviton and dilaton propagators respectively. These are given by
\begin{equation}
    \begin{aligned}
        \D_{\m\n,\r\s}(p) & = \fr{1}{2 p^2}\left(\h_{\m\r}\h_{\n\s}+\h_{\n\r}\h_{\m\s} - \fr{2}{D-2}\h_{\m\n}\h_{\r\s}\right) = \int d^{10} x e^{-i p x} G_{\m\n,\r\s}(x), \\
        \D(p) & =  \fr{1}{p^2} = \int d^D x e^{-i p x} G(x).
    \end{aligned}
\end{equation}
Using these expressions we obtain the following result for the amplitude of the one-particle exchange between two Dp-branes placed at $X_A$ and $X_B$
\begin{equation}
\label{eq:1-loop_grav}
    \tilde{\mc{A}}^{eff}_{Dp,Dp} = \fr{D-2}{4}\big(T_p \tilde{\k}\big)^2 V_{p+1}G^{\perp}(X_A-X_B).
\end{equation}
Here $V_{p+1}$ denotes world-volume of the Dp-brane and 
\begin{equation}
    G^{\perp}(x-y) = \int \fr{d^{D-p-1}p}{(2\p)^{D-p-1}}\fr{e^{i p_{\perp}(x-y)^{\perp}}}{p_{\perp}{}^2}
\end{equation}
is the transverse Green's function.

This result is then to be compared to the amplitude of a closed string exchange between two Dirichlet surfaces in the $\a'\to 0$ limit, that keeps only the massless states. Technically, what one has to do is to calculate the cylinder partition function in the closed string theory that is by the open-closed string duality is the same as the one-loop partition function for the open string (see Fig. \ref{fig:openclosed}). To calculate the latter we will do the following trick. Consider the 1-loop vacuum amplitude $\mc{A} = \log Z_{vac} = -1/2 \Tr \log (\Box + M^2)$ for a particle of mass $M$ living in $D$ dimensions and rewrite it as follows
\begin{equation}
    \begin{aligned}
        -\fr{1}{2}\Tr\log (\Box + M^2) = V_D \int \fr{d^D k}{(2\p)^D}\int_0^\infty \fr{dt}{2t}e^{-\fr12(k^2 +M^2)t}.
    \end{aligned}
\end{equation}
The expression $1/2(k^2 +M^2)$ is nothing but the Hamiltonian of the theory and when turning to the theory of open string this is to be replaced by $L_0/\a'$. Therefore to arrive at the open string 1-loop amplitude one simply uses the whole mass spectrum
\begin{equation}
    \a' M^2 = \sum_{n=1}^\infty\a^i_{-n}\a_n^i - 1 + \fr{\D X^2}{4\p^2 \a'}
\end{equation}
and sums over all states. Here we are working in the light-cone gauge from the very beginning and $\D X^2 = (X_A-X_B)^2$ denotes the distance between two Dirichlet surfaces. First we are working in the bosonic string theory, therefore $i=1,\dots,24$. The reason is that after the GSO projection in the superstring case contributions from the NS-NS and the R-R sectors precisely cancel each other meaning that two Dp-branes do not interact, as BPS objects indeed should. Since each oscillatory state for the open string with $26-p-1$ Dirichlet boundary conditions can have momentum in $p+1$ directions, the 1-loop amplitude takes the following form
\begin{equation}
    \mc{A}_{Dp,Dp}^{eff} =2 V_{p+1} \int \fr{d^{p+1} k}{(2\p)^{p+1}}\int_0^\infty \fr{dt}{2t}\sum_{I}e^{-2 \p\a'(k^2 +M_I^2)t},
\end{equation}
where the sum is performed over all states in the spectrum. The extra $4\p\a'$ factor in the exponents comes from the definition of the zero level Virasoro operators. The overall factor of two comes from taking into account strings moving in both directions.

Denoting $q = e^{-2\p t}$ with $t\in \mathbb{R}$ we perform summation over the oscillatory modes in one direction as follows
\begin{equation}
    \Tr q^{L_0} = \Tr q^{\sum_{n=1}^\infty\a_{-n}\a} = \prod_{n=1}^{\infty}(1-q^n)^{-1},
\end{equation}
and the same for all 24 directions. The function
\begin{equation}
    \h(q) = q^{\fr1{24}}\prod_{n=1}^{\infty}(1-q^n)^{-1}
\end{equation}
is the first Dedekind's function. This gives the following expression for the amplitude
\begin{equation}
    \mc{A}_{Dp,Dp}^{eff} = 2 V_{p+1}\int_0^\infty \fr{dt}{2t} \big(8 \p^2 \a' t \big)^{-\fr{p+1}{2}}\h(q)^{-24} e^{-\fr{\D X^2t}{2\p\a'}}
\end{equation}
By calculation this result contains contributions from all modes, while to compare to \eqref{eq:1-loop_grav} we need to keep only the contribution from the massless sector. Expanding the Dedekind's function we get
\begin{equation}
    \mc{A}_{Dp,Dp}^{eff} = 2 V_{p+1}\int_0^\infty \fr{dt}{2t} \big(8 \p^2 \a' t \big)^{-\fr{p+1}{2}}\left(e^{\fr{2\p}{t}} + 24 + \dots\right) e^{-\fr{\D X^2t}{2\p\a'}},
\end{equation}
where the second term in the parentheses corresponds precisely to the massless pole. The first term is the tachyon divergence as its existence indicates that the theory is pathological, or more concretely --- the vacuum is unstable. We will drop it for a while and comment on that later. Hence, the result is
\begin{equation}
\label{eq:1-loop_string}
    \mc{A}_{Dp,Dp}^{eff} \simeq V_{p+1} \fr{24 \p}{2^{10}}\big(4\p^2 \a'\big)^{11-p}G_{\perp}(\D X).
\end{equation}
Comparing to \eqref{eq:1-loop_grav} with $D=26$ we obtain the following expression for the coupling between a $p+1$-dimensional  Dirichlet surface and closed string
\begin{equation}
    T_p = \fr{\sqrt{\p}}{16 \tilde{\k}}\big(4\p^2 \a'\big)^{11-p}.
\end{equation}
Going back to the action \eqref{eq:sugra+brane} we see that the effective brane coupling is
\begin{equation}
    \t_p = T_p e^{-\F_0}.
\end{equation}
On the other hand $e^{\F_0} = g_s$ is the coupling of the closed string perturbation theory responsible for ranging contributions from amplitudes of different genera. Therefore, when $g_s <<1 $, i.e. when the string perturbation theory is valid, Dirichlet surfaces have effectively large tension and back reaction can be omitted. This is precisely one one does when replacing dynamical Dp-branes by static Dirichlet boundary conditions.

Let us now turn to the superstring case, where at the massless level in the R-R sector one additionally has $p$-forms. The effective field theory action is that of $D=10$ Type IIA or IIB supergravity and can be schematically written as
\begin{equation}
    S_{II} = \fr{1}{2 \tilde{\k}^2} \int d^{D}x \sqrt{-G} \Big[e^{-2\Phi} \big(R[G] + 4 \dt_\m \Phi \dt^\m \Phi  + \fr{1}{12}H^2\big)+ \sum_{p}F_p^2\Big],
\end{equation}
where $H=dB$ and $F_p$ denote field strengths for the Kalb--Ramond and R-R potentials respectively and the sum goes over even $p$ in the IIA case and odd $p$ in the IIB case. Interaction of the Dp-brane with R-R potentials naturally starts with the minimal term
\begin{equation}
    \m_p \int_{\S_p+1} C_{p+1},
\end{equation}
 and has to be extended by various terms to ensure gauge invariance. We will return to that in Section \ref{sec:actions}. 

 For the Dp-Dp 1-particle exchange amplitude one follows the same steps as before but with additional summation over oscillators in the Ramond sector
 \begin{equation}
     \mc{A}_{Dp,Dp}^{eff} = \int_0^\infty \fr{dt}{2t}\Tr\left[\fr{1+(-1)^F}{2}e^{-2\p L_0 t}\right],
 \end{equation}
where $F$ counts the number of fermionic oscillators. Performing the same calculation, now in 10 dimensions, one obtains the following result
\begin{equation}
    \mc{A}_{Dp,Dp}^{eff} = 2 V_{p+1}\int_0^\infty \fr{dt}{2t} \big(8 \p^2 \a' t \big)^{-\fr{p+1}{2}}e^{-\fr{\D X^2t}{2\p\a'}}\fr12f_1(q)^{-8}\left(-f_2(q)^8 + f_3(q)^8 - f_4(q)^8\right),
\end{equation}
where 
\begin{equation}
    \begin{aligned}
        f_1(q)&=q^{\fr1{24}}\prod_{n=1}^{\infty}(1-q^n)\equiv \h(q), \\
        f_2(q)&=\sqrt{2}q^{\fr1{24}}\prod_{n=1}^{\infty}(1+q^n),\\
        f_3(q)&=q^{-\fr1{48}}\prod_{n=1}^{\infty}(1+q^{n-\fr12}),\\
        f_4(q)&=q^{-\fr1{48}}\prod_{n=1}^{\infty}(1-q^{n-\fr12}).
    \end{aligned}
\end{equation}
By Jacobi's abstruse identity the three terms in the parentheses sum up identically to zero. This means, first, that two Dp-branes do not interact with each other, that is naturally related to their BPS properties. Second, what we observe is actually a precise cancellation between NS-NS and R-R contributions in the closed string cylinder partition function, meaning that the R-R charge $\m_p$ is precisely equal to the brane's tension $  \m_p = T_p.$

The non-trivial coupling of Dp-branes to closed string states has two obvious consequences. First, Dp-branes experience back reaction when emitting closed strings and therefore have to be described as dynamical objects, i.e. by an action. Second, beyond the string perturbation theory but at low energies, when the closed string massless sector is described by $D=10$ supergravity,  Dp-branes must act as a source of a non-trivial metric and R-R fields. These two effects will be discussed in details in Sections \ref{sec:actions} and \ref{sec:solutions} respectively.

\section{Branes as dynamical objects}
\label{sec:actions}

As it has been discussed above Dp-branes understood as surfaces of Dirichlet boundary conditions for open strings possess an effective tension $T_p$ and R-R charge $\m_p=T_p$ resulting in non-trivial interaction with closed string states. The non-vanishing tension makes Dp-branes dynamical objects requiring an effective action for their description (in contrast to orientifold planes, that only have a R-R charge). The most straightforward way to derive the Dp-brane effective action is to start with the partition function for the open string theory and integrate out dynamics of the string body leaving only the quantum effective action for the open ends. This approach has been used in \cite{Fradkin:1985qd} where it has been shown that the resulting action is nothing but the DBI action (Dirac--Born--Infeld). This action has been suggested in earlier works by Born and Infeld \cite{Born:1934gh} and by Dirac \cite{Dirac:1962iy} to deal with the self-interaction of a charged particle in electrodynamics.

Here we will follow a different approach advocated in \cite{Abouelsaood:1986gd}, that uses the same construction as the one that relates D=10 supergravity equations to the closed string partition function. For the latter one starts with the standard Polyakov action
\begin{equation}
    S_1 = T_1 \int d^2 \s \dt_\a X^\m \dt_\b X^\n (\sqrt{-h} h^{\a\b} g_{\m\n} + 2\pi \a' B_{\m\n}),
\end{equation}
and considers the fields $G_{\m\n}$, $B_{\m\n}$ and $\f$ as couplings for $X^\m$. Requiring vanishing of the beta function for these couplings gives equations of motion for Type II supergravity. The interpretation of this result is as follows. On one hand we know that perturbatively the closed string has the graviton, Kalb--Ramond field and the dilaton excitations in its spectrum. Therefore it interacts with a classical background given by non-trivial profiles of these fields. Finally, consistency of quantum formulation of the closed string theory requires this classical background to satisfy known D=10 supergravity equations. Since the gravity, Kalb--Ramond field and the dilaton are the lowest mass excitations in the closed string spectrum, one concludes that the low energy dynamics of the string (on backgrounds with large enough curvature radius compared to the Planck length)  is described by D=10 Type II supergravity.

\subsection{DBI action of a Dp-brane}

Let us consider now the open string theory whose dynamics in the flat space-time with no Kalb--Ramond field is defined by the action
\begin{equation}
\label{eq:open0}
    S_{open} = \fr{1}{2\p \a'}\bigg[\int_\S d^2z \dt X^\m \bar{\dt}X_\m + i \int_{\dt \S} d\t A_\m \dt_\t X^\m  \bigg].
\end{equation}
Here $\dt \S$ denotes boundary of the world-sheet $\S$ with complex coordinates $z = \t + i \s$. The vector field $A_\m$ is the classical background of the massless vector excitations in the open string spectrum. It naturally lives on the Dp-brane and interacts with the open string ends, that are effectively point-like particles on the Dp-brane world-volume. Similarly to how effective dynamics of the closed string background given by the metric, the Kalb--Ramond field and the dilaton is given by the  corresponding beta-functions, effective dynamics of the Dp-brane is given by beta-function for the vector field $A_\m$ understood as an open string coupling. This statement deserves more discussion.

To start with it is worth to notice that the boundary term in \eqref{eq:open0} is written for space-time filling D-brane ($p=9$ for the superstring). For $p<9$ one should split the space-time index as $\m=\{\a,m\}$ with $\a=0,\dots,p$ corresponding to Neumann directions and the boundary term is 
\begin{equation}
    S_B = \fr{1}{2\p\a'}i \int_{\dt \S} d\t \big(A_\a \dt_\t X^\a + \F_m \dt_\s X^m\big).
\end{equation}
The vector field $A_\a$ interacts with the open string ends moving on the $p+1$-dimensional world-volume, and the scalar fields $\F_m$ encode fluctuations of the brane in transverse directions. The same can be seen from the open string spectrum with $10-p$ Dirichlet boundary conditions. The corresponding supermultiplet is that of $D=p+1$ super Yang--Mills (SYM) theory that can be obtained by dimensional reduction from the SYM theory in $D=10$. Therefore indeed, dynamics of the vector field $A_\m$ following from the beta-function of \eqref{eq:open0} describes low energy dynamics of the open superstring ends, equivalently, of the D9-brane in 10 dimensions.

Derivation of the beta-function for $A_\m$ is straightforward and has been first performed in \cite{Abouelsaood:1986gd} in the background field approach and in the approximation of slowly varying fields. Expanding the fields around a classical background $X^\m = \bar{X}^\m(z) + \z^\m(z)$ we have for the action
\begin{equation}
\label{eq:open_expanded}
    \begin{aligned}
        S[\bar{X}+\z] = &\ S[\bar{X}] + \fr{1}{2\p \a'} \bigg[\int_\S d^2z  \dt \z^\m \bar\dt \z_\m\\
        & + i \int_{\dt \S} d\t \Big( \fr 12 \nabla_\n F_{\m\l}\, \z^\n \z^\l\dt_\t \bar X^\m   +\fr12 F_{\m\n}\z^\n \dt_\t \z^\m + \fr13 \nabla_\n F_{\m\l}\z^\n \z^\l \dt_\t \z^\m + \dots  \Big) \bigg], 
    \end{aligned}
\end{equation}
where $F_{\m\n}=\dt_\m A_\n - \dt_\n A_\n$ (for $p<9$ the derivatives are taken along the world-volume coordinates). Slowly varying fields approximation can schematically be defined as 
\begin{equation}
    \sqrt{2\p\a'}\left|\fr{\dt F}{F}\right|<<1.
\end{equation}
The background $\bar X^\m(z)$ satisfies classical equations of motion
\begin{equation}
    \begin{aligned}
        \dt \bar{\dt}\bar{X}^\m & = 0,\\
        \dt_\s \bar X^\m + i F^\m{}_\n \dt_\t \bar X^\n \Big|_{\dt \S} & = 0,
    \end{aligned}
\end{equation}
that is the reason for terms of the first order in $\bar X^\m $ and zero order in derivatives of $F_{\m\n}$ are absent in the expansion \eqref{eq:open_expanded}. Contributions to the beta-function for the field $A_\m$ come from the counter-term of the form
\begin{equation}
    S_c = \fr{i}{2\p \a'}\int_{\dt \S} d\t \G_\m \dt_\t \bar X^\m,
\end{equation}
where in 1-loop we have
\begin{equation}
    \G_\m = -\fr12 \nabla_\n F_{\m\l} G^{\n\l}(\t,\t).
\end{equation}
Here $G^{\m\n}(\t,\t)$ is the 1-loop exact Green's function for the quantum fluctuations $\x^\m$ in the presence of the background field $F_{\m\n}$. The exact Green's function must satisfy the following boundary condition
\begin{equation}
    \dt_\s G_{\m\n}(z,z') + i F_\m{}^\l \dt_\t G_{\l\n}(z,z')\Big|_{\s=0}=0.
\end{equation}
Introducing a cut-off $\L$ one obtains the following expression for the Green's function in the limit $\z \to \z'$ with $\s=0$
\begin{equation}
    G_{\m\n}(\t,\t) = -2\a' \log \L \, (1-F^2)^{-1}_{\m\n}.
\end{equation}
The vanishing beta-function condition then gives the following equation
\begin{equation}
\label{eq:beta_A}
    \b^{(A)}{}_\m = \L \fr{\dt}{\dt \L}\G_\m = \nabla^\n F_{\m}{}^\l(1-F^2)^{-1}_{\l\n}=0.
\end{equation}
This equation is all orders in $2\p\a'$ since this factor is already inside the vector field $A$ (see \eqref{eq:open0}). 

The equations \eqref{eq:beta_A} do not follow from variation of any action, however, it is possible to write an action with extrema at solution to the equations \eqref{eq:beta_A}. This action was derived in \cite{Fradkin:1985qd} as an effective action for the field $A_\m$ and is nothing but the Dirac--Born--Infeld action for non-linear electrodynamics:
\begin{equation}
\label{eq:DBI0}
    S_{DBI} = \int d^{10}X  \sqrt{\det(1+F)}.
\end{equation}
Field equations are the following
\begin{equation}
    \sqrt{\det(1+F)}(1-F^2)^{-1}_{\m\n}\b^{(A)\n}=0.
\end{equation}
Since the matrix $(1+F)$ is never degenerate and square of the antisymmetric matrix $F_{\m\n}$ cannot cancel the unity in the second factor, the only option to solve the equation is \eqref{eq:beta_A}. Therefore, solutions to the vanishing beta-function equations are the same as those to the DBI field equations. The same action \eqref{eq:DBI0} was obtained in \cite{Fradkin:1985qd} from a completely different approach of the quantum effective action for the open string in the presence of a constant field strength $F_{\m\n}$.

The DBI action \eqref{eq:DBI0} restricts the allowed configuration of the D9-brane world-volume vector field $A_\mu$ similarly to how the $D=10$ Type IIA/B supergravity action restricts allowed configurations of the background metric, B-field, dilaton the R-R fields and their superpartners. In the latter case the fields correspond to massless excitation of the closed string. In the case in question what we obtained is the effective action for the massless sector of the open string in the limit of slowly varying fields in all orders in $\a'$. The action in in the form \eqref{eq:DBI0} is actually the gauge fixed version of the full action, which can easily be restored by requiring general covariance on the Dp-brane world-sheet. This gives the following result
\begin{equation}
    S_{DBI1} = \int d^{p+1}\x \sqrt{\det \left(g_{\a\b} + F_{\a\b} \right)}.
\end{equation}
Here $\{\x^\a\}$ with $\a=0,\dots p$ denote world-volume coordinates, $F_{\a\b} =\dt_\a A_\b - \dt_\b A_\a$ and 
\begin{equation}
    g_{\a\b} = g_{\m\n}\dt_\a X^\m \dt_\b X^\n
\end{equation}
is the induce metric. Fixing the static gauge in the flat space-time $X^\a = \x^\a$ one ends up with $9-p$ scalar fields $\F^m =X^m$, where $m=1,\dots,9-p$. These correspond to excitations of the open strings transverse to the Dirichlet surface and encode small movements of the Dp-brane. 

As we have seen in Section \ref{sec:endpoints} the Dp-brane interacts not only with the metric in the NS-NS sector, but also with the Kalb--Ramond field $B_{\m\n}$. It is straightforward to obtain the corresponding contribution to the DBI action by starting with the open string action \eqref{eq:open0} in the flat space-time with a constant B-field. This results in shifting $F \to F+B$. The same result can be obtained by considering gauge transformations of the open string and of the DBI action. This approach will be useful in deriving interactions with R-R fields. Start with the open string action on a general background
\begin{equation}
    S_{open}= \fr{1}{2\p\a'} \int d^2 \s \dt_\a X^\m \dt_\b X^\n \sqrt{-h} h^{\a\b} g_{\m\n} + \int_\S  \bar{B} +  \fr{1}{2\p\a'}\int_\S dA, 
\end{equation}
where $\bar{B}$ denotes pull-back of the Kalb--Ramond field to the open string world-volume $\S$. Dynamics of the background field configuration is provided by supergravity field equations which are invariant under $B \to B+ d\L$, where $\L$ is a 1-form gauge parameter. By the open-closed string duality one concludes that the open string action must also be invariant under the same symmetry, which requires the following gauge transformation for the vector field
\begin{equation}
    A \to A - 2\p \a' \bar{\L}.
\end{equation}
Therefore the full gauge invariant field strength to appear in the DBI action must be defined as
\begin{equation}
    \mc{F}_{\a\b} = 2 \dt_{[\a}A_{\b]} + 2\p\a' B_{\a\b}.
\end{equation}
Taking into account the effective Dp-brane coupling we obtain the following action
\begin{equation}
\label{eq:DBI}
    S_{DBI} = T_p \int d^{p+1}\x \,e^{-\F}\sqrt{\det \left(g_{\a\b} + 2\p\a' B_{\a\b} + F_{\a\b}  \right)},
\end{equation}
that we will refer to as the kinetic part of the full Dp-brane action. 

The last comment in this Section is that expanding the action \eqref{eq:DBI0} in powers of $F_{\a\b}$ one obtains the standard abelian Yang--Mills action in the quadratic order. We leave this as a simple exercise and move further to constructing interactions with R-R fields.

\subsection{Gauge invariant Wess-Zumino actions}

Since the string does not interact minimally with R-R fields one cannot follow the same route as above to derive a Dp-brane effective action that includes interactions with R-R backgrounds. However, it is still possible to recover the relevant terms in addition to the DBI action following two simple rules: i) the Dp-brane interacts minimally with the $C_{(p+1)}$ R-R potential, ii) the interaction must respect the same gauge symmetries as the full Type IIA/B supergravity action. Therefore, we proceed we analysis of the latter and start with the full bosonic Type IIA (for concreteness) supergravity action:
\begin{equation}
\label{eq:IIA}
\begin{aligned}
    S_{IIA} = &\fr{1}{2\tilde{\k}}\int e^{-2\Phi}\left(R *1 - 4 d\Phi \wedge * d\Phi + \fr12 H\wedge * H\right) \\
    &-G_2 \wedge *G_2  - G_4 \wedge * G_4 + d C_3 \wedge dC_3 \wedge B_2.
\end{aligned}
\end{equation}
Here the integration is performed over the $D=10$ space-time. The field strengths $G_2$ and $G_4$ are defined as
\begin{equation}
    \begin{aligned}
        & G_2 = dC_1, && G_4 = dC_3 + H_3\wedge C_1.
    \end{aligned}
\end{equation}
This form of the bosonic action is dictated by supersymmetry and most easily can be obtained by dimensional reduction of the $D=11$ supergravity action (see \cite{Ortin:2015hya} for the details). Gauge transformations respected by the action \eqref{eq:IIA} are the following
\begin{equation}
    \begin{aligned}
        B_2 & \to B_2 + d\L_1, \\
        C_1 & \to C_1 + d\l_0, \\
        C_3 & \to C_3 + d\l_2 + H_3 \l_0.
    \end{aligned}
\end{equation}
From these we deduce that the Wess--Zumino part of the D0-brane action is pretty much straightforward
\begin{equation}
    S_{WZ}^{D0} = T_0 \int_{\S_1} C_1,
\end{equation}
where $\S_1$ is the world-volume of the D0-brane.

For the D2-brane interacting minimally with the $C_3$ potential things get more complicated and the gauge invariant action takes the form
\begin{equation}
    S_{WZ}^{D2} = T_2 \int_{\S_3} C_3 + B_2 \wedge C_1 + d\tilde{c}_2 .
\end{equation}
Meaning of the potential $c_2$ will become clear momentarily. For that let us check the invariance explicitly
\begin{equation}
    \begin{aligned}
        \d S_{WZ}^{D2} & = T_2 \int_{\S_3} d\l_2 + H_3 \l_0 + B_2 \wedge d\l_0 + d\delta \tilde{c}_2 \\
        & T_3 \int_{\S_3} d\l_2+ d(B_2 \l_0)  + d\d \tilde{c}_2 .
    \end{aligned}
\end{equation}
As in the case of the fundamental string, if the D2-brane is closed the result vanishes, while is it is open one has to require
\begin{equation}
    \d \tilde{c}_2 = -\l_2 - B_2 \l_0.
\end{equation}
Certainly, here $B_2$ means pullback of the space-time Kalb--Ramond gauge potential to the D2-brane world-volume. Therefore we conclude that the potential $c_2$ is necessary for the open D2-brane and it interacts with its boundaries. This potential is conceptually the same as the Born--Infeld vector field $A_\a$ for the open string. As we will see later, such potentials collectively denoted $c_p$ appear in Wess--Zumino actions for any branes (except D0) and imply that Dp-branes end on each other and on the NS5-brane to be introduced later.

\subsection{Democratic formulation}

Understood as Dirichlet surfaces Type IIA Dp-branes can have any even dimension from 0 up to 8, while the action \eqref{eq:IIA} contains only the $C_1$ and $C_3$ gauge potentials. The other gauge potential interacting with Dp-branes of higher dimension are introduced into the theory by the dualization procedure. For that it is convenient to turn to what is called democratic formulation where all R-R potentials enter on equal footing \cite{Townsend:1995gp,Bergshoeff:1996ui,Bergshoeff:2001pv}. This allows to write both Type IIA and IIB supergravity actions in a unique form and do the same for Wess--Zumino actions for all Dp-branes.  

To do so start with the equations of motion for the R-R potentials that follow from the action \eqref{eq:IIA}
\begin{equation}
    \begin{aligned}
        d*G_2 - H_3 \wedge *G_4 & = 0,\\
        d*G_4 - dC_3 \wedge H_3 & = 0.
    \end{aligned}
\end{equation}
Since $dH_3 = 0$ the second line can be written as
vanishing of a full derivative
\begin{equation}
    d(*G_4 - C_3 \wedge H_3)=0.
\end{equation}
Locally this defines a 5-form potential $*G_4 - C_3 \wedge H_3 = - dC_5$. The corresponding dual field strength is defined as
\begin{equation}
    *G_4 = - G_6, \quad G_6 = dC_5 +H_3\wedge C_3.
\end{equation}
Upon this dualization equations of motion for the field strength $G_4$ can be written as (generalized) Bianchi identities for the field strength $G_6$:
\begin{equation}
    d G_6 + H_3 \wedge G_4 =0.
\end{equation}
Similarly dualization of the 2-form field strength leads to a 7-form gauge potential
\begin{equation}
    *G_2 = G_8, \quad G_8 = dC_7 + H_3\wedge C_5.
\end{equation}
In uniform notations we have the following for the field strengths
\begin{equation}
    G_{2n} = dC_{2n-1} + H_3 \wedge C_{2n-3}, \quad n = 1,2,3,4, \quad C_{-1}\equiv 0,
\end{equation}
for Bianchi identities
\begin{equation}
\label{eq:BIs}
    dG_{2n} + H_3 \wedge G_{2n-2}=0,
\end{equation}
and for equations of motion
\begin{equation}
    d*G_{2n} - H_3\wedge *G_{2n-2}=0. 
\end{equation}
These equations can  be derived from the so-called democratic Type IIA pseudo-action, whose R-R part reads
\begin{equation}
\label{eq:demoIIA}
    S_{R-R}^{IIA} = \fr{1}{4 \tilde{\k}^2}\int  G_2 \wedge *G_8 - G_4\wedge *G_6 - G_6\wedge *G_4 - G_8\wedge *G_2
\end{equation}
endowed with the following dualization rules
\begin{equation}
    * G_{2n} = (-1)^{n+1} G_{10-2n}.
\end{equation}
It is straightforward to check, that substituting the dualization rules into the action \eqref{eq:demoIIA} explicitly one ends up with identical zero. This is the reason the action \eqref{eq:demoIIA} is called a pseudo-action: it is crucial to first vary w.r.t. the gauge potentials $C_1,\dots,C_7$ and only afterwards impose the dualization conditions.

Now gauge transformations can be written uniformly as
\begin{equation}
    \d C_p = d\l_{p-1} + H_3 \wedge \l_{p-3},
\end{equation}
and the Dp-brane  Wess--Zumino  action reads
\begin{equation}
    \begin{aligned}
        S_{WZ}^{Dp} & = T_p \int_{\S_{p+1}} C_{p+1} + B_2 \wedge C_{p-1} + \fr12 B_2\wedge B_2 \wedge C_{p-3} + \dots +dc_p\\
       & = T_p \int_{\S_{p+1}}\left[e^{B_2}\wedge \mc{G}\right]_{p+1} 
    \end{aligned}
\end{equation}
Here $\mc{G} = \mc{G}_1 + \mc{C}_3 + \dots +\mc{C}_7$ denotes formal sum of gauge invariant world-volume field strengths
\begin{equation}
    \mc{G}_{p+1} =  C_{p+1} + dc_p + H_3\wedge c_{p-2},
\end{equation}
and  $[]_{p+1}$ denotes projection of the result to the space of $p+1$-forms. Gauge transformation for the fields $c_p$ is simply $\d c_p = -\l_p$. This potentials are related $\tilde{c}_p$ introduce above as
\begin{equation}
    c_p = \left[e^{B_2}\wedge \tilde{c}\right]_p,
\end{equation}
where $\tilde{c}=\tilde{c}_0 + \dots \tilde{c}_6$, and prove more convenient.

It appears that this is not the end of the story as the term $e^{B_2}$ is not invariant under gauge transformation of the Kalb--Ramond field $B_2 \to B_2 + d\L$. However, we have already learned that $\mc{F}_2 = 2\pi \a' B_2 + dA$ is gauge invariant, therefore one must simply replace $B_2$ by $(2\p\a')^{-1}\mc{F}_2$ everywhere. Collecting everything together we end up with the following action for a single Dp-brane
\begin{equation}
\label{eq:Dp}
    S_{Dp} = T_p \int_{\S_{p+1}} e^{-\F}\sqrt{\det \left(g_{\a\b} + \mc{F}_{\a\b}  \right)} + \left[e^{(2\p \a')^{-1}\mc{F}_2}\wedge \mc{G}\right]_{p+1}
\end{equation}
It is easy to see, that for the IIB theory one gets the same answer with the only difference, that the potentials are even forms.

A few comments are in place here. First, note that a Dp-brane action includes potentials $c_n$ with $n < p$ in its Wess--Zumino part. This means, that Dp-branes are allowed to end on each other. Going slightly ahead of the narrative, let us mention that this can be easily seen by S-dualising the system of a fundamental string ending on a D3-brane, that gives D1 ending on the D3. This is the reason for using $c_p$ rather than $\tilde{c}_p$, which appear less natural for Dp-brane dynamics.

For the second comment we emphasize that the derived action describes dynamics of a single Dp-brane, that for small world-volume fields reproduces U(1) Yang--Mills theory. To describe 

\subsection{S-duality of Type IIB theory. NS5-brane} 
\label{sec:NS5}

Looking at the massless spectrum of the Type IIB string \eqref{eq:massless_spectrum} it is easy to notice that some states come in doublets. This is the linear manifestation of the full non-linear non-perturbative symmetry of the Type IIB string theory called S-duality. At the level of $D=10$ Type IIB supergravity this results in the fact that the fields \eqref{eq:massless_fields} can be combined into multiplets of $\rmSL(2)$ (see e.g. \cite{Ortin:2015hya}). S-duality transformation rules in terms of the Type IIB supergravity fields can be expressed as follows
\begin{equation}
    \begin{aligned}
        \t & \to \fr1\t, \quad \t = C_0 + i e^{-\Phi}, \\
        \tilde{g}_{\m\n} & \to \tilde{g}_{\m\n}, \quad \tilde{g}_{\m\n} = e^{-\fr\Phi2}g_{\m\n}, \\
        B_2 & \to -C_2,\\
        C_2 & \to B_2, \\
        A_4 & \to A_4, \quad A_4 = C_4 +\fr12C_2\wedge B_2.
    \end{aligned}
\end{equation}
Here the metric in the Einstein frame $\tilde{g}_{\m\n}$ is invariant as well as the 4-form $A_4$. 

From the duality between the Kalb--Ramond and R-R 2-forms one infers that the fundamental string and the D1-brane must also map into each other. Let us show this explicitly at the level of effective action starting with the D1-brane action
\begin{equation}
    S_{D1} = T_1 \int d^2 \s e^{-\f} \sqrt{-\det(G_{\a\b} + \mF_{\a\b})} + T_1 \int C_2 + C_0 \, \mF_2.  
\end{equation}
Here the integration is performed along the D1-brane world-volume, $\mF_2= F_2-B_2$ with $F_2 = dA$ and $\a,\b=0,1$ are world-volume indices. To map this to the action of the fundamental string we first must get rid of the gauge field strength $\mF_2$ under the square root. For that we follow the approach of \cite{Tseytlin:1996it,Aganagic:1997zk} and introduce a Lagrange multiplier $H^{\a\b}$ that allows to rewrite the action as
\begin{equation}
    \begin{aligned}
        S'_{D1} = &\ T_1 \int d^2\s \left[\sqrt{-\det(G_{\a\b} + \mF_{\a\b})} + \fr12 \e^{\a\b}C_{\a\b} + \fr12 C_0 \e^{\a\b}(F_{\a\b} - B_{\a\b}) - \fr12 \L \e^{\a\b}B_{\a\b}  \right.\\
        &\left. + \fr12 H^{\a\b}(F_{\a\b} - 2\dt_\a A_\b)\right].
    \end{aligned}
\end{equation}
Here $\L$ is an arbitrary constant to be fixed later and we define the epsilon-symbol as $\e_{01}=1$. In Minkowski signature, that we adopt here, $\e^{01}=-1$. The action $S'_{D1}$ is equivalent to the action $S_{D1}$ upon integrating out $H^{\a\b}$ that fixes its value to
\begin{equation}
    H^{\a\b} = \e^{\a\b} \L.
\end{equation}
Now it is possible to integrate out $F_{\a\b}$ and write the action in terms of $\L$ instead. Since $F_{\a\b}$ enters the action without derivatives the integration fixes its value according to
\begin{equation}
    \mF_{\a\b} = - e^{\f}\sqrt{-\det(G+\mF)}(C_0 +\L)\e_{\a\b}.
\end{equation}
This implies
\begin{equation}
    -\det(G+\mF) = -\det G \left(1 + e^{2\f}(C_0 + \L)^2\right)^{-1}.
\end{equation}
Substituting everything into the action $S'_{D1}$ we obtain
\begin{equation}
    S'_{D1} = T_1 \int d^\s \sqrt{e^{-2\f} + (C_0 + \L)^2}\sqrt{-\det G} + \fr12 \e^{\a\b C_{\a\b}} - \fr12 \L \e^{\a\b}2 \dt_\a A_\b. 
\end{equation}
The next step is to perform a shift $C_0+\L \to C_0$ that is actually a part of the modular SL(2) symmetry group of the Type IIB theory. The action then becomes
\begin{equation}
    S'_{D1} = T_1 \int d^2\s |\t|\sqrt{-\det G} + T_1\int C_2 - \L dA.
\end{equation}
Upon the S-duality transformation and rescaling $-\L A \to A $ this reproduces precisely the action of the open string in the Nambu--Goto form. Therefore, we conclude that the fundamental string is indeed S-dual to the D1-brane.

The D3-brane is mapped to itself under S-duality. Let us now apply the S-duality transformation to the D5-brane action that must give an action for an object that is charged w.r.t. to the S-dual of the R-R 6-form $C_6$. This is the NS5 brane of Type IIB theory (NS5B). Let us endow the above rules by
\begin{equation}
    \begin{aligned}
        C_6 & \to -B_6 + \fr12 B_2\wedge C_2\wedge C_2, \\
        A_1 & \to -c_1. 
    \end{aligned}
\end{equation}
At this point the first line can be understood as a definition of $B_6$, which however will be consistent with dualization of the Kalb--Ramond potential $B_2$. The resulting action reads
\begin{equation}
    \label{eq:NS5}
    \begin{aligned}
        S_{NS5B} = &-T_{5}\int_{\S_6} d^6\xi e^{-2\Phi}\sqrt{1+e^{2\Phi}C_0^2}\sqrt{-\det\left( G_{ij} - \fr{e^{\Phi}}{\sqrt{1+e^{2\Phi}C_0^2}} \mathcal{G}_{ij} \right)}\\
    &- B_6 -\fr12 B\wedge C_2\wedge C_2 + (C_4 + C_2 \wedge B_2 ) \wedge \mG_2 - \fr12 B_2 \wedge \mG_2 \wedge \mG_2 \\
    &+ \fr1{3!} |\t|^{-2} C_0 \,\mG_2 \wedge \mG_2 \wedge \mG_2 .
    \end{aligned}
\end{equation}
The tension of the NS5-brane is defined by $T_{NS5} = g_s^{-1}T_{D5}$. Note that the Wess--Zumino part of this action cannot be written in a similar nice form as in \eqref{eq:Dp}. The 1-form potential $c_1$ living on the NS5B-brane world-volume by construction interacts with boundaries of the D1-branes, the same is true for the $C_0$ potential. The world-volume electro-magnetic duality, described e.g. in \cite{Musaev:2022yqh}, gives the world-volume potential $c_3$ that interacts with boundaries of the D3-brane. This allows to conclude that Dp-branes can end on the NS5-brane (both in IIA and IIB theory). This plays crucial role in the Hanany--Witten effect \cite{Hanany:1996ie} to be discussed in Section \ref{sec:hanany-witten}

In a sense the NS5-brane can be understood as an effective description of dynamics of Dp-brane boundaries. However, most straightforwardly it can be understood as a magnetic monopole for the fundamental string. For that return to the democratic Type IIA supergravity action \eqref{eq:demoIIA} and write field equations for the Kalb--Ramond field $B_2$. Taking into account the duality relations we obtain
\begin{equation}
    d*H_3 +d \left(C_1\wedge G_6 - C_3 \wedge G_4 + C_5 \wedge G_2\right) = 0.
\end{equation}
As before this allows (at least locally) to introduce a 6-form potential $B_6$. Defining the 7-form field strength as $H_7 = * H_3$ we have
\begin{equation}
\label{eq:H7A}
    H_7 = dB_6 -C_1\wedge G_6 + C_3 \wedge G_4 - C_5 \wedge G_2.
\end{equation}
The corresponding expression in Type IIB theory will be
\begin{equation}
\label{eq:H7B}
    H_7 = dB_6 -C_0\wedge G_7 + C_2 \wedge G_5 - C_6 \wedge G_1.
\end{equation}
Now, on the one hand we obtain the 6-form potential $B_6$ by dualization of the 2-form potential $B_2$. On the other hand the latter  is S-dual to the R-R field $C_2$ whose magnetic dual is $C_6$. Therefore, the potential $B_6$ in \eqref{eq:H7B} is precisely the one that is S-dual to $C_6$ and hence interact minimally with the NS5-brane. We conclude that the NS5-brane, that is S-dual to the D5-brane, is nothing but the magnetic monopole for the fundamental string.

Historically, this brane was called solitonic and denoted ``S5''. Its tension is proportional to $g_s^{-2}$ implying that its dynamics gets frozen earlier that that of Dp-branes in the $g_s \to 0$ limit. This is part of the reason why we do not see excitations of $B_6$ in the closed string spectrum directly and why the standard brane-scan of \cite{Achucarro:1987nc,Blencowe:1988gj} does not reveal the Type II NS5-brane. Speaking more correctly, the old brane-scan did contain a $p=5$, $D=10$ fivebrane slot, however, the Type II NS5-brane  was not identified there as a solitonic magnetic brane of string theory. The name new brane-scan  usually refers to the later picture of \cite{Duff:1993ye,Duff:1994an,Townsend:1995af,Stelle:1998xg} that emerged once one includes solitonic supergravity p-branes, D-branes, and worldvolume vector/tensor multiplets.

\section{Branes as solutions to supergravity equations}
\label{sec:solutions}

Information about gauge potential and symmetries gained from D=10 supergravity has proven useful in deriving Wess--Zumino actions for Dp-branes and for the NS5-brane. As we will see in this section this utility goes much more beyond and allows, for example, to recover full non-perturbative geometry around branes. Analyzing coupling of closed string states to Dirichlet surfaces in Section \ref{sec:endpoints} we have seen that the latter interact with both NS-NS and R-R closed string excitations. Of interest is the massless sector which is able to condense into a background geometry defined by classical profiles of the metric, the dilaton, the Kalb--Ramond and R-R fields. The idea for searching such backgrounds is precisely the same as searching the electromagnetic field configuration created by an electric charge or the space-time metric of a massive body. These are the famous Coulomb potential and Schwarzschield space-time.

Backgrounds created by (standalone) branes of string theory (including the fundamental string itself) differ from these examples in that they preserve 1/2 of the total supersymmetry. Various intersections of branes may preserve smaller fraction of supercharges up to none. This indeed is expected if one recalls that massless excitations of open string ends on the D9-brane form $\mc{N}=1$ vector multiplet, that is invariant under 16 supercharges.

Black hole like solutions to supergravity field equations have been known from the very early years of supergravity (see e.g. \cite{Dabholkar:1990yf,Strominger:1990et,Duff:1990xz,Horowitz:1991cd}) and called black branes. The fact that their extremal limit is precisely the background created by D-brane has been realized in \cite{Polchinski:1995mt} and further discussed in more details in \cite{Lifschytz:1996iq,Klebanov:1997kc,Callan:1996dv,Maldacena:1996ky}. Here we will follow a direct approach, that is to start with the action for a brane interacting with supergravity
\begin{equation}
    S_{Full} = S_{SUGRA} + S_{brane}.
\end{equation}
Varying this w.r.t. both supergravity fields and embedding functions $X^\m = X^\m(\x)$ of the brane leads to a system of equations with a source. Their solution for a given brane movement will a supergravity background. Doing this for GR plus a massive particle indeed produces the Schwarzschild space-time however moving of the massive  particle is not in consistency with the background. This inconsistency disappears if preservation of 1/2 of supersymmetry is required, and the black hole is extremal. We start with discussing this example.

\subsection{Critical R-N black holes and $\mc{N}=2$ SUSY }

Consider a theory of a point-like object with mass $M$ and charge $Q$ interacting with the gravitational and electromagnetic fields in 4 dimensions. The action will be given by
\begin{equation}
    \begin{aligned}
        S_{Full} & = \fr{1}{8 \p \k^2} \int d^4x \sqrt{-g}R + \fr{1}{16 \p} \int d^4x \sqrt{-g}F_{\m\n}F^{\m\n}\\
        & + m \int_\g ds - q \int_\g A_\m dx^\m,
    \end{aligned}
\end{equation}
where the one-dimensional integrals are taken along world-line of the particle and $ds^2 = g_{\m\n}dx^\m dx^\n$ denotes square of the interval. Varying the action w.r.t. the fields $g_{\m\n}$, $A_\m$ and the functions $x^\m(s)$ that define embedding of the world-line we obtain the following equations
\begin{equation}
\label{eq:EMpeqns}
    \begin{aligned}
        G_{\m\n}=R_{\m\n} - \fr12 g_{\m\n} R - \k^2 T_{\m\n} & =  4\p\k^2 J_{\m\n}, \\
        \nabla_\m F^{\m\n} & = 4\p J^\n, \\
        m \fr{d u^{\m} }{ds}+ m \G_{\r\s}{}^\m u^\r u^\s &= -\fr12 Q F^{\m\n}u_\n.
    \end{aligned}
\end{equation}
Where the following conventions have been adopted
\begin{equation}
    \begin{aligned}
        T_{\m\n} & = F_{\m\r}F_\n{}^\r -\fr14 F_{\r\s}F^{\r\s}g_{\m\n}, \\
        J_{\m\n} & = m \int ds \fr{1}{\sqrt{-g}}u_\m(s)u_\n(s)\d^{(4)}(x - z(s)),\\
        J^{\m} & = q \int ds \fr{1}{\sqrt{-g}}u^\m(s)\d^{(4)}(x - z(s)),
    \end{aligned}
\end{equation}
and $u^\m = dz(s)^\m/ds$ is the covariant velocity. Note the difference in the delta-functions between the point of observation with coordinates $x^\m$ and the trajectory $z^\m(s)$ of the particle.

Suppose now that the particle is at rest at the origin $\vec{z}(s)=0$ and attempt to solve the equations \eqref{eq:EMpeqns} consistently. Reparametrization invariance along the world-line allows to gauge fix $z^0(s) = x^0$, that gives $s = \sqrt{g_{00}} z^0$ and
\begin{equation}
    u^\m = \fr{dz^\m}{ds} = (g_{00}{}^{-\fr12},\vec{0}).
\end{equation}
It is convenient to work in spherical coordinates and to choose the following ansatz
\begin{equation}
    \begin{aligned}
        ds^2 & = -f(r) dt^2 + f(r)^{-1}dr^2 + r^2 (d\q^2 + \sin^2 \q d\f^2), \\
        A & = a(r)dt,
    \end{aligned}
\end{equation}
where $t=x^0$. In this case Maxwell equations boil down to the Poisson equation $ \Delta_{(3)}a(r) = - Q \d^{(3)}(\vec{x})$. In spherical coordinates this reads
\begin{equation}
    \fr{1}{r^2}\dt_r (r^2 \dt_r a(r)) = -\fr{1}{r^2 \sin \q} q \d^{(3)}(r),
\end{equation}
where the delta-function $ \d^{(3)}(r)$ is defined as $1 = \int dr  \d^{(3)}(r)$. We arrive at the familiar Coulomb potential 
\begin{equation}
    A = \fr q r dt.
\end{equation}

It is crucial to notice two related properties of the solution: i) the potential has  a singularity only at the origin, where the source is placed; ii) the singularity is time-like, corresponding to movement of a massive particle. Both properties break for the Schwarzschild and Reissner--Nordstr\"om solutions. To see that write the non-vanishing components of the LHS of the Einstein equations (the first line in \eqref{eq:EMpeqns}):
\begin{equation}
    \begin{aligned}
        G_{00} & = -\fr{1}{2 r^4}f(r)\Big[2r^2(f(r)-1)+ 2 r^3 f'(r) + \k^2(a(r) - r a'(r))^2\Big] ,\\
        G_{rr} & = -f(r)^{-2} G_{00}\\
        G_{\q\q} & = \fr{1}{2}\Big[r(2f'(r)+r f''(r)) + \k^2r^{-2}(a(r) - r a'(r))^2\Big] ,\\
        G_{\f\f} & = \sin^2\q G_{\q\q}.
    \end{aligned}
\end{equation}
To illustrate the problem let us consider the simplest case where $q=0$ and $f(r) = 1 - \fr{2 M}{r}$, i.e. the Schwarzschild solution. In this case it is easy to see that all components $G_{\m\m}$ are proportional to delta-function, meaning that the source must have not only the time-like but also the space-like singularity. This breaks the intuition behind movement of a massive particle. 

Another issue preventing finding a consistent solution to the problem comes from the equation of motion of the particle (the third line in \eqref{eq:EMpeqns}). For our ansatz we have
\begin{equation}
    \dt_r \Big( q a(r) + 2 m f^{\fr12}(r)\Big)=0,
\end{equation}
evaluated at $r=0$. This equation however suggest a solution to the whole problem. Indeed, consider the Reissner--Nordstr\"om solution defined by
\begin{equation}
    f(r) = 1 - \fr{2 m \k^2}{r} + \fr{q^2 \k^2}{r^2},
\end{equation}
and require $q \, a(r) + 2 m \sqrt{f(r)} = c=$const. This implies $c=2 m$ and $q=\sqrt{2}m\k$, and the latter condition has deep consequences. Indeed, given  $q=\sqrt{2}m\k$ the function $f(r)$ becomes
\begin{equation}
    f(r)= \fr{(r-\k^2 m)^2}{r^2},
\end{equation}
and two horizons of the charged black hole solution coincide. This means that the black hole is critical in the sense, that it has the maximal charge possible. 

It is convenient to introduce a coordinate system where the radial coordinate $\r$ takes value $\r=0$ at the horizon, i.e. $r = \k^2 m + \r$. Then the full solution can be written as
\begin{equation}
\label{eq:ERN}
    \begin{aligned}
        ds^2 & = -H^{-2}(\r)dt^2 + H^2(\r)\big(d\r^2 + \r^2 d\q^2  + \r^2 \sin^2 \q  d\f^2\big), \\
        A& = \pm\sqrt{2}\k^{-1}\big(H^{-1} - 1 \big)dt, \\
        H & = 1 + \fr{\k^2 m}{\r}.
    \end{aligned}
\end{equation}
Explicit check shows that in this extremal limit only $G_{00}$ is proportional to $\d^{(3)}(\vec{x})$, meaning that the background has only time-like singularity that is consistent with a massive source.

Remarkable is that the condition $q=\sqrt{2}\k\,m$ is actually the 1/2BPS of the $\mc{N}=2$ extended supersymmetry algebra, when the central charge is equal to mass of the multiplet. This implies that the solution \eqref{eq:ERN} is uplifted to a solution to $\mc{N}=2$ pure supergravity theory that preserves half of all supersymmetries. The field content of the theory is given by the metric, encoded in the vielbein $e_\m{}^\a$, the vector field $A_\m$ and the doubled of gravitini $\w_\m{}^I$, where $I=1,2$. In the notations of \cite{Ortin:2015hya} supersymmetry transformations are given by
\begin{equation}
    \begin{aligned}
        \fr{1}{2\sqrt{2}\k}\d_\e e_\m{}^\a & = - i \bar{\e}_I \g^\a \y_\m{}^I,\\
        \fr14 \d_{\e} A_\m& = -i \bar{\e}_I (\s^2)^I{}_J\y_\m{}^J,\\
        \d_\e \y_\m{}^I & =\left(\dt_\m - \fr14 \w_\m{}^{\a\b}\right)\e^I +\fr{\k^2}{4}\left(F_{\r\s} +2 i \bar{y}_\r \s^2 \y_\s\right)\g^{\r\s}(\s^2)_J{}^I\e^J,
    \end{aligned}
\end{equation}
where $\w_\m{}^{\a\b}$ is the standard spin-connection. To check preservation of (a portion of) supersymmetry we substitute the solution \eqref{eq:ERN} together with $\y_\m{}^I = 0$ into these transformation rules. The first two lines trivially vanish, while the third line renders a non-trivial equation on the spinors $\e^I$ referred to as the BPS equation. Its solution is straightforward and reads
\begin{equation}
    \e^I = H(\r)^{\fr12}\exp\left(\fr12 \g^{\r\q}\q\right)\exp\left(\fr12 (\cos \q \g^{\q\f} + \sin\q \g^{\r\f})\f\right)\ve^I,
\end{equation}
where $\gamma$-matrices in the exponents are flat and the spinors $\ve^I$ satisfy the condition
\begin{equation}
    \left(\d_I{}^J + \g^0 (\s^2)_I{}^J\right)\ve^J=0,
\end{equation}
that leaves precisely half of the total number of components. Since the solution does not change under SUSY transformation with such chosen spinor parameters, it is said to preserve half of supersymmetries and is called a 1/2BPS solution.

We conclude, that for 1/2BPS solutions the full set of equations with a source can be solved consistently with the correct time-like singularity at the place of the source. Moreover, two such extremal R-N black holes placed at an arbitrary distance of each other will not interact, that can be seen by checking energy of a state with $q=\sqrt{2}\k\, m$ in the background \eqref{eq:ERN}. As we will learn in further sections all Dp-branes and the NS5-brane of string theory generate 1/2BPS backgrounds.

\subsection{Extended objects in supergravity}

Similarly to how a massive charged particle interacting with the gravitational and electromagnetic field generates a background in the surrounding space-time, all fundamental objects of string theory generate a background of supergravity fields.  In Section we have learned that a Dp-brane interacts with the metric, the dilaton and minimally with the R-R $(p+1)$-form $C_{(p+1)}$. Therefore, given the Reissner--Nordstr\"om extremal black hole background it is natural to expect the background of a Dp-brane to be of the form
\begin{equation}
\label{eq:Dp_bg}
    \begin{aligned}
        ds^2 & = f_1(r)\big(-(dx^0)^2 + \dots + (dx^p)^2\big) + g_1(r)\big(dr^2 + r^2 d\W_{8-p}\big),\\
        C_{p+1} & = f_2(r)dx^0\wedge \dots \wedge dx^{p},\\
        e^{2 \phi} & = g_2(r).
    \end{aligned}
\end{equation}
Here $d\W_{8-p}$ denotes volume element of the $8-p$-dimensional sphere surrounding the brane in the $9-p$ transverse space, the coordinate $r$ is transverse to the branes' world-volume. The functions $f_{1,2,3}(r)$ and $g_{1,2}(r)$ are to be determined from the supergravity field equations with the RHS determined by the Dp-brane action. 
Directly substituting this ansatz and solving the resulting the equations one obtains the following
\begin{equation}
    \begin{aligned}
        ds^2  &= H_p^{-\fr12}\, dx^2 + H_p^{\fr12}\,\big(dr^2 + r^2 d\W_{8-p}\big),\\
        e^\phi &= g_s\, H_p^{\frac{3-p}{4}}, \\
        C_{p+1} & = \pm \,\bigl(H_p^{-1}-1\bigr)\,dx^0 \wedge \dots \wedge dx^{p+1}, 
    \end{aligned}
\end{equation}
where 
\begin{equation}
\label{eq:harmonic}
     H_p(r) =1+\frac{d_p\, g_s N\, (\alpha')^{\fr{7-p}{2}}}{r^{7-p}},
\qquad
d_p=(2\sqrt{\pi})^{\,5-p}\Gamma\!\left(\frac{7-p}{2}\right).
\end{equation}
is the so-called defining function of the solution and the $
\pm$ sign corresponds to Dp-brane and Dp-antibrane. It is a harmonic function in the space transverse to the Dp-branes' world-volume. Since a sum of harmonic functions is again a harmonic function with the same asymptotics, its is straightforward to write a background of multiple parallel Dp-branes placed at points with coordinates $\vec{x}_i$. The 
background will be precisely the same as \eqref{eq:Dp_bg} but with a different defining function
\begin{equation}
    H(r) = 1 + \sum_{i=1}^N \fr{1}{|\vec{x} - \vec{x}_i|^{7-p}}.
\end{equation}
This is the field theoretical manifestation of the fact that parallel Dp-branes do not interact, that is the gravitational attraction is compensated by repulsion due to R-R fields. The cases $p=7$ and larger are special in many sense and are a window to an infinite set of the so-called exotic branes. We do not cover this huge subject in this review, and the interested reader is referred to \cite{deBoer:2012ma}.

The solution \eqref{eq:Dp_bg} should be thought of as an effective closed string background created by a (stack of $N$) Dp-brane(s) in the 10-dimensional space-time. This is completely analogous to how a charged particle creates an electric-field around it. From QED, that is a second-quantized theory, we know that elementary excitations of electro-magnetic fields are photons. Similarly here massless elementary excitations in the theory are given by massless modes of closed strings, while the classical description of their collective behavior is given by the Type II supergravity theory. The Dp-brane is a 1/2BPS object, that we have seen in he fact that the massless open string spectrum has twice less supersymmetries than that of the massless closed string. At the classical level this manifests in the fact that the solution \eqref{eq:Dp_bg} preserves half of Killing spinors $\e$. Their space-time dependence is given by $\epsilon = H_p^{-1/8}\, \epsilon_0$, and $\e_0$ is further restricted by
\begin{equation}
\label{eq:BPS_Dp}
    \left(1\pm\,\Gamma^{0\cdots p}\,\c^{\frac p2+1}\right)\epsilon_0=0,
\end{equation}
where $\c = \G_0\dots \G_9$ is the $D=10$ chirality operator.

Similarly to how the extremal BPS black hole appears as a limit of the non-supersymmetric Reissner--Nordstr\"om black hole, the Dp-brane solution \eqref{eq:Dp_bg} is a limit of the so-called black brane solution. This has an additional function $f(r)$ called the blackening function, and can be written in the following form
\begin{equation}
\label{eq:Dp_thermal}
    \begin{aligned}
        ds_{\text{str}}^2 & = H_p(r)^{-\fr12}\Bigl(-f(r)\,dt^2 + d\vec x_p^{\,2}\Bigr) + H_p(r)^{\fr12}\Bigl(f(r)^{-1}dr^2 + r^2 d\Omega_{8-p}^2\Bigr) ,\\
        e^\phi & = g_s\, H_p(r)^{\frac{3-p}{4}}, \\
        C_{p+1} &= \pm \,\bigl(H_p^{-1}(r)-1\bigr)\coth\alpha\, dx^0\wedge \dots \wedge dx^p
    \end{aligned}
\end{equation}
where the harmonic function $H(r)$ and the blackening function $f(r)$ are conveniently written in the following form
\begin{equation}
    \begin{aligned}
        H_p(r)&=1+\left(\frac{r_0}{r}\right)^{7-p}\sinh^2\alpha,\\
        f(r)&=1-\left(\frac{r_0}{r}\right)^{7-p}, 
    \end{aligned}
\end{equation}
Such solutions are often called thermal Dp-branes since they have non-vanishing area of the horizon and therefore Bekenstein--Hawking entropy and temperature, that is
\begin{equation}
    T_H=\frac{7-p}{4\pi r_0 \cosh\alpha}.
\end{equation}
Hence, the parameter $\alpha$ controls the corresponding temperature associated with the horizon. In the extremal limit $r_0 \to 0$, $\a \to \infty$, such that $r_0^{7-p}\sinh^2 =Q_p$ is fixed, gives the extremal Dp-brane background. The extremal object has zero temperature that is consistent with its stability, as a supersymmetric object, under Hawking radiation. The $p=7$ case here seems special since the harmonic function in 2 transverse dimensions depends on logarithm of the radial direction, and the general expression simply does not apply.

Certainly, the fundamental string itself and its magnetic monopole, the NS5-brane, also interact with massless closed string states, and therefore generate non-trivial space-time backgrounds. For the fundamental string one has
\begin{equation}
    \label{eq:F1_bg}
    \begin{aligned}
        ds^2&= H_{\text{F1}}(r)^{-1}\,(-dt^2+dx_1^2) + dy^m dy^m \\
        e^\phi & = g_s\, H_{\text{F1}}(r)^{-\fr12} ,\\
        B_{2}& = \pm\,\bigl(H_{\text{F1}}(r)^{-1}-1\bigr)\,dt\wedge dx^1,
    \end{aligned}
\end{equation}
where as before the $\pm$ corresponds to positive/negative string charge and
\begin{equation}
    H_{\text{F1}}(r)=1+\frac{32\pi^2\, g_s^2\, N\, (\alpha')^3}{r^6}.
\end{equation}
The Killing spinor depends on the radial coordinate $r$ as $\epsilon = H_{\text{F1}}(r)^{-1/4}\,\epsilon_0$ and is restricted by
\begin{equation}
\label{eq:BPS_F1}
    \left(1\pm\,\Gamma^{01}\c\right)\epsilon_0=0.
\end{equation}

Background of a stack of (anti-)NS-branes is most straightforwardly written in terms of the magnetically dual potential $B_6$ introduced in Section \ref{sec:NS5}, and reads
\begin{equation}
    \begin{aligned}
        ds^2 & = - dt^2 + dx_1^2+\cdots+dx_5^2 +H_5(r)\,bigl(dr^2 + r^2 \d\W_3\bigr), \\
        e^{\phi}&=g_s\,H_5(r)^{\fr12}, \\
        B_6 &= \pm\, g_s^{-2}\,\bigl(H_5^{-1}(r)-1\bigr)\, dt\wedge \dots \wedge dx^5,
    \end{aligned}
\end{equation}
where
\begin{equation}
    H_5(r)=1+\frac{N\alpha'}{r^2}.
\end{equation}
The corresponding Killing spinor is constant $\e = \e_0$ and satisfies
\begin{equation}
    \label{eq:BPS_NS5}
        \left(1\pm\, \Gamma^{0\dots 5}\right)\epsilon_0=0.
\end{equation}

Historically, black branes were investigated at the very early stages of research on supergravity as the natural desire to obtain black hole-like solutions in this supersymmetric gravitational theory. Classical black $p$-brane solutions were constructed before the D-brane interpretation was available, notably in work on black strings, membranes, and higher-dimensional solitons in the late 1980s and early 1990s \cite{Horowitz:1991cd,Duff:1990xz,Gueven:1992hh,Duff:1994an}.
In \cite{Polchinski:1995mt} a suggestion has been made that extremal black branes of Type II supergravity with non-vanishing R-R fields are nothing but the Dirichlet branes of string theory. Since Type IIA supergravity can be obtained as a consistent reduction of the $ D=11 $ $\mc{N}=1$ supergravity, all its brane solutions descend from brane solutions in 11 dimensions. Gauge fields of $D=11$ supergravity are given by the 3-form $C_{(3)}$ and its dual 6-form $C_{(6)}$. Therefore we will have a two-dimensional membrane, the M2-brane, and a 5-dimensional membrane, the M5-brane. Wrapping the M2-brane along the compact circle when reducing to Type IIA, we obtain the fundamental string, while reduction without wrapping gives the D2-brane. Similarly, we get NS5 and D4-branes from the M5-brane The D0-brane comes from reduction of the spherical M2-brane. The D6-brane of Type IIA is the reduction of the so-called KK6-monopole in 11 dimensions, along the special Taub--NUT circle. Reduction along a longitudinal direction gives the KK5-monopole of Type IIA, whose background we present below in Section \ref{sec:bps_states}.

\subsection{BPS states of extended superalgebra representations}
\label{sec:bps_states}

From the point of view of $\mc{N}=2$ superalagebra the critical R-N black hole solution that preserves half of supersymmetries is a state that saturates the so-called BPS bound (Bogomolmy--Prasad--Sommerfeld). Such states belong to short massive supermultiplets of the SUSY algebra with non-vanishing central charge. The BPS bound is saturated when mass is equal to central charge (up to a sign).

Let us write the $\mc{N}=2$ superalgebra with a central charge in the following, rather unconventional, form
\begin{equation}
    \{Q_\a,Q_\b\} = C^{-1}P^\m (\G_\m)_{\a\b} + i Z \d_{\a\b}.
\end{equation}
Here $\a,\b$ are $D=4$ spinorial indices, the charge conjugation matrix $C=i \G_0\G_2$ and the gamma-matrices are chosen in the Weyl basis:
\begin{equation}
    \G_\m = 
        \begin{bmatrix}
            0 & \s_\m \\
            \bar{\s}_\m & 0
        \end{bmatrix},
\end{equation}
with $\bar{\s}_\m = (1,-\s_i)$. The precise form of the RHS comes from the fact that the LHS is a symmetric product of two spinors. Decomposing the product of two spinorial irreps we find the vector $P^\m$ and the singlet  $Z$ irreps among the symmetric ones. Choose now the rest frame of a massive particle where $P^\m=(m,0,0,0)$. The algebra is then
\begin{equation}
    \{Q_\a,Q_\b\} = i 
        \begin{bmatrix}
            Z & -\s_2 m \\
            \s_2 m & Z
        \end{bmatrix}_{\a\b}
\end{equation} 

Defining now supercharges  $Q_\pm = Q_L \pm \s_2 Q_R$ we arrive at the following commutation relations
\begin{equation}
    \{Q_\pm,Q_\pm\} = 2 i (Z \mp m) \mathbf{1},
\end{equation}
and $\{Q_+,Q_-\}=0$. This implies that half of the supercharges act trivially on  a massive  particle state $|\y\rangle$ with $Z=\pm m$:
\begin{equation}
    Q_\pm |\y\rangle_{Z=\pm m} = 0.
\end{equation}
As a result the multiplet has twice less states that a full massive multiplet (hence the name short). Comparing this to the R-N black hole solution we conclude, that the central charge of a SUSY algebra, when realized on supergravity solutions, is related to charges of extremal black objects. To generalize this to extended algebras in higher dimensions we notice that the trivial action of half of the supercharges follows from degeneracy of the matrix
\begin{equation}
    M= C^{-1}P^\m\G_\m + i Z \mathbf{1}.
\end{equation}
Therefore, the condition $\det M = 0$ will serve as a different way of stating the BPS saturation condition.

In ten space-time dimensions spinors can be both Majorana and Weyl. This means, that to build a supersymmetry algebra we may take i) one supercharge of any chirality, iia) two supercharges of the opposite chiralities, iib) two supercharges of the same chirality. The corresponding algebras are $\mc{N}=(1,0)$, $\mc{N}=(1,1)$, $\mc{N}=(2,0)$ and are referred to as the Type I, Type IIA and Type IIB respectively. Closed string states are classified as multiplets of Type IIA/B supersymmetry, and we will be interested in such states that are invariant under a subset of the total of 32 supercharges. For concreteness we will go through details of the calculation for the Type IIA algebra.

With all possible central charges taken into account the Type IIA supersymmetry algebra can be written as follows
\begin{equation}
\label{eq:IIA_susy}
    \{Q^t,Q\} = C\G^\m P_\m + C \c Z + C\G_\m \c Z^\m+ \fr12 C\G_{\m\n}Z^{\m\n}    + \fr1{4!}\G_{\m_1\dots \m_4}\c Z^{\m_1\dots \m_4} + \fr1{5!}\G_{\m_1\dots \m_5}Z^{\m_1\dots \m_5} 
 , 
\end{equation}
where $C=\G^0$ is the charge conjugation matrix and $\c = \G_0\dots \G_9$ is the chirality matrix. The charge $Q$ here is a 32 component real spinor. This set of central charges has a simple and straightforward origin at the Fierz rules for two Majorana 32-component spinors. The easiest way to track it is to investigate the irreducible representation structure of the bispinor in the LHS of \eqref{eq:IIA_susy}. To start with we note that each supercharge belongs to the $\mathbf{16 + \ol{16}}$ representation of $\frso(1,9)$. The product of two such representations will the decompose according to 
\begin{equation}
    \begin{aligned}
        \mathbf{16 \times 16 } & = \mathbf{10 + 120 + \ol{126}}, \\
        \mathbf{16 \times \ol{16}}  & = \mathbf{1 + 45 + 210}, \\
        \mathbf{\ol{16} \times \ol{16} } & = \mathbf{10 + 120 + 126}. \\
    \end{aligned}
\end{equation}
The Majorana condition results in mutual cancelation of two $\mathbf{120}$, and the remaining representations can be realized as tensors with space-time indices as follows
\begin{equation}
    \begin{aligned}
        & \mathbf{1} && \mathbf{10} && \mathbf{10} && \mathbf{45}  && \mathbf{126 + \ol{126}} && \mathbf{210} \\
        & Z && P_\m && Z^\m && Z^{\m\n} && Z^{\m\n\r\s\t} && Z^{\m\n\r\s}.
    \end{aligned}
\end{equation}
It is worth to notice that from the point of view of the representation theory there is no difference in the two $\mathbf{10}$'s that we distinguish as $P_\m$ and $Z^\m$. However, they have different meaning when the algebra is realized as algebra of states of massive particles. 

As the central charge $Z$ of the $\mc{N}=2$ supergravity in $D=4$ gives charge of a point-like supersymmetric objects: the extremal Reissner--Nordstr\"om black hole,  central charges of the Type IIA SUSY algebra correspond to charges of branes in string theory. The most explicit way to see that is to investigate conditions that shortening of multiplets implies for the supercharges $Q$. As before consider massive states and the rest frame, that is $P^\m = (m,\vec0)$, and start with the case where $Z^\m$ is the only non-vanishing central charge.

Since $C=\G^0$ it appears to be crucial to distinguish between the cases $Z^0 =$ and $Z^0 \neq 0$. We start with the former and chose orientation of the space frame such that only $Z^1 \neq 0$. Then the algebra becomes
\begin{equation}
    \{Q^t,Q\} = m + Z^1 \G_0 \G_1 \c.
\end{equation}
For states when $Z^1 = \pm m$ the supercharges satisfy the the following condition
\begin{equation}
    \big(1 \pm \G_{01}\c\big)Q = 0.
\end{equation}
This is precisely the 1/2BPS condition for the fundamental string, therefore we conclude that $Z^i$ corresponds to the F1 charge. If $Z\neq 0$, while $Z^i=0$, we obtain a different condition
\begin{equation}
    \big(1 \pm \G_{0\dots 9}\big)Q=0.
\end{equation}
This is BPS condition for a 9-brane, which we didn't encounter in the analysis of extremal solutions. Note, that there is no Dirichlet 9-branes in Type IIA theory. This brane is space-time filling, non-dynamical and does not have a corresponding gauge potential. 

Similarly, for $Z^{\m\n}$ we obtain the following conditions depending on whether $Z^{0i}$ or $Z^{ij}$ is non-vanishing:
\begin{equation}
    \begin{aligned}
        & Z^{09}: && \big(1 \pm \G_{0\dots 8}\c\big)Q =0, \\
        & Z^{12}: && \big(1 \pm \G_{012} \big)Q=0. 
    \end{aligned}
\end{equation}
In the first line we insert $1 = \c^2$ to arrive at a product of gamma-matrices involving $\G_0$. The first BPS condition has precisely the form of that for the D8-brane, while the second corresponds to the D2-brane. Note, that D2 and D8 are not magnetically dual. Proceeding further along the same lines we arrive at results, that can be compactly collected into Table \ref{tab:bps_states}

\begin{table}[h]
    \centering
    \begin{tabular}{r|c|c|c|c|c|c|c|c|c}
         central charge & $Z$ & $Z^0$ & $Z^i$ & $Z^{0i}$ & $Z^{ij}$ & $Z^{0ijk}$ & $Z^{ijkl}$ & $Z^{0ijkl}$ & $Z^{ijklp}$  \\
         \hline
        brane & D0  & 9-brane & F1 & D8 & D2 & D6 & D4 & KK5 & NS5 
    \end{tabular}
    \caption{Caption}
    \label{tab:bps_states}
\end{table}

The state KK5 in Table \ref{tab:bps_states} corresponds to what is called the Kaluza--Klein monopole. This is a 5-dimensional brane whose background is not asymptotically flat, but is rather asymptotically Taub--NUT. Such a solution in D=5, now called the Gross--Perry monopole, has first been found in \cite{Gross:1983hb}. In Type II theory the solution for $N$ parallel coinciding (anti-)KK monopoles has the following form
\begin{equation}
\label{eq:KK5}
    \begin{aligned}
        ds^2 =- dt^2 + dx_1^2+\cdots+dx_5^2 + H_{\text{KK}}(r)\, d\vec y\,^2 + H_{\text{KK}}(r)^{-1}\bigl(d\psi+\omega\bigr)^2,
    \end{aligned}
\end{equation}
where $ \nabla\times \vec\omega = \pm \nabla H_{\text{KK}} $ with the harmonic function given by
\begin{equation}
    H_{\text{KK}}(r)=1+\frac{N R_9}{2r},
\end{equation}
and $R_9$ is the radius of the circular direction $\y$. The corresponding BPS condition is
\begin{equation}
    \left(1\pm,\Gamma^{0\dots 5}\c\right)\epsilon_0=0.
\end{equation}
The background does not depend on six coordinates $(t,x^1,\dots,x^5)$, that are world-volume coordinates, and on the circular coordinate $\y$. The latter is not a world-volume and is referred to as the Taub--NUT circle. KK5-monopole is a dynamical object, but it interacts magnetically with the component $g_{i\y}$, which cannot be defined in the full non-linear gravitational theory. For a discussion of KK5-monopole, its relation to NS5-brane via T-duality and its coupling to the magnetic dual of graviton we refer to \cite{Hull:1997kt,Bergshoeff:1997ak}. For a review see \cite{Duff:1994an,Stelle:1998xg}, for a more modern perspective see \cite{Hull:2023iny,Hohm:2018qhd}

\section{Interacting branes}
\label{sec:interact}

In the previous sections we have seen that branes of string theory have their own dynamics, defined by an effective action, non-trivially interact with string states and possess BPS properties under supersymmetry transformations. All this is equally true for Dp-branes, the NS5-brane and for the fundamental string itself, that makes it tempting to put branes at the same level of fundamental objects as the string. In other words, it is natural to consider string theory as a theory of not only strings, but of Dp- and NS5-branes \cite{Johnson:1995bf, Townsend:1996xj}. Interactions between branes as dynamical object develop various effects, which are essentially non-perturbative in nature. The effects related to brane interactions usually manifest themselves in creation of various states of new branes, not present in the system, or binding of branes into a state of lower energy. In this section we consider several such effects and start with supergravity description of brane bound states.

\subsection{Brane bound state backgrounds}
\label{sec:bound}

Bound states of branes arise when more than one conserved charge is present in a single supersymmetric background. They can be extremal or non-extremal, and they may be viewed either as intersections, as branes with world-volume fluxes, or as duality images of simpler parent solutions \cite{Tseytlin:1996bh,Johnson:2003gi}. As we have seen, D-branes are charged with respect to R-R fields of supergravity. However, the Wess-Zumino term of the Dp-brane action in addition to the R-R potential $C_{(p+1)}$ contains R-R form fields of lower rank and the Kalb--Ramond form. This implies, that in addition to its own charge a D-brane may carry the charge of lower-dimensional branes or of fundamental strings.

In many cases this additional charge is not associated with a separate localized constituent, but is instead encoded in a nontrivial gauge-field configuration on the brane world-volume. It is in this precise sense that one speaks of dissolved charge. As an explicit example consider the D3-brane action
\begin{equation}
\label{eq:D3_action1}
S_{D3}
=
-T_3 \int d^4\xi \, e^{-\phi}
\sqrt{-\det\bigl(G+ \mc{F}\bigr)}
+
\mu_3 \int
\left(
C_4 + C_2 \wedge \mathcal F + \frac12 C_0 \, \mathcal F \wedge \mathcal F
\right),
\end{equation}
where $\mF = B_2 + 2\pi\alpha' F $, and $G$ and $B_2$ are understood as pull-backs of the corresponding space-time fields  to the D3-brane world-volume Equation \eqref{eq:D3_action1} makes manifest that the same object may simultaneously couple to several bulk fields, and therefore carry several charges at once. In supergravity this is reflected in the existence of bound-state backgrounds, whereas in the world-volume description the same phenomenon is encoded in electric or magnetic fluxes \cite{Polchinski:1995mt,Douglas:1995bn,Witten:1995im,Johnson:2003gi}.

To write the supergravity solution corresponding to the D3-F1 bound state let the D3-brane extend along the spatial directions $(x^1,x^2,x^3)$, and let the string charge be oriented along $x^1$. In the string frame the solution reads
\begin{equation}
\label{eq:D3F1}
\begin{aligned}
ds^2_{\mathrm{str}} &= H^{-1/2}
\Bigl[
h_F^{-1}\bigl(-dt^2+dx_1^2\bigr)
+dx_2^2+dx_3^2
\Bigr]
+
H^{1/2}
\Bigl(
dr^2+r^2 d\Omega_5^2
\Bigr), \\
e^{2\phi} &= g_s^2\, h_F^{-1}, \\
B_{01} &= \sin\varphi \, \bigl(H^{-1}-1\bigr)\, h_F,\\
C &= \frac{\cos\varphi}{g_s}\, \bigl(H^{-1}-1\bigr)\, h_F dx^0\wedge \cdots dx^3 
\end{aligned}
\end{equation}
with
\begin{equation}
H(r)=1+\frac{r_0^4}{r^4},
\qquad
h_F^{-1}
=
\cos^2\varphi + \sin^2\varphi \, H^{-1}.
\label{eq:D3F1_harmonic_review}
\end{equation}
The associated five-form field strength is
\begin{equation}
F_5 = (1+*)\, dC_4 .
\label{eq:D3F1_F5_review}
\end{equation}
The angle $\varphi$ determines the relative admixture of D3 and F1 charge: for $\varphi=0$ one recovers the pure D3-brane solution, while nonzero $\varphi$ turns on a nontrivial NS-NS two-form potential and hence fundamental-string charge \cite{Tseytlin:1996bh,Lu:1999uca,Johnson:2003gi}.

The physical interpretation of this geometry is particularly instructive. The D3-brane charge is carried in the standard way by the self-dual five-form flux and is measured by
\begin{equation}
Q_{D3} \propto \int_{S^5} F_5 .
\label{eq:D3_charge_review}
\end{equation}
The string charge, however, is of a different nature. It is not represented by a separate macroscopic string placed alongside the D3-brane, but by an electric flux on the D3-brane world-volume. This is why one says that the F1 charge is dissolved in the D3-brane.

To see that, note first that the fundamental string couples electrically to the NS-NS two-form $B_2$. Accordingly, the F1 charge carried by the D3-brane is obtained by varying the world-volume action with respect to $B_{01}$, or equivalently with respect to $\mathcal F_{01}$. The corresponding charge density is the electric displacement,
\begin{equation}
\Pi^1
=
\frac{\partial \mathcal L}{\partial \mathcal F_{01}} .
\label{eq:F1_displacement_review}
\end{equation}
For a flat D3-brane with a purely electric world-volume field, the DBI Lagrangian density takes the form
\begin{equation}
\mathcal L_{\mathrm{DBI}}
=
-T_3 \sqrt{1-\mathcal F_{01}^2},
\label{eq:DBI_electric_review}
\end{equation}
so that
\begin{equation}
\Pi^1
=
\frac{T_3 \, \mathcal F_{01}}{\sqrt{1-\mathcal F_{01}^2}} .
\label{eq:F1_density_review}
\end{equation}
This quantity is proportional to the density of dissolved strings running along $x^1$. This charge is electric because it is associated with the component $\mathcal F_{01}$, which contains one time index and one spatial index. In other words, the D3--F1 bound state is realized by an electric field on the D3-brane world-volume, and the fundamental strings appear as electric flux lines spread across the brane rather than as individually separated objects \cite{Witten:1995im,Polchinski:1995mt,Johnson:2003gi}. It is therefore important to distinguish this mechanism from the one responsible for dissolved lower-dimensional D-brane charge, to which we now turn.

As an example of dissolved lower-dimensional D-brane charge consider the D3-D1 system, that is S-dual to the D3-F1 system. Let the D3-brane again extend along $(x^1,x^2,x^3)$, and let the D1-brane be aligned along $x^1$. A convenient form of the extremal D3-D1 solution is
\begin{equation}
\label{eq:D3D1}
\begin{aligned}
ds^2_{\mathrm{str}} &= H^{-\fr12}
\Bigl[-dt^2+dx_1^2 +h_1 \bigl(dx_2^2+dx_3^2\bigr)\Bigr]
+ H^{\fr12}\Bigl(dr^2+r^2 d\Omega_5^2\Bigr),\\
e^{2\phi} &= g_s^2\, h_1,\\
C_{01} &= \frac{\sin\theta}{g_s}\, \bigl(H^{-1}-1\bigr),\\
B_{23} &= \tan\theta \, \bigl(H^{-1} h_1 -1 \bigr),\\
C_{0123} &= \frac{\cos\theta}{g_s}\, \bigl(H^{-1}-1\bigr)\, h_1 ,
\end{aligned}
\end{equation}
where
\begin{equation}
\label{eq:D3D1_harm}
H(r)=1+\frac{r_0^4}{r^4},
\qquad
h_1^{-1}
=
\sin^2\theta\, H^{-1}+\cos^2\theta .
\end{equation}
The parameter $\theta$ controls the relative proportion of D3 and D1 charge. For $\theta=0$ the solution reduces to the ordinary D3-brane, whereas nonzero $\theta$ turns on a nontrivial R-R two-form potential $C_2$ and hence D1-brane charge \cite{Tseytlin:1996bh,Breckenridge:1996tt,Lu:1999uca}. Since the D1-brane couples electrically to the R-R two-form $C_2$, its charge is measured by the corresponding three-form field strength,
\begin{equation}
q_{D1} \propto \int_{S^5} *F_3 ,
\qquad
F_3=dC_2 ,
\label{eq:D1_charge_review}
\end{equation}
whereas the D3 charge remains
\begin{equation}
Q_{D3} \propto \int_{S^5} F_5 .
\label{eq:D3_charge_again_review}
\end{equation}

Thus the supergravity solution clearly describes a single BPS geometry carrying both D3 and D1 charge. As in the D3-F1 case the world-volume description makes int manifest that the D1 charge is induced by a flux on the D3-brane, magnetic this time. Indeed, the relevant term in the D3-brane Wess--Zumino action is
\begin{equation}
S_{\mathrm{WZ}}
=
\mu_3 \int_{\mathcal W_4}
\left(
C_4 + C_2 \wedge \mathcal F + \frac12 C_0 \, \mathcal F \wedge \mathcal F
\right).
\label{eq:D3_WZ_review}
\end{equation}
If one turns on a world-volume field strength in the $(x^2,x^3)$ plane,
$\mathcal F_{23}\neq 0$, then the term linear in $\mathcal F$ becomes
\begin{equation}
\mu_3 \int_{\mathcal W_4} C_2 \wedge \mathcal F
=
\mu_3 \int dt \wedge dx^1 \wedge dx^2 \wedge dx^3 \;
C_{01}\, \mathcal F_{23}.
\label{eq:D1_from_flux_review}
\end{equation}
This has precisely the structure of a D1-brane coupling along the $(t,x^1)$ directions. If the $(x^2,x^3)$ directions are compact, flux quantization implies
\begin{equation}
\frac{1}{2\pi}\int_{\Sigma_{23}} F = n \in \mathbb Z ,
\label{eq:flux_quantization_review}
\end{equation}
and one finds
\begin{equation}
\mu_3 \int_{\mathcal W_4} (2\pi\alpha' F)\wedge C_2
=
n\, \mu_1 \int_{(t,x^1)} C_2 .
\label{eq:n_units_D1_review}
\end{equation}
Hence $n$ units of magnetic flux induce $n$ units of D1-brane charge. This is the precise sense in which the D1-brane is dissolved in the D3-brane: its charge is carried by the flux sector of the D3 world-volume theory rather than by a separately localized stringlike object. In contrast to the D3--F1 system, where the relevant quantity is an electric displacement, the D3--D1 system is governed by a purely spatial, and therefore magnetic, world-volume field.

Therefore, the D3--F1 and D3--D1 systems illustrate two complementary realizations of dissolved charge. In the former, the additional charge is that of the fundamental string and is carried by an electric world-volume field, as encoded in the DBI coupling to $B_2$. In the latter, the additional charge is the D1-brane charge and is carried by a magnetic world-volume flux, as dictated directly by the Wess--Zumino coupling $\int C_2 \wedge \mathcal F$. More generally, the expansion of $\int C \wedge e^{\mathcal F}$ explains why magnetic flux on a D$p$-brane induces lower-dimensional D-brane charges. The D3--D1 system is simply the first nontrivial example of a much more general mechanism. For a D$p$-brane, the Wess--Zumino term takes the universal form
\begin{equation}
S_{\mathrm{WZ}}^{(p)}
=
\mu_p \int_{\mathcal W_{p+1}} C \wedge e^{\mathcal F},
\qquad
C=\sum_q C_q .
\label{eq:general_WZ_review}
\end{equation}
Expanding the exponential yields
\begin{equation}
S_{\mathrm{WZ}}^{(p)}
=
\mu_p \int
\left(
C_{p+1}
+\mathcal F \wedge C_{p-1}
+\frac12 \mathcal F \wedge \mathcal F \wedge C_{p-3}
+\cdots
\right).
\label{eq:general_WZ_expand_review}
\end{equation}
If $\mathcal F$ is purely spatial, then it is naturally interpreted as a magnetic flux on the brane world-volume. The term $\mathcal F \wedge C_{p-1}$ shows that a magnetic two-form flux induces D$(p-2)$-brane charge, while the term $\mathcal F \wedge \mathcal F \wedge C_{p-3}$ induces D$(p-4)$-brane charge, and so forth. In this way lower-dimensional D-branes may be regarded as flux defects or topological sectors inside higher-dimensional D-branes \cite{Douglas:1995bn,Witten:1995im,Johnson:2003gi}. The supergravity bound-state backgrounds are the macroscopic realization of this world-volume statement: a single geometry carries several R-R charges because the underlying D-brane configuration supports nontrivial quantized flux.

We conclude that the spectrum of supergravity fields and the central charges of the BPS algebra do not simply encode existence of standalone branes. Instead they also carry information on how multiple charges are packed into a single BPS object. A non-zero R-R potential with legs on the world-volume indicates D-brane charge, a non-zero $B$-field may indicate dissolved strings, and off-diagonal metric components may indicate momentum or KK structure. The resulting bound states are therefore not ad hoc constructions but natural solutions of the low-energy equations, tightly constrained by supersymmetry and duality. In this sense the classical backgrounds already contain a large portion of the physical information later captured more microscopically by D-brane effective actions and string dualities \cite{Polchinski:1995mt,Tseytlin:1996bh,Johnson:2003gi,Harmark:2000wv}.

\subsection{Branes ending on branes}
\label{sec:intersect}

Gauge invariance of Wess--Zumino action under field transformations requires introduction of additional gauge potentials. This is similar to how the Born--Infeld field $A$ entering the DBI action ensures invariance under gauge transformation $B_2 \to B_2 + d\L_1$. And similar to how this field interacts with endpoints of the open string, the other additional gauge fields on Dp-branes and the NS5-brane interact with endpoints of other D-branes. From the supergravity point of view systems, where one brane is ending on another, correspond to intersecting brane field configurations. These systems are described by harmonic superpositions of elementary brane solutions and preserve a fraction of the underlying supersymmetry determined by the compatibility of the constituent projection conditions \cite{Tseytlin:1996bh,Stelle:1998xg}. Orthogonal pairwise intersections of the standard 1/2BPS objects typically preserve one quarter of the maximal supersymmetry. More complicated webs or intersections at angles preserve one eighth or less. This reduction of supersymmetry is precisely what makes such systems useful in the construction of lower-dimensional effective theories with nontrivial matter content. In particular, intersecting D-brane systems and five-brane configurations have played a central role in brane engineering of supersymmetric gauge theories and, after compactification and orientifolding, in string-motivated particle physics models with chiral matter \cite{Hanany:1996ie,Giveon:1998sr,Uranga:2003pz,Blumenhagen:2005mu}.

One can follow a simple line of arguments to see that Dp-branes can end on each other. Start with F1 ending on D3 and perform S-duality to arrive at D1 ending on the same D3, that is self-dual. A chain of T-dualities can change dimensions of both D1 and D3, leading us to the desired conclusion. Starting now with the D3-brane ending on the D5-brane and performing S-duality again we arrive at the NS5-D3 system, where again D3 can be T-dualized into any Dp-brane. We conclude, that Dp-brane can end on the NS5-brane. The corresponding supergravity backgrounds are most straightforwardly obtained by the harmonic superposition rule and are most naturally written in a partially delocalized form \cite{Tseytlin:1996bh,Stelle:1998xg,Duff:1994an,Johnson:2003gi}. As a concrete example we consider the D4-D2 and NS5-D2 systems. Both preserve one quarter of the total of 32 supercharges, corresponding to eight unbroken supersymmetries, and both admit a direct interpretation in terms of branes intersecting over the common directions $(t,x^1)$. From the world-volume point of view these correspond to the D2-brane ending on the D4 or the NS5-brane.

Start with the D4-D2 system where the D4 and D2 branes are oriented in the 10D space-time according to the intersection Table \ref{tab:D4D2_orientation}, so that the D4-brane spans $(x^1,x^2,x^3,x^4)$, the D2-brane spans $(x^1,x^5)$, and the common intersection is the string $(t,x^1)$.
\begin{table}[h]
    \centering
    \begin{tabular}{c|cccccccccc}
        &0&1&2&3&4&5&6&7&8&9\\ 
        \hline
        D4&$\times$&$\times$&$\times$&$\times$&$\times$&$-$&$-$&$-$&$-$&$-$\\
        D2&$\times$&$\times$&$-$&$-$&$-$&$\times$&$-$&$-$&$-$&$-$
    \end{tabular}
    \caption{Orientation of intersecting D4 and D2 branes: $\times$ denotes world-volume directions, $-$ denotes transverse directions.}
    \label{tab:D4D2_orientation}
\end{table}
The relative transverse directions are $(x^2,x^3,x^4)$ for the D2-brane and $x^5$ for the D4-brane, while the overall transverse space is $\mathbb R^4_{6789}$. The corresponding partially delocalized $1/4$BPS solution in the Type IIA string frame is
\begin{equation}
\label{eq:D4D2_inter}
\begin{aligned}
ds^2_{\mathrm{str}} &= H_2^{-\fr12}H_4^{-\fr12}\left(-dt^2+dx_1^2\right)
+
H_2^{\fr12}H_4^{-\fr12}\left(dx_2^2+dx_3^2+dx_4^2\right)
+ H_2^{-\fr12}H_4^{\fr12}\,dx_5^2\\ 
& \quad + H_2^{\fr12}H_4^{\fr12}\left(dr^2+r^2 d\Omega_3^2\right),\\
e^{\phi} &= g_s\, H_2^{\fr14}H_4^{-\fr14},\\
C_3 &= \left(H_2^{-1}-1\right)\, dt\wedge dx^1\wedge dx^5,\\
C_5 &= \left(H_4^{-1}-1\right)\, dt\wedge dx^1\wedge dx^2\wedge dx^3\wedge dx^4,
\end{aligned}
\end{equation}
with harmonic functions
\begin{equation}
H_2(r)=1+\frac{Q_2}{r^2},
\qquad
H_4(r)=1+\frac{Q_4}{r^2},
\qquad
r^2=x_6^2+x_7^2+x_8^2+x_9^2 .
\label{eq:D4D2_harmonics_uniform}
\end{equation}
The corresponding electric R-R field strengths may be written as
\begin{equation}
F_4 = dt\wedge dx^1\wedge dx^5 \wedge d(H_2^{-1}),
\qquad
F_6 = dt\wedge dx^1\wedge dx^2\wedge dx^3\wedge dx^4 \wedge d(H_4^{-1}),
\label{eq:D4D2_fluxes_uniform}
\end{equation}
with $F_6$ dual to the magnetic four-form description of the D4-brane. The preserved supersymmetries are determined by two mutually compatible projector conditions
\begin{equation}
\Gamma_{015}\,\epsilon = \epsilon,
\qquad
\Gamma_{01234}\c\,\epsilon = \epsilon.
\label{eq:D4D2_projectors}
\end{equation}
Since the two conditions are independent and commute, the configuration preserves one quarter of the supersymmetry. 

Partial delocalization of the solution \eqref{eq:D4D2_inter} means that both harmonic functions depend only on the common overall transverse radius $r$ in $\mathbb R^4_{6789}$. Thus the D2-brane is smeared along the directions $(x^2,x^3,x^4)$, while the D4-brane is smeared along the  direction $x^5$. This smearing is what allows the simple harmonic superposition form. Fully localized orthogonal intersections are substantially more complicated and, in many cases, no comparably simple closed form is known. Nevertheless, the delocalized solution correctly captures the long-distance charges, supersymmetry, and asymptotic geometry. It is also closely related to the world-volume description in which D2 charge is dissolved in a D4-brane by magnetic flux through the coupling $\int_{W_5} C_3\wedge \mathcal F$.

Our second example is the orthogonal NS5--D2 intersection encoded by the intersection Table \ref{tab:NS5D2_orientation}, 
where the NS5-brane spans $(x^1,x^2,x^3,x^4,x^5)$, the D2-brane spans $(x^1,x^6)$, and the common intersection is again the string $(t,x^1)$. 
\begin{table}[h]
    \centering
    \begin{tabular}{c|cccccccccc}
        &0&1&2&3&4&5&6&7&8&9\\ \hline
        NS5&$\times$&$\times$&$\times$&$\times$&$\times$&$\times$&$-$&$-$&$-$&$-$\\
        D2&$\times$&$\times$&$-$&$-$&$-$&$-$&$\times$&$-$&$-$&$-$
    \end{tabular}
    \caption{Orientation of intersecting NS5 and D2 branes: $\times$ denotes world-volume directions, $-$ denotes transverse directions.}
    \label{tab:NS5D2_orientation}
\end{table}
The D2-only relative transverse direction is $x^6$, the NS5-only relative transverse directions are $(x^2,x^3,x^4,x^5)$, and the overall transverse space is $\mathbb R^3_{789}$. The corresponding delocalized supergravity solution  takes the form
\begin{equation}
\label{eq:NS5D2_intersect}
\begin{aligned}
ds^2_{\mathrm{str}} &= H_2^{-\fr12}\left(-dt^2+dx_1^2\right)
+ H_2^{\fr12}\left(dx_2^2+dx_3^2+dx_4^2+dx_5^2\right)
+H_2^{-\fr12}H_5\, dx_6^2\\
&\quad +H_2^{\fr12}H_5\left(dr^2+r^2 d\Omega_2^2\right),\\
e^{\phi} &= g_s\, H_2^{\fr14}H_5^{\fr12},\\
C_3 &= \left(H_2^{-1}-1\right)\, dt\wedge dx^1\wedge dx^6,
\end{aligned}
\end{equation}
where the R-R four-form is
\begin{equation}
F_4 = dt\wedge dx^1\wedge dx^6 \wedge d(H_2^{-1}),
\label{eq:NS5D2_F4_uniform}
\end{equation}
and the NS-NS three-form sourced by the NS5-brane may be written as
\begin{equation}
H_3 = dx^6 \wedge \star_3 dH_5 .
\label{eq:NS5D2_H3_uniform}
\end{equation}
Here $\star_3$ denotes the Hodge dual on the overall transverse $\mathbb R^3_{789}$, and the harmonic functions are
\begin{equation}
H_2(r)=1+\frac{Q_2}{r},
\qquad
H_5(r)=1+\frac{Q_5}{r},
\qquad
r^2=x_7^2+x_8^2+x_9^2 .
\label{eq:NS5D2_harmonics_uniform}
\end{equation}
As in the D4--D2 case, the unbroken supersymmetries are selected by two compatible projectors
\begin{equation}
\Gamma_{016}\,\epsilon = \epsilon,
\qquad
\Gamma_{012345}\,\epsilon = \epsilon.
\label{eq:NS5D2_projectors}
\end{equation}
These conditions are independent and compatible, so the intersection preserves one quarter of the supersymmetry. In physical terms, this is the expected amount for two orthogonal branes intersecting over a string. 

The background \eqref{eq:NS5D2_intersect} is again partially delocalized. The D2 harmonic function and the NS5 harmonic function both depend only on the common overall transverse radius in $\mathbb R^3_{789}$. Consequently, the D2-brane is smeared along the NS5-only directions $(x^2,x^3,x^4,x^5)$, while the NS5-brane is smeared along the D2-only direction $x^6$. This is the natural harmonic-superposition solution associated with the NS5--D2 intersection. In a fully localized picture the D2-brane may end on the NS5-brane, and its boundary appears as a string-like defect on the NS5 world-volume; the delocalized supergravity background should be viewed as the large-distance BPS geometry carrying the corresponding charges. Dynamics of such string-like objects is known as Little String Theory whose existence was suggested in \cite{Seiberg:1997zk,Losev:1997hx} (a review can be found e.g. in \cite{Aharony:1999ks,Kutasov:2001uf,Gubser:2010zz})

Intersecting brane background have close relation to the so-called Hanany--Witten effect, to be discussed in Section \ref{sec:hanany-witten}, that states that a brane stretched between two branes is created, when they are moved through each other. For example, if one moves the D4-brane through the NS5-brane a D2-brane stretched between the two is created. The statement that the created D2-brane can consistently end on both the NS5-brane and the D4-brane has a precise world-volume origin. In each case, the endpoint of the D2-brane is realized as a charged defect for an appropriate field living on the host brane. For the NS5-brane, the endpoint string is electrically charged under the world-volume two-form potential. For the D4-brane, the same endpoint string is most naturally described either as electrically charged under the dual two-form potential or, equivalently, as a magnetic source for the Born--Infeld gauge field. These two descriptions are the local world-volume counterparts of the supergravity statement that D2 Page charge is generated when the NS5-brane crosses the D4-brane. Let us discuss both cases in more details.

\textbf{D2-brane ending on the NS5-brane.} The Type IIA NS5-brane world-volume contains a two-form gauge potential $b_2$ inherited from the M5-brane under reduction, where it acts as the analogue of the Born--Infeld gauge field interacting with endpoints of the M2-brane. The gauge-invariant three-form field strength is then
\begin{equation}
h_3 = db_2 - P[C_3] ,
\label{eq:NS5_h3_def}
\end{equation}
where $P[C_3]$ denotes the pullback of the R-R three-form potential to the NS5 world-volume. The precise completion of \eqref{eq:NS5_h3_def} may involve additional background fields, but this simplified form is sufficient for the present purpose. Now, recall that the open D2-brane action contains the Wess--Zumino coupling
\begin{equation}
S_{D2}^{\mathrm{WZ}}
=
\mu_2 \int_{M_3} P[C_3] ,
\label{eq:open_D2_WZ_NS5}
\end{equation}
where $M_3$ is the D2 world-volume. If the D2-brane is open and ends on the NS5-brane, then $\partial M_3=\Sigma_2$ lies inside the NS5 world-volume. Under an R-R gauge transformation
\begin{equation}
\delta C_3 = d\Lambda_2 ,
\label{eq:C3_gauge_var}
\end{equation}
the variation of \eqref{eq:open_D2_WZ_NS5} is
\begin{equation}
\delta S_{D2}^{\mathrm{WZ}}
=
\mu_2 \int_{M_3} d\Lambda_2
=
\mu_2 \int_{\Sigma_2} \Lambda_2 .
\label{eq:open_D2_boundary_var}
\end{equation}
Gauge invariance is restored provided one adds to the NS5-brane world-volume action the boundary coupling
\begin{equation}
S_{\partial D2-\mathrm{NS5}}
=
-\mu_2 \int_{\Sigma_2} b_2 ,
\label{eq:boundary_b2_coupling}
\end{equation}
and assigns the transformation law
\begin{equation}
\delta b_2 = -\Lambda_2 .
\label{eq:b2_gauge_var}
\end{equation}
Then the variation of \eqref{eq:boundary_b2_coupling} precisely cancels \eqref{eq:open_D2_boundary_var}. This is the direct world-volume proof that an open D2-brane may consistently end on the NS5-brane. Also we see the direct analogy between the field $b_2$ and the Born--Infeld field $A_1$ in the Dp-brane action.

The boundary $\Sigma_2$ is therefore a string-like defect electrically charged under the NS5 world-volume two-form $b_2$. Equivalently, it acts as a source in the field equation for $h_3$. Schematically one has
\begin{equation}
d(*_6 h_3) = 2\pi\, \delta_4(\Sigma_2) ,
\label{eq:NS5_eom_with_string}
\end{equation}
where $\delta_4(\Sigma_2)$ is the four-form Poincar\'e dual to the string world-sheet inside the six-dimensional NS5 world-volume. Since in the chiral theory on the Type IIA NS5-brane the three-form field strength is self-dual, the same object may also be viewed as magnetically charged. Integrating \eqref{eq:NS5_eom_with_string} over a four-ball $B^4$ whose boundary is a linking three-sphere $S^3$ around the endpoint string gives
\begin{equation}
\int_{S^3} *_6 h_3 = 2\pi N_{D2},
\label{eq:NS5_endpoint_charge}
\end{equation}
or, using self-duality, equivalently
\begin{equation}
\int_{S^3} h_3 = 2\pi N_{D2} .
\label{eq:NS5_endpoint_charge_selfdual}
\end{equation}
Thus the endpoint of the D2-brane is measured by the flux of the NS5 world-volume three-form through the $S^3$ linking the string. This is the direct analogue, on the five-brane world-volume, of the familiar statement that the endpoint of a fundamental string on a D-brane is an electric source for the Born--Infeld field \cite{Strominger:1995ac,Townsend:1995kk,Howe:1997ue}.

\textbf{D2-brane ending on the D4-brane.} The world-volume derivation for the D4-brane is slightly different and is closely tied to the mechanism of dissolved D2 charge discussed earlier. The D4-brane supports a Born--Infeld one-form gauge potential $A_1$ with field strength
\begin{equation}
\mathcal F = P[B_2] + 2\pi\alpha' F_2 ,
\qquad
F_2=dA_1 .
\label{eq:D4_calF_def}
\end{equation}
Its Wess--Zumino action contains the coupling
\begin{equation}
S_{\mathrm{WZ}}^{D4}
=
\mu_4 \int_{W_5}
\left(
C_5 + C_3 \wedge \mathcal F + \frac12 C_1 \wedge \mathcal F \wedge \mathcal F
\right),
\label{eq:D4_WZ_action}
\end{equation}
where $W_5$ is the D4-brane world-volume. The term linear in $C_3$ implies that magnetic flux on the D4-brane induces D2-brane charge. Varying \eqref{eq:D4_WZ_action} with respect to $C_3$ gives the induced D2 current on the D4-brane,
\begin{equation}
*j_{D2}^{(D4)} = \mu_4 \, \mathcal F .
\label{eq:D2_current_from_D4_flux}
\end{equation}
Thus D2-brane charge dissolved in the D4-brane is measured by the flux of $\mathcal F$ through a two-cycle. In particular, for a two-sphere $S^2_{\mathrm{link}}$ linking the endpoint string inside the spatial D4 world-volume, one has
\begin{equation}
N_{D2}
=
\frac{1}{2\pi} \int_{S^2_{\mathrm{link}}} F_2 ,
\label{eq:D2_charge_from_D4_flux}
\end{equation}
up to the standard normalization conventions for $\alpha'$ and $\mu_p$. This already suggests that the endpoint of a D2-brane on a D4-brane should behave as a magnetic source for the D4 world-volume gauge field.

Indeed, suppose that an open D2-brane ends on the D4-brane along a string world-sheet $\Sigma_2 \subset W_5$. Then the boundary of the D2-brane acts as a magnetic defect for the world-volume field strength:
\begin{equation}
dF_2 = 2\pi\, \delta_3(\Sigma_2) .
\label{eq:D4_magnetic_source}
\end{equation}
Here $\delta_3(\Sigma_2)$ is the three-form Poincar\'e dual to the string world-sheet inside the five-dimensional D4-brane world-volume. Integrating \eqref{eq:D4_magnetic_source} over a three-ball $B^3$ transverse to the string gives
\begin{equation}
\frac{1}{2\pi}\int_{S^2_{\mathrm{link}}} F_2
=
\int_{B^3} \delta_3(\Sigma_2)
=
N_{D2} ,
\label{eq:D4_magnetic_charge_quantization}
\end{equation}
where $S^2_{\mathrm{link}}=\partial B^3$ links the endpoint string. Thus the endpoint of the D2-brane carries one unit of magnetic charge with respect to the D4 world-volume gauge field. This is the precise world-volume statement that a D2-brane can end on a D4-brane.

It is useful to phrase the same result in a dual language. In five dimensions a one-form gauge field is dual to a two-form gauge potential $\widetilde A_2$, whose gauge-invariant field strength $\widetilde{\mathcal F}_3$ is Hodge dual to the variation of the D4 action with respect to $F_2$. In that dual formulation, the endpoint string couples electrically,
\begin{equation}
S_{\partial D2-\mathrm{D4}}
=
\mu_2 \int_{\Sigma_2} \widetilde A_2 ,
\label{eq:D4_dual_2form_coupling}
\end{equation}
which makes the analogy with the NS5-brane case manifest. The description in terms of $A_1$ and the magnetic source equation \eqref{eq:D4_magnetic_source}, however, is usually more convenient for discussing dissolved D2 charge via the Wess--Zumino term \eqref{eq:D4_WZ_action}.

\subsection{Hanany--Witten effect}
\label{sec:hanany-witten}

As dynamical objects Dp-branes can be created or annihilated in certain processes. However, due to their nature, non-perturbative in the string coupling constant $g_s$, one cannot see these processes in the standard string perturbation theory (probably, except the brane--anti-brane instability, that manifests itself in the presence of tachyons). In this section and in Section \ref{sec:myers} we consider in details two effects, the Hanany--Witten effect and Myers polarization, where brane creation and restructuring can be traced using the world-volume and supergravity description.

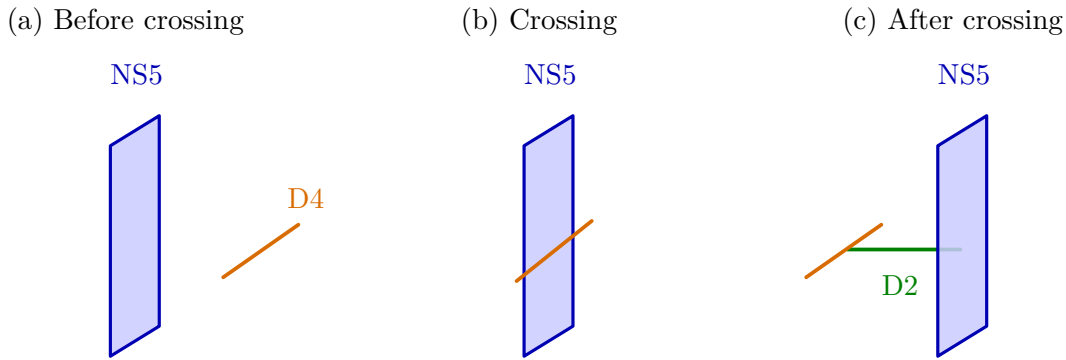
\begin{figure}[h]
\centering
\begin{tikzpicture}[scale=1.0, line cap=round, line join=round]

  \tikzset{
    nsfive/.style={draw=blue!70!black, fill=blue!20, very thick, fill opacity=0.85},
    dfour/.style={draw=orange!85!black, fill=orange!22, line width=0.5mm, fill opacity=0.85},
    dtwo/.style={draw=green!50!black, fill=green!25, line width=0.5mm, fill opacity=0.9},
    move/.style={-{Latex[length=2.5mm]}, very thick, red!70!black},
    lab/.style={font=\small}
  }

  \def\xA{0}
  \def\xB{5.5}
  \def\xC{11.0}

  \node[lab] at (\xA,2.85) {(a) Before crossing};
  \node[lab] at (\xB,2.85) {(b) Crossing};
  \node[lab] at (\xC,2.85) {(c) After crossing};

  \filldraw[nsfive]
    (\xA-0.2,-1.6) -- (\xA+0.45,-1.2) -- (\xA+0.45,1.6) -- (\xA-0.2,1.2) -- cycle;
  \node[lab, blue!70!black] at (\xA+0.15,2.15) {NS5};

  \filldraw[dfour]
    (\xA+1.3,-0.55) -- (\xA+2.3,0.15);
 
  \node[lab, orange!85!black] at (\xA+2.4,0.5) {D4};


  \filldraw[nsfive]
    (\xB-0.2,-1.6) -- (\xB+0.45,-1.2) -- (\xB+0.45,1.6) -- (\xB-0.2,1.2) -- cycle;
  \node[lab, blue!70!black] at (\xB+0.15,2.15) {NS5};

  \filldraw[dfour]
    (\xB-0.3,-0.6) -- (\xB+0.7,0.2);

  \filldraw[dtwo]
    (\xC-1.4,-0.18) -- (\xC+0.1,-0.18);

  \filldraw[nsfive]
    (\xC-0.2,-1.6) -- (\xC+0.45,-1.2) -- (\xC+0.45,1.6) -- (\xC-0.2,1.2) -- cycle;
  \node[lab, blue!70!black] at (\xC+0.15,2.15) {NS5};

  \filldraw[dfour]
    (\xC-1.95,-0.55) -- (\xC-0.95,0.15);

  \node[lab, green!50!black] at (\xC-0.7,-0.65) {D2};


\end{tikzpicture}
\caption{Schematic depiction of the Hanany--Witten effect: when the D4-brane passes through the NS5-brane, a D2-brane is created stretched between them.}
\label{fig:hanany3d}
\end{figure}

The Hanany--Witten effect manifests itself in creation of a brane when a Dp-brane is moved through the NS5-brane. It was first observed in \cite{Hanany:1996ie,Giveon:1998sr} for the Type IIB system where moving of the D5-brane through the NS5-brane creates a D3-brane stretched between the two. To make contact to the previous sections we consider here its Type IIA counterpart and will move the D4-brane across the NS5-brane. Schematics of this effect is depicted on Fig. \ref{fig:hanany3d} in three stages: before the crossing, during the crossing and after the crossing. 
Before turning to calculational details, it is instructing to have in mind the overall brane-creation process described as a step-by-step process.

\textbf{Step 1: initial configuration.}
Suppose that the NS5-brane is located at some value $x^6=x^6_{\mathrm{NS}}$ and the D4-brane at $x^6=x^6_{\mathrm{D4}}$, with $x^6_{\mathrm{NS}}<x^6_{\mathrm{D4}}$. Assume initially that no D2-brane stretches between them. The configuration is then static, supersymmetric, and characterized only by the NS5 and D4 charges.

\textbf{Step 2: adiabatic motion.}
Now move the NS5-brane slowly in the positive $x^6$ direction. Because the brane system preserves supersymmetry, the process can be followed adiabatically and no instability need develop before the actual crossing. At this stage  there is still no stretched D2-brane, and the NS5 and D4 are simply approaching one another in the transverse $x^6$ direction.

\textbf{Step 3: crossing.}
When the NS5-brane reaches the position of the D4-brane, the naive picture of two independent source branes ceases to be adequate. At the crossing point the magnetic source for the NS-NS three-form $H_3$ carried by the NS5-brane meets the R-R flux sourced by the D4-brane. As we explain below, the modified Bianchi identities imply that D2 brane-source charge is generated precisely at this event. Equivalently, the topological data encoded in the linking numbers would jump discontinuously unless a D2-brane were created.

\textbf{Step 4: final configuration.}
After the NS5-brane has moved to $x^6_{\mathrm{NS}}>x^6_{\mathrm{D4}}$, a D2-brane stretched between the D4 and the NS5 must be present. If one reverses the motion and moves the NS5-brane back through the D4-brane, the D2-brane disappears. If $N$ coincident D4-branes are crossed, then $N$ D2-branes are created. The effect is therefore discrete and quantized, as expected for brane charge, and moreover completely reversal and thus symmetric.

Apparently, this process is non-perturbative and one cannot use the string perturbation theory for its analysis. Instead, we rely on supergravity description, that involves generalized Bianchi identities \eqref{eq:BIs}, and knowledge of world-volume fluxes. Start with the brane orientation  given by the intersection Table \ref{tab:hanany}.
\begin{table}[h]
    \centering
    \begin{tabular}{c|cccccccccc}
& 0 & 1 & 2 & 3 & 4 & 5 & 6 & 7 & 8 & 9 \\ 
\hline
{NS5} & $\times$ & $\times$ & $\times$ & $\times$ & $\times$ & $\times$ & $-$ & $-$ & $-$ & $-$ \\
{D4}  & $\times$ & $\times$ & $-$ & $-$ & $-$ & $-$ & $-$ & $\times$ & $\times$ & $\times$ \\
{D2}  & $\times$ & $\times$ & $-$ & $-$ & $-$ & $-$ & $\times$ & $-$ & $-$ & $-$
    \end{tabular}
    \caption{Orientation of the D4-brane, the NS5-brane, and of the D2-brane that is created in the Hanany--Witten process.}
    \label{tab:hanany}
\end{table}
The NS5-brane and the D4-brane intersect over the $(x^0,x^1)$ directions. When present, the D2-brane stretches along $x^6$ between them. This system preserves one quarter of the supersymmetry and is T-dual to the original Type IIB NS5--D5--D3 construction.

The generalized Bianchi identities relevant for the system in question are given by
\begin{equation}
\label{eq:BIsHW}
    \begin{aligned}
        dF_4 & = *j_{D4}^{\mathrm{bs}}, \\
        dF_6 - H_3 \wedge F_4 & = *j_{D2}^{\mathrm{bs}},\\
        dH_3 & = *j_{NS5}^{\mathrm{bs}},
    \end{aligned}
\end{equation}
where $j^{\mathrm{bs}}$ denote the brane-source current of each of the brane. Taking an exterior derivative of the second line in \eqref{eq:BIsHW} gives
\begin{equation}
d *j_{D2}^{\mathrm{bs}}
=
-\, dH_3 \wedge F_4
+
H_3 \wedge dF_4 .
\label{eq:nonconservation_D2_bs_1}
\end{equation}
Now using the first and the third lines, one obtains
\begin{equation}
d *j_{D2}^{\mathrm{bs}}
=
-\, *j_{NS5}^{\mathrm{bs}} \wedge F_4
+
H_3 \wedge *j_{D4}^{\mathrm{bs}} .
\label{eq:nonconservation_D2_bs_2}
\end{equation}
Equation \eqref{eq:nonconservation_D2_bs_2} is the local supergravity statement of brane creation: in the simultaneous presence of NS5 and D4 sources, the D2 brane-source current is not separately conserved. Integrating \eqref{eq:nonconservation_D2_bs_2} over a seven-dimensional region enclosing the crossing event yields precisely the jump in D2 charge. For one unit of NS5 charge and one unit of D4 charge, the change is one unit of D2 brane-source charge. Let us see this explicitly, but first discuss a subtlety, already important in the original literature and clarified systematically later \cite{Marolf:2000cb}. It is that in the presence of Chern--Simons couplings there is no unique notion of brane charge. Instead, one has three different notions of a charge that differ in locality, gauge  invariance and conservation properties.

The \textbf{brane-source charge} is the localized source appearing directly on the right-hand side of the Bianchi identities, such as $j_{D2}^{\mathrm{bs}}$ in \eqref{eq:BIsHW}. It is local and gauge invariant, but in general it is not conserved, as shown by \eqref{eq:nonconservation_D2_bs_2}. It is this nonconservation that directly encodes the local creation of a D2-brane at the crossing.

The \textbf{Maxwell charge} is defined by integrating gauge-invariant field strengths over spheres at infinity. For D2-brane charge one may schematically write
\begin{equation}
Q_{D2}^{\mathrm{Maxwell}}
\propto
\int_{S^6_\infty} F_6 .
\label{eq:Maxwell_D2_charge}
\end{equation}
This quantity is conserved and gauge invariant, but it is not localized: part of it can be carried by smooth bulk flux rather than by identifiable brane world-volumes. For local questions such as brane creation at a crossing, Maxwell charge is therefore too coarse.

The \textbf{Page charge} current combines R-R fluxes with the $B$-field in such a way that it is localized and quantized, though not invariant under large gauge transformations of $B_2$. In the present case one defines
\begin{equation}
*j_{D2}^{\mathrm{Page}}
=
d\left(
F_6 - B_2 \wedge F_4
\right),
\label{eq:Page_current_D2}
\end{equation}
so that the corresponding Page charge is
\begin{equation}
Q_{D2}^{\mathrm{Page}}
=
\frac{1}{(2\pi l_s)^5 g_s}
\int_{S^6}
\left(
F_6 - B_2 \wedge F_4
\right).
\label{eq:Page_D2_charge}
\end{equation}
Similarly, the D4 Page charge is
\begin{equation}
Q_{D4}^{\mathrm{Page}}
=
\frac{1}{(2\pi l_s)^3 g_s}
\int_{S^4}
F_4 ,
\label{eq:Page_D4_charge}
\end{equation}
where we have again specialized to the case $F_0=F_2=0$. Away from NS5 sources, one may choose a patch in which $H_3=dB_2$, and then
\begin{equation}
d\left(
F_6 - B_2 \wedge F_4
\right)
=
*j_{D2}^{\mathrm{bs}} - B_2 \wedge *j_{D4}^{\mathrm{bs}} ,
\label{eq:Page_current_expanded}
\end{equation}
so the Page current is localized, i.e. does not vanish only on a brane. However, because $B_2$ is not globally defined in the presence of NS5-branes, the Page charge depends on the choice of gauge patch or, equivalently, on the placement of the NS5 Dirac surface (a generalization of the Dirac string for the magnetic monopole of electrodynamics).

This is precisely where the ambiguity in the definition of charge becomes physically important. If, during the motion of the NS5-brane, the Dirac surface associated with $B_2$ is swept across a D4-brane, then $B_2$ undergoes a large gauge transformation
\begin{equation}
B_2 \longrightarrow B_2 + \Delta B_2 ,
\qquad
\frac{1}{2\pi l_s^2}\int_{S^2} \Delta B_2 \in \mathbb Z .
\label{eq:large_gauge_B}
\end{equation}
Under such a transformation, the D2 Page charge changes by
\begin{equation}
\Delta Q_{D2}^{\mathrm{Page}}
=
-\frac{1}{(2\pi l_s)^5 g_s}
\int_{S^6}
\Delta B_2 \wedge F_4 .
\label{eq:Page_shift_D2}
\end{equation}
Now the important step is to note that because the branes are taken to be locally flat and mutually orthogonal, the space-time geometry near the crossing is approximately split into independent normal directions adapted to the two objects. This implies that the angular dependence of the NS5 and D4 fields separates, and the resulting flux integral factorizes. Usually this is referred to as factorization of the $\SS^6$ so that the $\SS^2$ links the NS5 Dirac surface and the $\SS^4$ links one D4-brane. This is not completely correct statement, but it allows to rewrite  \eqref{eq:Page_shift_D2} as
\begin{equation}
\Delta Q_{D2}^{\mathrm{Page}}
=
-
\left(
\frac{1}{2\pi l_s^2}\int_{S^2} \Delta B_2
\right)
\left(
\frac{1}{(2\pi l_s)^3 g_s}\int_{S^4} F_4
\right)
=
-\, Q_{D4}^{\mathrm{Page}} .
\label{eq:Page_shift_factorized}
\end{equation}
Thus, for one unit of large gauge transformation and one unit of D4 Page charge, the D2 Page charge jumps by one unit. This is the Page-charge version of the Hanany--Witten effect.

\subsubsection*{Dirac surface for the NS5-brane}

A useful way to understand the Hanany--Witten effect is to make explicit the role of the Dirac surface associated with the NS5-brane, which is a higher dimensional analogue of the Dirac string stretching off the magnetic monopole. The basic reason such an object is needed is that the NS5-brane is a magnetic source for the NS-NS two-form potential $B_2$. Its gauge-invariant field strength $H_3$ satisfies
\begin{equation}
dH_3 = 2\pi\, \delta_4(W_6^{\mathrm{NS5}}),
\label{eq:dH_NS5_source}
\end{equation}
where $W_6^{\mathrm{NS5}}$ is the six-dimensional NS5-brane world-volume and $\delta_4(W_6^{\mathrm{NS5}})$ is its Poincar\'e-dual delta-current in ten dimensions. Because the right-hand side is nonzero, $H_3$ cannot be written globally as $dB_2$ with a smooth globally defined potential $B_2$. This is completely analogous to the fact that the field strength of a Dirac monopole cannot be globally written as $dA$ with a single smooth vector potential.

It is thus instructive to analyze the toy model of crossing of the magnetic monopole and the electric charge. The monopole example makes completely explicit what a Dirac surface is, why it is gauge-dependent, and why one may choose it differently before and after a crossing process even though one cannot, in general, interpolate between these choices by a single smooth family. 

Consider ordinary Maxwell theory in 1+3 dimensions with a magnetic monopole of charge $g$ at the origin. The magnetic field is
\begin{equation}
\vec B = g \frac{\vec r}{r^3},
\label{eq:monopole_Bfield}
\end{equation}
or, in the differential-form language,
\begin{equation}
F_2 = g \sin\theta \, d\theta \wedge d\phi .
\label{eq:monopole_F2}
\end{equation}
Its flux through a sphere surrounding the origin is
\begin{equation}
\int_{S^2} F_2 = 4\pi g .
\label{eq:monopole_flux}
\end{equation}
Because this flux is nonzero, $F_2$ cannot be written globally as $dA_1$ with a single smooth vector potential on all of $S^2$. The standard resolution is to cover the sphere by two patches. On the northern (southern) patch $U_N$ ($U_S$), regular away from the south (north) pole, one may take
\begin{equation}
A_{N(S)}= g(1\mp\cos\theta)\, d\phi .
\label{eq:A_NS_monopole}
\end{equation}
Both give the same field strength, $dA_N = dA_S = F_2$
On the overlap $U_N\cap U_S$ they differ by a gauge transformation,
\begin{equation}
A_N-A_S = 2g\, d\phi = d\lambda,
\qquad
\lambda = 2g\, \phi .
\label{eq:AN_AS_difference}
\end{equation}
Despite the notation $d\l$ is not an exact form as it is not a differential of a 0-form. Indeed, for $\l$ to be a function on $\SS^1$ it must be single valued under $\phi\to \phi+2\pi$, which is however not the case.  The requirement that a charged wavefunction transform consistently implies the Dirac quantization condition. In the conventions used here,
\begin{equation}
e^{\, i e \lambda(\phi+2\pi)} = e^{\, i e \lambda(\phi)}
\qquad\Longrightarrow\qquad
2e g \in  \mathbb Z .
\label{eq:Dirac_quantization_raw}
\end{equation}

The same monopole may be described by a single potential that is singular along a chosen half-line, the Dirac string. For example, the potential $A_N$ is singular only at the south pole, corresponding to a string running down the negative $z$-axis, while $A_S$ is singular only at the north pole, corresponding to a string along the positive $z$-axis.

In space-time it is more natural to speak of the world-sheet swept out by the Dirac string. Let $W_1$ denote the monopole world-line in four-dimensional spacetime. Then one may introduce a two-dimensional Dirac surface $V_2$ such that
\begin{equation}
\partial V_2 = W_1 .
\label{eq:Dirac_surface_monopole_boundary}
\end{equation}
If $\delta_2(V_2)$ denotes the Poincar\'e-dual two-form current associated with $V_2$, then
\begin{equation}
d\,\delta_2(V_2) = \delta_3(W_1) ,
\label{eq:delta2_boundary_monopole}
\end{equation}
and the monopole field strength may be written as
\begin{equation}
F_2 = dA_1 + 4\pi g \, \delta_2(V_2) .
\label{eq:F_with_Dirac_surface}
\end{equation}
Taking the exterior derivative gives
\begin{equation}
dF_2 = 4\pi g \, \delta_3(W_1) ,
\label{eq:dF_monopole_source}
\end{equation}
which is precisely the spacetime form of the magnetic-source equation, and is the analogue of \eqref{eq:BIsHW}.

The Dirac surface, as an artificial construction, is not unique. If $V_2$ and $V_2'$ have the same boundary, then their difference is the boundary of a three-chain $X_3$,
\begin{equation}
V_2' - V_2 = \partial X_3 .
\label{eq:V2_difference_X3}
\end{equation}
Correspondingly,
\begin{equation}
\delta_2(V_2') = \delta_2(V_2) + d\, \delta_1(X_3) .
\label{eq:delta2_shift_monopole}
\end{equation}
To keep $F_2$ unchanged, the vector potential must transform as
\begin{equation}
A_1 \longrightarrow A_1' = A_1 - 4\pi g \, \delta_1(X_3) .
\label{eq:A_shift_monopole}
\end{equation}
Thus changing the Dirac surface is equivalent to a singular gauge transformation. This is the space-time version of the patching condition \eqref{eq:AN_AS_difference}.

\begin{figure}[h]
\centering
\begin{tikzpicture}[>=Latex, line cap=round, line join=round, scale=1.1]

  \draw[->, thick] (-4.8,0) -- (4.8,0) node[right] {$z$};
  \draw[->, thick] (0,-0.2) -- (0,6.2) node[above] {$t$};

  \node at (-4.3,-0.35) {$z<0$};
  \node at (4.1,-0.35) {$z>0$};

  \draw[very thick, blue] (0,0.3) -- (0,5.8);
  \fill[blue] (0,0.3) circle (1.5pt);
  \fill[blue] (0,5.8) circle (1.5pt);
  \node[blue, right] at (0,1.2) {\small electric charge world-line $W_e$};

  \draw[very thick, red] (-3.7,0.3) -- (3.2,5.4);
  \fill[red] (-3.7,0.3) circle (1.4pt);
  \fill[red] (3.2,5.4) circle (1.4pt);
  \node[red, above left, rotate=36.5] at (0.1,2.5) {\small monopole world-line $W_m$};

  \draw[dashed, gray] (-4.2,3.05) -- (4.2,3.05);
  \node[left] at (-4.2,3.05) {$t=t_\ast$};

  \fill[orange!35, opacity=0.7]
    (-3.7,0.3) -- (3.2,5.4) --  (3.2,5.8) -- (-3.7,5.8) -- cycle;

  \node[orange!80!black, align=center] at (-2.2,4.4)
    {Dirac surface $V_2$};

  \fill[black] (0,3.05) circle (1.4pt);
  \node[align=left] at (2.65,3)
    {\small Dirac surface sweeps across\\ \small the electric world-line};

  \draw[decorate,decoration={brace,mirror,amplitude=5pt}]
    (-3.9,0.2) -- (-0.2,0.2)
    node[midway,below=6pt] {before crossing};

  \draw[decorate,decoration={brace,mirror,amplitude=5pt}]
    (0.2,0.2) -- (3.9,0.2)
    node[midway,below=6pt] {after crossing};

\end{tikzpicture}\
\caption{Schematic spacetime picture of a magnetic monopole moving past an electric charge. The shaded surface is the Dirac world-sheet $V_2$ attached to the monopole world-line $W_m$. Although one may choose Dirac strings that avoid the electric world-line $W_e$ on individual time slices, a single smooth interpolation of Dirac surfaces through the full process is obstructed; equivalently, one must perform a large gauge transformation. }
\label{fig:hanany}
\end{figure}
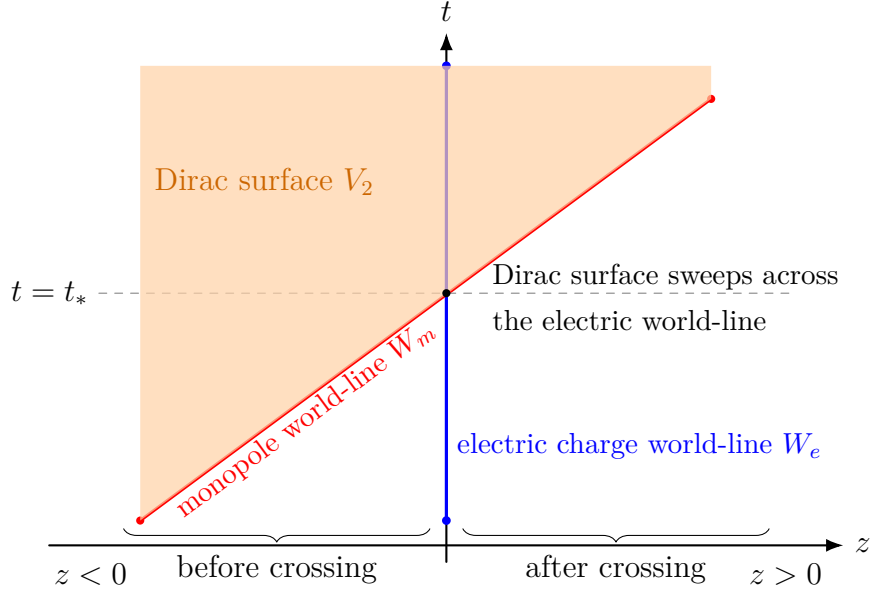

Now consider a toy dynamical process that captures the idea of the Hanany--Witten effect, schematically depicted on Fig. \ref{fig:hanany}. Place an electric charge  $e$ at the spatial origin, so that its world-line is the time axis. Let a magnetic monopole move along the $z$-axis from $z<0$ to $z>0$, crossing the plane $z=0$ at some time $t=t_*$. On each time slice one may choose a Dirac string for the monopole that avoids the electric charge:
\begin{itemize}
\item for $z_m(t)<0$, choose the string to run downward toward $z=-\infty$;
\item for $z_m(t)>0$, choose the string to run upward toward $z=+\infty$.
\end{itemize}
At each fixed time this is perfectly allowed. However, these two choices cannot, in general, be connected by a single smooth Dirac world-sheet in space-time that avoids the electric world-line throughout the whole process. If one drags the initial Dirac surface continuously along with the monopole, then at some stage the surface sweeps across the electric world-line. Alternatively, one may switch discontinuously to a new Dirac surface after the crossing; in that case one has changed gauge by the singular transformation \eqref{eq:A_shift_monopole}.

The charged particle couples through the world-line term
\begin{equation}
S_e = e \int_{W_e} A_1 .
\label{eq:electric_particle_action}
\end{equation}
Under the gauge transformation \eqref{eq:A_shift_monopole}, the action shifts by
\begin{equation}
\Delta S_e = e \int_{W_e} d\lambda ,
\label{eq:delta_Se_monopole}
\end{equation}
which is nontrivial precisely when the interpolating three-chain $X_3$ links the electric world-line. The requirement that this shift be physically invisible is exactly the Dirac quantization condition. The fact, that nothing additional is created in this process has a simple explanation, that the theory does not have string-like fundamental objects. Equivalently, the Bianchi identity of electrodynamics does not receive contributions from a Chern--Simons-like coupling. Therefor, the Hanany--Witten effect in this theory is simply the Dirac quantization condition.

The translation of this discussion to the NS5-brane is straightforward, and one may follow the formal dictionary
\begin{equation}
A_1 \;\longleftrightarrow\; B_2,
\qquad
F_2=dA_1 \;\longleftrightarrow\; H_3=dB_2,
\qquad
W_1 \;\longleftrightarrow\; W_6^{\mathrm{NS5}},
\qquad
V_2 \;\longleftrightarrow\; V_7 .
\label{eq:monopole_NS5_dictionary}
\end{equation}
The NS5-brane is a magnetic source for the NS-NS three-form field strength,
\begin{equation}
dH_3 = 2\pi\, \delta_4(W_6^{\mathrm{NS5}}) .
\label{eq:NS5_source_again}
\end{equation}
Introducing a seven-dimensional Dirac surface $V_7$ with
\begin{equation}
\partial V_7 = W_6^{\mathrm{NS5}},
\label{eq:V7_boundary_NS5}
\end{equation}
one may write
\begin{equation}
H_3 = dB_2 + 2\pi\, \delta_3(V_7) .
\label{eq:H3_Dirac_NS5}
\end{equation}
If $V_7$ is changed to another Dirac surface $V_7'$ with the same boundary, then
\begin{equation}
V_7' - V_7 = \partial X_8 ,
\label{eq:V7_difference_X8}
\end{equation}
and hence
\begin{equation}
\delta_3(V_7') = \delta_3(V_7) + d\, \delta_2(X_8) .
\label{eq:delta3_shift_NS5}
\end{equation}
To keep $H_3$ fixed, one must shift
\begin{equation}
B_2 \longrightarrow B_2' = B_2 - 2\pi\, \delta_2(X_8) .
\label{eq:B2_shift_NS5}
\end{equation}
This is the exact analogue of the monopole gauge transformation \eqref{eq:A_shift_monopole}.

The new ingredient in the NS5--D4 system is that the lower-dimensional D2 Page charge depends explicitly on $B_2$. In Type IIB one has, schematically,
\begin{equation}
Q_{D2}^{\mathrm{Page}}
=
\frac{1}{(2\pi l_s)^5 g_s}
\int_{\SS^6}
\left(
F_6 - B_2 \wedge F_4
\right).
\label{eq:D3_Page_charge_again}
\end{equation}
Now, at an initial time one chooses a Dirac surface $V_7^{\mathrm{in}}$ attached to the NS5-brane and extending to infinity in some chosen direction. As the NS5-brane moves, one may continuously drag this Dirac surface along with it. If the NS5-brane does not cross any D4-brane, this deformation can be carried out without changing the topology of the Dirac surface relative to the other branes, and no jump in Page charge occurs. However, when the NS5-brane passes through a D5-brane, the dragged Dirac surface necessarily sweeps through the D4-brane world-volume. Equivalently, one may choose instead to jump to a new Dirac surface $V_7^{\mathrm{out}}$ that again avoids the D4-brane after the crossing. The two choices differ by an eight-manifold $X_8$ whose associated current implements the large gauge transformation 
\begin{equation}
B_2 \longrightarrow B_2' = B_2 - 2\pi\, \delta_2(X_8).
\label{eq:B2_large_gauge_shift}
\end{equation}

It is important to emphasize that the Dirac surface itself is not physical. What is physical is the fact that $B_2$ cannot be globally defined in the presence of NS5-brane magnetic charge, and that changing the auxiliary Dirac surface changes the Page-charge bookkeeping by a large gauge transformation. The Hanany--Witten effect may therefore be viewed as the statement that when the NS5-brane is moved through the D5-brane, the necessary rearrangement of the NS5 Dirac surface changes the Page charge by an integer, and this integer shift is realized physically by the creation of a D3-brane.

A potential source of confusion might be that, at any fixed time before or after the crossing, one may indeed choose a Dirac surface for the NS5-brane that avoids the D5-brane. The obstruction is not local in time, but global in spacetime: one cannot in general choose a \emph{single smooth family} of Dirac surfaces attached to the moving NS5-brane that avoids the D5-brane throughout the entire crossing process. Equivalently, the Dirac surfaces chosen before and after the crossing differ by an eight-chain, and the corresponding change in the singular representative of $H_3$ must be compensated by a large gauge transformation of $B_2$. Since the D3 Page charge depends explicitly on $B_2$, this large gauge transformation shifts the Page charge by an integer. Thus the Hanany--Witten effect does not require that a particular Dirac surface literally intersect the D5-brane in one chosen gauge; rather, it follows from the fact that the initial and final gauge choices for the NS5 Dirac surface are not globally equivalent during the crossing. A formal treatment of Dirac branes/surfaces, magnetic sources, and their description in generalized differential cohomology is given e.g. in \cite{Freed:2000ta,Henneaux:1986tt,Teitelboim:1985yc,Henneaux:1986ht}. Related global aspects of higher-form gauge fields in brane theory are discussed in \cite{Freed:1998tg,Moore:1999gb,Minasian:1997mm}.

\subsection{Dielectric brane polarization, Myers effect}
\label{sec:myers}

Another example of brane interactions is provided by the dielectric, or Myers, effect. Previously we discussed brane creation in the Hanany--Witten effect in Section \ref{sec:hanany-witten} and the formation of brane bound states in Section \ref{sec:bound}. The dielectric effect belongs to the same general class of phenomena, where background fields and interactions between branes  reorganize the low-energy degrees of freedom of a brane system into a new extended object. In this case --- carrying different multipole moments and admitting dual microscopic and macroscopic descriptions. More concretely: a collection of lower-dimensional D-branes expands into a higher-dimensional fuzzy brane in the presence of background R-R flux \cite{Myers:1999ps}.

Below we will follow the conventions of \cite{Myers:1999ps} and work in type IIA string theory, in string frame. We take the dilaton to be constant and denote $\lambda = 2\pi \alpha'$. The Levi--Civita symbol is defined by $\epsilon_{123}=+1$. For a stack of $N$ D-branes an open string may end on any of the branes and therefore its states are labeled by additional indices effectively turning it into an $N\times N$ matrix. These are known as Chan--Paton factors and in the effective theory of a stack of $N$ coincident Dp-branes manifest themselves as matrix-valuedness of the transverse coordinates,
\begin{equation}
X^i=\lambda \Phi^i,
\qquad
\Phi^i\in \mathfrak{u}(N),
\qquad
i=p+1,\ldots,9,
\end{equation}
so that the notion of brane position becomes intrinsically non-Abelian. When the matrices $\Phi^i$ commute they may simultaneously be diagonalized, and their eigenvalues recover the usual interpretation as transverse positions of individual branes. When they do not commute, the configuration is more naturally interpreted as a non-commutative, or fuzzy, geometry. The fuzzy two-sphere that will appear below is the finite-matrix approximation to $S^2$ introduced in \cite{Madore:1991bw}; for a more general discussion of non-commutative geometry and matrix-valued coordinates in string theory see \cite{Connes:1997cr,Douglas:2001ba,Szabo:2001kg}.

The appropriate effective action for coincident D-branes is the non-Abelian Born--Infeld plus Wess--Zumino action \cite{Tseytlin:1997csa,Myers:1999ps},
\begin{equation}
\label{eq:Myers}
\begin{aligned}
S_{\rm DBI}
&=
-T_p \int d^{p+1}\sigma\;
\mathrm{STr}\!\left[
e^{-\phi}
\sqrt{
-\det\!\Bigl(
P\!\left[E_{ab}+E_{ai}\bigl(Q^{-1}-\delta\bigr)^{ij}E_{jb}\right]
+\lambda F_{ab}
\Bigr)\,
\det\!\bigl(Q^i{}_j\bigr)
}
\right], \\
S_{\rm WZ}
&=
\mu_p \int
\mathrm{STr}\!\left[
P\!\left(
e^{\,i\lambda\,\iota_\Phi\iota_\Phi}
\sum_q C^{(q)} e^B
\right)e^{\lambda F}
\right],
\end{aligned}
\end{equation}
where
\begin{equation}
E_{\mu\nu}=G_{\mu\nu}+B_{\mu\nu},
\qquad
Q^i{}_j=\delta^i{}_j+i\lambda[\Phi^i,\Phi^k]E_{kj}.
\end{equation}
To construct this action one begins with the D9-brane action, promotes world-volume fields to matrices, and then uses T-duality to infer the dependence on transverse scalars and background R-R potentials \cite{Myers:1999ps}. In particular, the  inner product insertions $\iota_\Phi\iota_\Phi$  in the Wess--Zumino action in \eqref{eq:Myers} are required by T-duality and imply that lower-dimensional branes couple to higher R-R forms. At linear order in weak supergravity backgrounds, the same couplings are also reproduced from matrix-theory considerations and open-string consistency arguments \cite{Taylor:1999gq}. Finally, the symmetrized-trace prescription is an effective low-energy prescription valid to the order relevant for the dielectric solution, rather than a complete all-orders definition of the non-Abelian Born--Infeld action \cite{Tseytlin:1997csa,Myers:1999ps}.

The basic mechanism of dielectric polarization is already visible in the scalar potential. In flat space one expands the square root in the Born--Infeld action and as we did earlier in Section \ref{sec:actions} producing the quartic commutator term
\begin{equation}
V_{\rm quartic}\sim -\mathrm{Tr}\,[\Phi^i,\Phi^j][\Phi^i,\Phi^j].
\end{equation}
If no other contributions to the scalar potential present, it can be minimized by taking commuting configurations, that correspond to the same $N$ Dp-branes placed at points determined by the diagonal values of $\Phi^i$. In the presence of background R-R flux, however, the Wess--Zumino term induces cubic couplings schematically of the form
\begin{equation}
V_{\rm R-R}\sim -F^{(p+4)}\,\mathrm{Tr}(\Phi^i\Phi^j\Phi^k).
\end{equation}
The competition between these two terms can make a non-commuting configuration energetically favorable. In the simplest case, the matrices satisfy an $SU(2)$ algebra and form a fuzzy two-sphere, that is the dielectric Myers effect \cite{Myers:1999ps}.

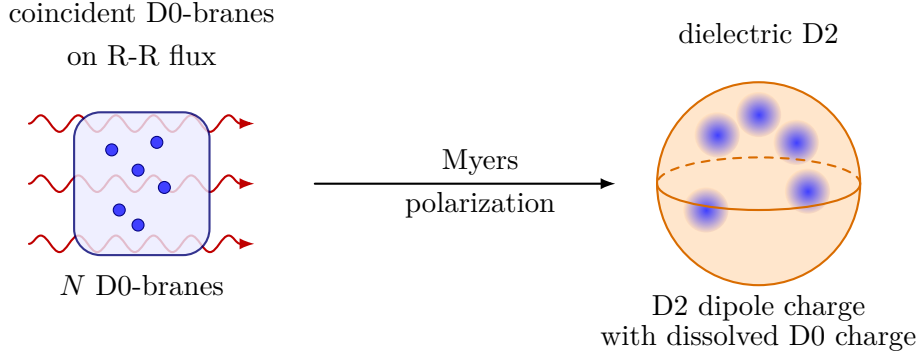
\begin{figure}[h]
\centering
\begin{tikzpicture}[scale=1.0, line cap=round, line join=round]

  \tikzset{
    brane/.style={draw=orange!85!black, fill=orange!20, thick},
    dzero/.style={circle, fill=blue!75, draw=blue!50!black, inner sep=1.6pt},
    flux/.style={red!75!black, thick, decorate,
      decoration={snake, amplitude=1.1mm, segment length=5mm}},
    every node/.style={font=\small}
  }

  \draw[flux, -{Latex[length=2mm]}] (-6,0.8) -- (-3,0.8);
  \draw[flux, -{Latex[length=2mm]}] (-6,0.0) -- (-3,0.0);
  \draw[flux, -{Latex[length=2mm]}] (-6,-0.8) -- (-3,-0.8);

  \node at (-4.5,2.0) {\parbox{4cm}{\centering coincident D0-branes \\ on R-R flux}};

  \draw[rounded corners=10pt, fill=blue!8, draw=blue!45!black, thick, opacity = 0.8]
    (-5.4,-0.95) rectangle (-3.6,0.95);

  \foreach \x/\y in {-4.9/0.45,-4.55/0.18,-4.2/-0.05,-4.8/-0.35,-4.3/0.55,-4.55/-0.55} {
    \node[dzero] at (\x,\y) {};
  }

  \node at (-4.5,-1.35) {$N$ D0-branes};

  \node[align=center] at (0,0) {\parbox{2cm}{\centering Myers\\polarization}};
  \draw[thick, -{Latex[length=2mm]}] (-2.2,0) -- (1.8,0);

  \node at (3.7,2.0) {dielectric D2};

  \shade[ball color=orange!25] (3.7,0) circle (1.35);
  \draw[brane] (3.7,0) circle (1.35);

  \foreach \x/\y in {4.2/0.55,3.15/0.65,4.35/-0.1,3.0/-0.35,3.7/0.92} {
    \shade [inner color = blue!75, outer color = orange!25] (\x,\y) circle [radius = 0.3]; 
  }

  \draw[orange!85!black, thick] (2.35,0)
    arc[start angle=180,end angle=360,x radius=1.35,y radius=0.35];
  \draw[orange!85!black, thick, dashed] (5.05,0)
    arc[start angle=0,end angle=180,x radius=1.35,y radius=0.35];

  \node[align=center] at (3.7,-1.65) {D2 dipole charge};
  \node[align=center] at (3.7,-2.05) {with dissolved D0 charge};

\end{tikzpicture}
\caption{Schematic illustration of the Myers effect: a stack of $N$ coincident D0-branes polarizes, in a background R-R four-form flux, into a spherical dielectric D2-brane. Microscopically, the transverse coordinates become non-commuting matrices; macroscopically, the configuration is described as a spherical D2-brane carrying dissolved D0-brane charge.}
\label{fig:myers}
\end{figure}

Start with a stack of $N$ D0-branes in a constant electric R-R four-form flat background given by
\begin{equation}
G_{\mu\nu}=\eta_{\mu\nu},
\qquad
B_{\mu\nu}=0,
\qquad
F_{ab}=0,
\qquad
D_t\Phi^i=0,
\end{equation}
and turn on
\begin{equation}
F^{(4)}_{tijk}=-2f\,\epsilon_{ijk},
\qquad i,j,k=1,2,3.
\label{eq:F4choice}
\end{equation}
To leading non-trivial order in $\lambda$, the static potential derived from \eqref{eq:Myers} is (see \cite{Myers:1999ps} for more details)
\begin{equation}
V(\Phi)
=
N T_0
-\frac{\lambda^2 T_0}{4}\,
\mathrm{Tr}\!\left([\Phi^i,\Phi^j][\Phi^i,\Phi^j]\right)
-\frac{i}{3}\lambda^2\mu_0\,
\mathrm{Tr}\!\left(\Phi^i\Phi^j\Phi^k\right)F^{(4)}_{tijk}.
\label{eq:D0potential}
\end{equation}
The cubic term is specific to the dielectric coupling: it originates from the $i\lambda\,\iota_\Phi\iota_\Phi C^{(3)}$ piece of the Wess--Zumino action and has no analogue in the abelian theory.

With the choice \eqref{eq:F4choice}, the potential becomes
\begin{equation}
V(\Phi)
=
N T_0
-\frac{\lambda^2 T_0}{4}\,
\mathrm{Tr}\!\left([\Phi^i,\Phi^j][\Phi^i,\Phi^j]\right)
+\frac{2if\lambda^2\mu_0}{3}\,
\epsilon_{ijk}\,
\mathrm{Tr}\!\left(\Phi^i\Phi^j\Phi^k\right),
\label{eq:D0potentialf}
\end{equation}
and the equations of motion read
\begin{equation}
[[\Phi^i,\Phi^j],\Phi^j]
+i f\,\epsilon_{ijk}[\Phi^j,\Phi^k]=0.
\label{eq:D0eom}
\end{equation}

To solve these equations, one makes the standard $SU(2)$ ansatz
\begin{equation}
\Phi^i=a\,J^i,
\qquad i=1,2,3,
\label{eq:SU2ansatz}
\end{equation}
where $a$ is a dimensionful constant to be determined dynamically, and $J^i$ are Hermitian generators of the $N$-dimensional irreducible representation of $SU(2)$,
\begin{equation}
[J^i,J^j]=i\epsilon^{ijk}J^k,
\qquad
\sum_{i=1}^3 J^iJ^i=C_2\,\mathbf{1}_N,
\qquad
C_2=j(j+1)=\frac{N^2-1}{4},
\qquad
N=2j+1.
\end{equation}
These generators satisfy the following useful identities
\begin{equation}
\sum_{i,j=1}^3 [J^i,J^j][J^i,J^j]=-2C_2\,\mathbf{1}_N,
\qquad
\epsilon_{ijk}J^iJ^jJ^k=iC_2\,\mathbf{1}_N.
\label{eq:SU2ids}
\end{equation}
Substituting \eqref{eq:SU2ansatz} and \eqref{eq:SU2ids} into \eqref{eq:D0potentialf} yields
\begin{equation}
V(a)
=
N T_0
+\lambda^2 T_0\,N C_2
\left(
\frac{a^4}{2}
-\frac{2fa^3}{3}
\right).
\label{eq:Va}
\end{equation}
The extremal condition for the potential is 
\begin{equation}
\frac{dV}{da}
=
2\lambda^2 T_0\,N C_2\,a^2(a-f)=0,
\end{equation}
so there are two stationary points:
\begin{equation}
a=0,
\qquad
a=f.
\end{equation}
The trivial solution $a=0$ is the collapsed commuting configuration, while the non-trivial solution
\begin{equation}
\Phi^i=f\,J^i,
\qquad
X^i=\lambda f\,J^i,
\label{eq:DielectricSolution}
\end{equation}
describes the dielectric fuzzy sphere. Its physical radius is
\begin{equation}
R^2
\equiv
\frac{\lambda^2}{N}\,\mathrm{Tr}(\Phi^i\Phi^i)
=
\lambda^2 f^2 C_2
=
\frac{\lambda^2 f^2}{4}(N^2-1),
\label{eq:Radius}
\end{equation}
so that
\begin{equation}
R=\lambda f\,\sqrt{C_2}
\sim \frac{\lambda f N}{2}
\qquad (N\gg 1).
\end{equation}
For static configurations the corresponding energy is given by the potential
\begin{equation}
V_{\rm fuzzy}
=
N T_0
-\frac{\lambda^2 T_0}{6}N C_2 f^4
=
N T_0
-\frac{\lambda^2 T_0}{24}N(N^2-1)f^4,
\label{eq:Vfuzzy}
\end{equation}
which is lower than $N T_0$, the energy of the collapsed configuration. Moreover,
\begin{equation}
\left.\frac{d^2V}{da^2}\right|_{a=f}
=
2\lambda^2 T_0 N C_2 f^2>0,
\end{equation}
showing that the dielectric sphere is locally stable and energetically favorable.

An important point, emphasized already in \cite{Myers:1999ps}, is that for fixed total D0 charge $N$ the preferred configuration is the irreducible representation. If the matrices decompose into irreducible blocks with spins $j_r$, such that $\sum_r (2j_r+1)=N$, then
\begin{equation}
V_{\rm red}
=
N T_0
-\frac{\lambda^2 T_0 f^4}{6}
\sum_r j_r(j_r+1)(2j_r+1).
\label{eq:Vreducible}
\end{equation}
For fixed $N$, this is minimized by a single irreducible block, corresponding to one maximal fuzzy sphere rather than a collection of smaller ones.

Schematics of the dielectric effect is presented in Fig. \ref{fig:myers}, where the resulting dissolved D0-brane charge is depicted by fuzzy circles. However, the dielectric configuration \eqref{eq:DielectricSolution} should not be visualized as $N$ classical point-like D0-branes arranged on an ordinary sphere. Rather, it is a genuinely matrix-valued configuration whose large-$N$ limit approximates the algebra of functions on $S^2$. In this sense, it provides a concrete realization of non-commutative geometry within D-brane dynamics \cite{Madore:1991bw,Connes:1997cr,Douglas:2001ba,Szabo:2001kg}. 

Stability of the resulting configuration has a clear physical interpretation. The quartic commutator term inherited from the Born--Infeld action disfavors non-commutativity and tends to collapse the branes, whereas the R-R-induced cubic term tends to polarize the stack into an oriented extended object. The equilibrium radius \eqref{eq:Radius} is the point at which these competing effects balance. The analogy with an ordinary dielectric medium in an external electric field motivates the terminology: although the system carries no net D2 monopole charge, it develops a non-vanishing D2 dipole moment in the background flux. However, since the dielectric D2-brane is spherical and contractible in $\mathbb{R}^3$, it does not carry a conserved D2 charge at infinity. Its conserved charge is the D0 charge $N$. What is induced by the R-R field is instead the local coupling appropriate to a D2-brane, or equivalently a D2 dipole moment. This distinction is crucial for the correct physical interpretation of the dielectric effect.

Let us check that the obtained fuzzy sphere configuration couples to the R-R three-form exactly as a spherical D2-brane should. On the D0-brane side, this coupling is already contained in the cubic term of the Wess--Zumino action. Evaluated on the solution \eqref{eq:DielectricSolution}, one finds
\begin{equation}
S^{\rm D0}_{C^{(3)}}
=
\int dt\;
\frac{2if\lambda^2\mu_0}{3}\,
\epsilon_{ijk}\,
\mathrm{Tr}(\Phi^i\Phi^j\Phi^k)
=
-\frac{2}{3}\lambda^2\mu_0 f^4 N C_2 \int dt.
\end{equation}
Since the overall sign depends on orientation conventions, it is more transparent to compare magnitudes:
\begin{equation}
\bigl|S^{\rm D0}_{C^{(3)}}\bigr|
=
\frac{2\mu_0}{3\lambda}\,
\frac{N}{\sqrt{C_2}}\,
f R^3 \int dt
\;\xrightarrow[N\gg1]{}\;
\frac{8\pi}{3}\,\mu_2 f R^3 \int dt,
\label{eq:D0C3}
\end{equation}
where we used $\mu_0=2\pi\lambda \mu_2$ and $N/\sqrt{C_2}\to 2$ at large $N$.

Now consider a spherical D2-brane of radius $R$ in the same background. Since
\begin{equation}
F_4=dC_3=-2f\,dt\wedge dx^1\wedge dx^2\wedge dx^3,
\end{equation}
Stokes' theorem implies
\begin{equation}
\left|
\mu_2 \int_{\mathbb{R}_t\times S^2_R} P[C_3]
\right|
=
\left|
\mu_2 \int_{\mathbb{R}_t\times B^3_R} F_4
\right|
=
\frac{8\pi}{3}\,\mu_2 f R^3 \int dt.
\label{eq:D2C3}
\end{equation}
Comparison of \eqref{eq:D0C3} and \eqref{eq:D2C3} shows that the large-$N$ dielectric sphere indeed couples to the R-R three-form with precisely the strength expected for a spherical D2-brane. This is the precise sense in which the D0-brane bound state acquires a D2 dipole moment.

The same object admits a complementary macroscopic description as a spherical D2-brane carrying $N$ units of D0 charge dissolved as world-volume magnetic flux as depicted on Fig. \ref{fig:myers}. The D2-brane action is
\begin{equation}
S_{\rm D2}
=
-T_2 \int d^3\sigma\,
\sqrt{-\det(P[G]+\lambda F)}
+
\mu_2 \int \left(P[C^{(3)}]+\lambda\,P[C^{(1)}]\wedge F\right).
\label{eq:D2action}
\end{equation}
We take the world-volume to be $\mathbb{R}_t\times S^2$, with angular coordinates $(\theta,\phi)$, and choose the standard monopole flux
\begin{equation}
F_{\theta\phi}=\frac{N}{2}\sin\theta,
\qquad
\frac{1}{2\pi}\int_{S^2}F=N.
\label{eq:FluxQuantization}
\end{equation}
Then the $C^{(1)}\wedge F$ term in \eqref{eq:D2action} gives
\begin{align}
S^{\rm D2}_{C^{(1)}}
&=
\mu_2\lambda
\int_{\mathbb{R}_t\times S^2} P[C^{(1)}]\wedge F
\nonumber\\
&=
\mu_2\lambda
\left(\int_{S^2}F\right)\int dt\,C_t^{(1)}
=
2\pi\lambda\mu_2\,N \int dt\,C_t^{(1)}
=
N\mu_0 \int dt\,C_t^{(1)}.
\label{eq:D0fromFlux}
\end{align}
This is exactly the coupling of $N$ D0-branes. Thus the D2-brane with magnetic flux \eqref{eq:FluxQuantization} carries the same conserved D0 charge as the original matrix configuration.  

For a round sphere of radius $R$, the DBI and Wess--Zumino terms combine to give for the macroscopic potential the following
\begin{equation}
V_{\rm D2}(R)
=
4\pi T_2
\sqrt{R^4+\frac{\lambda^2 N^2}{4}}
-
\frac{8\pi}{3}\mu_2 f R^3,
\label{eq:VD2}
\end{equation}
again up to the overall orientation sign of the Wess--Zumino term. In the large-$N$ regime this becomes
\begin{equation}
V_{\rm D2}(R)
=
N T_0
+\frac{2T_0}{\lambda^2 N}R^4
-\frac{8\pi}{3}\mu_2 f R^3
+\mathcal O(N^{-3}),
\label{eq:VD2largeN}
\end{equation}
whose stationary point lies at
\begin{equation}
R=\frac{\lambda f N}{2}.
\end{equation}
This agrees with the large-$N$ limit of the microscopic result \eqref{eq:Radius}. The matrix D0-brane description and the fluxed D2-brane description are therefore two equivalent descriptions of the same dielectric configuration.

The Myers effect provides a clean example of the dynamics nature of D-branes in string theory, that is: brane dimensionality is dynamical, charge can be dissolved into world-volume flux. Moreover, space-time geometry can emerge from non-Abelian matrix degrees of freedom. In this sense, the dielectric effect is an  illustration of how brane interactions reorganize the fundamental variables of the theory into new geometric phases.

\subsection{Supertubes}
\label{sec:supertubes}

Our final example of brane interactions is provided by the so-called supertubes, which are best viewed as a distinguished supersymmetric realization of brane polarization. Such states play an important role in understanding of supersymmetric bound states, black rings, and in constructions of black-hole microstate geometries. Supertubes are closely related to the branes polarized via the Myers effect. In the latter, coincident D-branes placed in an external R-R background polarize into a higher-dimensional fuzzy brane through non-Abelian couplings in the world-volume action \cite{Myers:1999ps}. Supertubes, first discovered in \cite{Mateos:2001qs}, arise when the initial configuration, say D2-branes carrying dissolved D0-brane and F1-string charges, expands into a tubular D2-brane without the need for an external R-R field. Instead, it is stabilized by electric and magnetic world-volume fluxes.  For the original papers and early discussions of the supertube effect see \cite{Mateos:2001qs,Emparan:2001ux}.

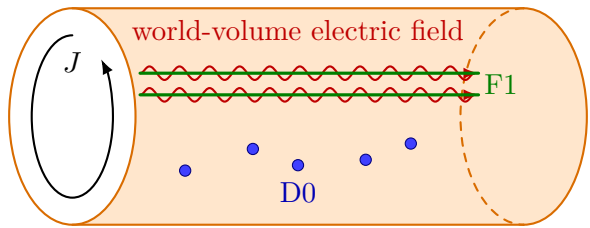
\begin{figure}[h]
    \centering
    \begin{tikzpicture}[scale=1.2, line cap=round, line join=round]

    \tikzset{
    tube/.style={draw=orange!85!black, fill=orange!20, thick},
    dzero/.style={circle, fill=blue!75, draw=blue!50!black, inner sep=1.5pt},
    fone/.style={green!50!black, very thick},
    flux/.style={red!75!black, thick, decorate,
      decoration={snake, amplitude=0.9mm, segment length=4mm}},
    note/.style={font=\small}
  }
  \def\L{2.5}
  \def\R{1.2}
  \def\rx{0.7}

  \fill[orange!20]
    (-\L,\R) -- (\L,\R)
    arc[start angle=90,end angle=-90,x radius=\rx,y radius=\R]
    -- (-\L,-\R)
    arc[start angle=-90,end angle=90,x radius=\rx,y radius=\R];

    \draw[draw=orange!85!black, thick] (-\L,\R) -- (\L,\R);
    \draw[draw=orange!85!black, thick] (-\L,-\R) -- (\L,-\R);
    \draw[draw=orange!85!black, thick] (-\L,0) ellipse [x radius=\rx, y radius=\R];
    \draw[draw=orange!85!black, thick,dashed] (\L,\R) arc[start angle=90,end angle=270,x radius=\rx,y radius=\R];
    \draw[draw=orange!85!black, thick] (\L,\R) arc[start angle=90,end angle=-90,x radius=\rx,y radius=\R];


    \draw[-{Latex[length=2mm]}, thick] (-\L,0.75*\R)
    arc[start angle=90,end angle=405,x radius=0.45,y radius=0.9];
    \node[note] at (-\L,\R/2) {$J$};


    \draw[flux, -{Latex[length=2mm]}] (-0.7*\L,0.2*\R) -- (0.8*\L,0.2*\R);
    \draw[flux, -{Latex[length=2mm]}] (-0.7*\L,0.4*\R) -- (0.8*\L,0.4*\R);
    \node[note, red!75!black] at (0,0.8*\R) {world-volume electric field};

    \draw[fone] (-0.7*\L,0.2*\R) -- (0.8*\L,0.2*\R);
    \draw[fone] (-0.7*\L,0.4*\R) -- (0.8*\L,0.4*\R);
    \node[note, green!50!black] at (0.9*\L,0.3*\R) {F1};

    \foreach \x/\y in {-0.5*\L/-0.5*\R,-0.2*\L/-0.3*\R,0.3*\L/-0.4*\R,0*\L/-0.45*\R,0.5*\L/-0.25*\R} {
    \node[dzero] at (\x,\y) {};
        }
    \node[note, blue!70!black] at (0,-0.7*\R) {D0};

\end{tikzpicture}

    \caption{Schematic depiction of the D2 supertube effect. The world-volume electric field, denoted by wavy lines, and magnetic field, denoted by dots, stabilizes the configuration generating effective angular momentum $J$. The dissolved F1 and D0 charges proportional to the electic and magnetic fuxes respectively.}
    \label{fig:supertube}
\end{figure}

The canonical supertube is a D2-brane in type IIA string theory carrying D0-brane and F1-string charge. One chooses world-volume coordinates $(t,z,\theta)$, where $z$ is the longitudinal direction of the tube and $\theta$ parametrizes its circular cross-section. For a tube of constant radius $R$ embedded in flat space, the induced metric is
\begin{equation}
ds^2_{\rm ind} = -dt^2 + dz^2 + R^2 d\theta^2.
\end{equation}
The world-volume field strength is taken to be
\begin{equation}
F = E\, dt\wedge dz + B\, dz\wedge d\theta,
\label{eq:supertubeF}
\end{equation}
where $E$ is an electric field along the tube and $B$ is a magnetic field on the $(z,\theta)$ directions. The electric field dissolves fundamental-string charge into the D2-brane, whereas the magnetic flux dissolves D0-brane charge.

The dynamics is governed by the Dirac--Born--Infeld action,
\begin{equation}
S_{\rm DBI}
=
-T_2 \int dtdzd\theta\,
\sqrt{-\det(g+\lambda F)},
\qquad
\lambda = 2\pi\alpha',
\end{equation}
and for the ansatz above, the Lagrangian density becomes
\begin{equation}
\mathcal{L}
=
-T_2
\sqrt{R^2(1-\lambda^2 E^2)+\lambda^2 B^2}.
\label{eq:supertubeL}
\end{equation}
The momentum conjugate to the electric field is the electric displacement
\begin{equation}
\Pi \equiv \frac{\partial \mathcal{L}}{\partial E}
=
\frac{T_2 \lambda^2 E R^2}
{\sqrt{R^2(1-\lambda^2 E^2)+\lambda^2 B^2}},
\label{eq:supertubePi}
\end{equation}
which measures the F1 charge density dissolved in the D2-brane. Likewise, the magnetic flux measures the D0 charge density, as we have see in Section \ref{sec:bound}. More precisely one has
\begin{equation}
Q_{\rm F1}= \fr{1}{2\p \a'}\int_0^{2\p} d\theta\, \Pi,
\qquad
Q_{\rm D0} = \fr{T_0}{2\p} \int_0^{2\p} d\theta\, B.
\end{equation}

To check energy of the configuration against the BPS bound we write the Hamiltonian density, that is
\begin{equation}
\mathcal{H}
=
\Pi E - \mathcal{L}
=
\frac{1}{R}
\sqrt{
\left(T_2^2 R^2 + \frac{\Pi^2}{\lambda^2}\right)
\left(R^2+\lambda^2 B^2\right)
}.
\label{eq:supertubeH}
\end{equation}
It follows  that for $B>0$ and $\Pi>0$ that the energy density satisfies the BPS bound
\begin{equation}
\mathcal{H}
\ge
\frac{\Pi}{\lambda}+T_2\lambda B,
\label{eq:supertubeBPS}
\end{equation}
with saturation if and only if
\begin{equation}
T_2 R^2 = \Pi B.
\label{eq:supertubeRadius}
\end{equation}
At the bound, the energy density becomes
\begin{equation}
\mathcal{H}_{\rm BPS}
=
\frac{\Pi}{\lambda}+T_2\lambda B,
\label{eq:supertubeBPSenergy}
\end{equation}
that is exactly the sum of the F1 and D0 contributions. Equivalently, one finds that the electric field reaches the critical value
\begin{equation}
|\lambda E|=1.
\label{eq:supertubeCriticalE}
\end{equation}

The physical interpretation is simple: the D2-brane tension tends to shrink the tube, but the world-volume electric and magnetic fields generate an angular momentum density along the cross-section, proportional to the Poynting vector of the world-volume gauge theory. This angular momentum stabilizes the tube at the finite radius \eqref{eq:supertubeRadius} and the corresponding brane configuration is schematically depicted on Fig. \ref{fig:supertube}. The resulting configuration is supersymmetric and its energy is entirely accounted for by the dissolved D0 and F1 charges. The D2-brane itself contributes only a dipole charge and does not appear as an independent conserved asymptotic charge. Note, that the circular profile chosen here is the simplest representative of a wider class. One of the remarkable features emphasized already in the original analysis \cite{Mateos:2001qs} is that supersymmetric supertubes admit more general profile functions. 

It is worth to note, that supertubes are naturally interpreted as dielectric branes because they realize the same pattern as the Myers effect: lower-dimensional charges expand into an extended higher-dimensional object carrying only a dipole charge of the latter. In the supertube case, D0-brane and F1-string charge are dissolved into the world-volume fields of a D2-brane, whose tubular shape is dynamically sustained by those fluxes. As in the dielectric D0 $\to$ D2 system, one may think of the higher-dimensional brane as emerging from the collective dynamics of lower-dimensional constituents. In this sense, the supertube is a polarized brane configuration.

However, at the same time, supertubes are more special than generic dielectric branes for several reasons. First, they are typically BPS and preserve supersymmetry, so their energy saturates a protected bound and depends only on the conserved D0 and F1 charges. Second, their stabilization does not require an external R-R field: it is achieved by internal world-volume fluxes and the associated angular momentum. Third, they are usually most naturally described in an abelian world-volume language rather than through non-Abelian matrix coordinates. Finally, they admit nontrivial moduli spaces of profiles, so the same conserved charges can be realized by a large family of distinct supersymmetric geometries. These properties make supertubes especially tractable and especially useful in applications to black-hole microphysics.

\section{Conclusions}
\label{sec:concl}

In this review we considered branes of string theory in  three different though complementary manifestations. 
First, Dirichlet (or D-)branes in the string perturbation theory are visible as surfaces of open string ends interacting with specific closed string states. In particular, we learned that Dp-branes carry charges with respect to the R-R fields of supergravity. This allowed to observe their second manifestation, that is as classical solutions of supergravity, where branes are described by charged, extended backgrounds supported by harmonic functions and antisymmetric tensor gauge fields. This point of view reveals complete analogy between branes of string theory and extremal Reissner--Nordstr\"om black holes: they carry conserved charges, saturate BPS bounds, and admit no-force conditions characteristic of supersymmetric states \cite{Horowitz:1991cd,Duff:1994an,Stelle:1998xg}. Finally, the third way to look at the branes is as at dynamical objects endowed with world-volume actions of Dirac--Born--Infeld and Wess--Zumino type, whose kinetic terms encode their geometry and fluctuations while the topological couplings explain how they source and absorb RR and NSNS fluxes \cite{Polchinski:1995mt,Johnson:2003gi}. Such a complicated description, from three different viewpoints, is a consequence of the fact that branes are essentially non-perturbative objects with tension and charge proportional to negative powers of $g_s$, the string coupling constant. Therefore, in the string perturbation theory, that is when $g_s \to 0$, branes simply freeze and no dynamics can be observed. 
To develop understanding of interaction between branes one has to translate between these three viewpoints and to take into account as much information as possible from all three. In Section \ref{sec:interact} we consider several examples of effects, that manifest such non-trivial interaction between branes. 

Brane bound states, considered in Sections \ref{sec:bound} and \ref{sec:intersect} represent the broadest class of interaction effects. Bound states such as Dp-F1, Dp-Dq, NS5-Dp etc.,  demonstrate that multiple charges can be carried by a single BPS configuration, either as explicit intersecting constituents or as dissolved world-volume fluxes \cite{Witten:1995im,Tseytlin:1996bh,Lu:1999uca}. In supergravity brane bound states provide the natural starting point for black-brane solutions, duality chains, and microscopic entropy counting. In gauge theory they serve as the fundamental microscopic description of non-commutative gauge theories, and defect sectors. In particular, near-critical electric-field limits of Dp-F1 bound states gave rise to non-commutative open-string theories (NCOS), clarifying how open-string sectors can survive decoupling limits with intrinsically stringy dynamics \cite{Seiberg:2000ms}. Closely related limits lead to non-relativistic closed-string theories (NRCS), where wound or near-critical string sectors dominate \cite{Gomis:2000bd}. In this context, the Dp-F1 bound states play the role of the natural D-brane-like charged objects of the non-relativistic theory, much as ordinary Dp-branes do in relativistic string theory. This viewpoint has been developed further in later formulations of non-relativistic string theory and its dualities, including work \cite{Harmark:2017rpg,Barakin:2025jwp,Barakin:2026mxz}. Therefor one should think about brane bound states as various example of many composite solutions. Instead, they define specific phases of string theory, special decoupling limits with effective notions of brane charge.

Using the supergravity and world-volume descriptions of branes it is possible to have a grasp on how brane can be created in (adiabatic) dynamics processes.  The Hanany--Witten effect is probably the most well known example of such a behavior \cite{Hanany:1996ie}. In Section \ref{sec:hanany-witten} we considered in details how a D2-brane is created when NS5 and D4 branes cross. The process is best understood either in terms of the Dirac surface of the NS5-brane or, more fundamentally, in terms of Page charges and modified Bianchi identities. Its main applications have been in particle physics and gauge-theory model building. In particular, it provided the geometric foundation for brane engineering of three- and four-dimensional supersymmetric gauge theories, Seiberg dualities, defect constructions, and quiver dynamics \cite{Hanany:1996ie,Witten:1997sc,Elitzur:1997fh,Giveon:1998sr}. In phenomenological settings, related brane-engineering ideas underlie intersecting-brane constructions with chiral matter and semi-realistic gauge sectors \cite{Uranga:2003pz,Blumenhagen:2005mu}. 

Two examples where interaction between brane and the background or world-volume fluxes results in a change of the ground state of the system are the Myers effect and the supertube effects. These where considered in Sections \ref{sec:myers} and \ref{sec:supertubes} respectively. The dielectric brane polarization, or the Myers effect \cite{Myers:1999ps}, states that in the presence of background fluxes, a stack of lower-dimensional D-branes can expand into a higher-dimensional non-commutative brane, with matrix-valued transverse coordinates realizing fuzzy geometry. In gauge/string duality the Myers effect underlies the polarization mechanisms of the Polchinski--Strassler type and helps to explain how fluxes, masses, and non-Abelian vacua are encoded geometrically \cite{Polchinski:2000uf}. In black-hole and AdS/CFT contexts it is closely related to giant gravitons and expanded branes stabilized by flux and angular momentum \cite{McGreevy:2000cw}. In cosmology and flux compactifications, dielectric polarization appears in brane-flux annihilation, metastable anti-brane states, and warped brane inflation scenarios, where world-volume and flux couplings strongly affect vacuum structure and tunneling channels \cite{Kachru:2002gs,Kachru:2003sx,Baumann:2014nda}. From the viewpoint of model building, the Myers effect also provides a natural mechanism for generating fuzzy extra dimensions and non-commutative internal structures in effective field theories descending from branes.

Supertubes originally found in \cite{Mateos:2001qs} as D2 configurations carrying dissolved D0 and F1 charge, are supersymmetric tubular bound states, whose stability follows from a balance between electric and magnetic world-volume fluxes. Such configuration appear to be crucially important  in the context of black-hole microphysics: supertubes can carry multiple conserved charges while admitting arbitrary profiles, and they therefore provide an important bridge between microscopic brane states and smooth, horizonless supergravity solutions \cite{Emparan:2001ux,Bena:2007kg}. In later developments, supertube transitions became a basic ingredient of microstate geometry programs, bubbling solutions, and fuzzball constructions. They also provide useful examples for world-volume supersymmetry, charge dissolution, and nontrivial topology in brane systems. 

The non-perturbative effects just summarized illustrate  more general principles of the non-perturbative dynamics of string theory. Hanany--Witten transitions show that the number of branes can change under smooth motions because the relevant conserved quantities are Page charges rather than naive source counts. The Myers effect shows that even the dimensionality of an effective brane configuration can become dynamical in flux backgrounds. Supertubes show that geometry and charge can reorganize into tubular or bubbling structures that are invisible in a purely point-particle intuition. Bound states show that lower-dimensional charges may be dissolved into flux and that the distinction between separate constituents and a single composite object is often frame-dependent. Together, these mechanisms clearly demonstrate the point, that string theory is much richer than what one sees at the perturbative level from the quantized Polyakov action. It is indeed a theory of not only strings but also of objects of other dimensions, that are not only described by dynamical actions, but can transform into each other, be created and annihilated.

This understanding can be naturally pushed forward to the level of M-theory, where no small coupling is present and whose dynamics is essentially non-perturbative. In this picture the Type IIA string theory appears as a particular phase of M-theory in which one spatial direction is compactified on a circle of radius $R_{11}$. In the weakly coupled Type IIA regime this radius is small in string units, and the effective ten-dimensional description is governed by the string coupling
\begin{equation}
R_{11}=g_s\, l_s ,
\qquad
l_p^3 = g_s\, l_s^3 ,
\end{equation}
with $l_s$ the string length and $l_p$ the eleven-dimensional Planck length \cite{Witten:1995ex,Townsend:1995kk}. In this relation one sees directly why the strong-coupling limit of Type IIA opens up an extra dimension. The fundamental objects of M-theory are the M2 and M5 branes and the KK6-monopole, which in the Type IIA phase give rise to the familiar fundamental string and branes. An M2-brane transverse to the KK circle reduces to a D2-brane, while an M2-brane wrapped on the circle reduces to the fundamental Type IIA string. An M5-brane transverse to the circle becomes an NS5-brane, whereas an M5-brane wrapped on the circle becomes a D4-brane. Momentum along the circle becomes D0-brane charge, and the KK6-monopole reduces to the D6-brane when the Taub--NUT circle is along the KK circle, and to the KK5-monopole when it is transverse \cite{Witten:1995ex,Townsend:1995kk}. Upon this reduction the membranes of M-theory, initially standing on the same footing, become objects that differ by tension in the Type IIA theory. This unification remains one of the deepest reasons why branes occupy such a central place in string theory: they tie together non-perturbative states, classical backgrounds, world-volume dynamics, dualities, and the emergence of spacetime itself \cite{Duff:1994an,Stelle:1998xg}.

\newpage
\bibliography{bib.bib}
\bibliographystyle{utphys.bst}

\end{document}